\documentclass[twocolumn]{aastex7}

\usepackage{CJK}

\defcitealias{Kerr21}{SPYGLASS-I}
\defcitealias{Kerr22b}{SPYGLASS-III}
\defcitealias{Kerr22a}{SPYGLASS-II}
\defcitealias{Kerr23}{SPYGLASS-IV}
\defcitealias{Kerr24}{SPYGLASS-V}
\defcitealias{Kerr25}{SPYGLASS-VI}
\defcitealias{Kerr25c}{SPYGLASS-VII-B}

\shorttitle{Demographics and Ages of Nearby Low-Mass Associations}
\shortauthors{Kerr et al.}
\graphicspath{{./}{figures/}}

\begin{document}

\begin{CJK*}{UTF8}{gbsn}

\title{SPYGLASS. VII-A.  The Demographics and Ages of Small Nearby Young Associations}

\correspondingauthor{Ronan Kerr}
\email{ronan.kerr@utoronto.ca}

\author[0000-0002-6549-9792]{Ronan Kerr}
\affiliation{Dunlap Institute for Astronomy \& Astrophysics, University of Toronto,
Toronto, ON M5S 3H4, Canada\\}
\affiliation{Department of Astronomy, University of Texas at Austin, 2515 Speedway, Stop C1400, Austin, Texas, USA 78712-1205\\}
\email{ronan.kerr@utoronto.ca}

\author[0000-0002-4128-7867]{Facundo Pe\'rez Paolino}
\affiliation{Department of Astronomy, California Institute of Technology, 1216 East California Blvd, Pasadena, CA 91125, USA\\}
\email{fperezpa@caltech.edu}

\author[0000-0002-3389-9142]{Jonathan C. Tan}
\affiliation{Dept. of Space, Earth \& Environment, Chalmers University of Technology, Gothenburg, Sweden}
\affiliation{Dept. of Astronomy \& Virginia Institute for Theoretical Astronomy, University of Virginia, Charlottesville, VA, USA}
\email{jctan.astro@gmail.com}

\author[0000-0003-2573-9832]{Joshua S. Speagle(沈佳士\ignorespacesafterend)}
\affiliation{Dunlap Institute for Astronomy \& Astrophysics, University of Toronto,
Toronto, ON M5S 3H4, Canada\\}
\affiliation{Department of Statistical Sciences, University of Toronto\\ 9th Floor, Ontario Power Building, 700 University Ave, Toronto, ON M5G 1Z5, Canada}
\affiliation{David A. Dunlap Department of Astronomy \& Astrophysics, University of Toronto\\ 50 St George Street, Toronto, ON M5S 3H4, Canada}
\affiliation{Data Sciences Institute, University of Toronto\\ 17th Floor, Ontario Power Building, 700 University Ave, Toronto, ON M5G 1Z5, Canada} 
\email{j.speagle@utoronto.ca}

\author[0000-0001-9811-568X]{Adam L. Kraus}
\affiliation{Department of Astronomy, University of Texas at Austin, 2515 Speedway, Stop C1400, Austin, Texas, USA 78712-1205\\}
\email{alk@astro.as.utexas.edu}

\author[0000-0003-3526-5052]{Jos\'e G. Fern\'andez-Trincado}
\affiliation{Instituto de Astronom\'ia, Universidad Cat\'olica del Norte, Av. Angamos 0610, Antofagasta, Chile}
\email{jose.fernandez@ucn.cl}

\author[0000-0002-3481-9052]{Keivan G. Stassun}
\affiliation{Department of Physics and Astronomy, Vanderbilt University, Nashville, TN 37235, USA\\}
\email{keivan.stassun@vanderbilt.edu}

\author[0000-0003-2481-4546]{Julio Chanam\'e}
\affiliation{Instituto de Astrof\'isica, Pontificia Universidad Cat\'olica de Chile, Avenida Vicu\~na Mackenna 4860, 782-0436 Macul, Santiago, Chile}
\email{jchaname@astro.puc.cl}



\begin{abstract}

Recent Gaia-based young stellar association surveys have revealed dozens of low-mass populations that have, until recently, been too small or sparse to detect. These populations represent a largely unstudied demographic with unknown origins, and their relative isolation may minimize gravitational disruptions that impact traceback, making them compelling targets for dynamical studies. In this paper, we survey 15 of these isolated young associations for the first time: Andromeda South (SCYA-97), Aquila East, Aries South (SCYA-104), Cassiopeia East (SCYA-43), Canis Major North, Leo Central (SCYA-2), Leo East (SCYA-3), Theia 72, Ophiuchus Southeast, Scutum North (SCYA-70), Taurus-Orion 1 (TOR1), Theia 78, Vulpecula East (UPK 88), SCYA-54, and SCYA-79. By combining Gaia astrometry and photometry with new ground-based spectroscopic measurements, we assess the membership of each population, search for substructure, analyze their demographics, and compute ages. We find that the smallest populations in our sample contain $<20$ $M_{\odot}$ of stellar mass, making them the smallest associations ever detected. Four host substantial substructure, including TOR1, where we discover TOR1B, a new \added{16} $M_{\odot}$ association with radial velocities inconsistent with an origin in the parent complex. Using PARSEC isochrones, we produce self-consistent ages for all populations supported by dynamical and lithium depletion ages, which range from 6.9 $\pm$ 0.5 Myr in \added{TOR1A} to 42.8 $\pm$ 2.4 in AndS. Our results provide the first detailed overview of the properties of these populations, characterizing a largely unknown category of young associations that may have an important role in tracing the processes that guide local star formation. 

\end{abstract}



\section{Introduction} \label{sec:intro}

Most nearby young stars are not found in isolation, but rather in co-moving star clusters and associations \citep{deZeeuw99, Krumholz19}. These populations inherit motions from their parent molecular cloud, and as such their positions, velocities and ages produce a record of the positions and dynamics of star-forming clouds long after their dispersal \citep[e.g.][]{Galli23, Swiggum24}. On small scales, young stellar populations are used to detect star formation patterns, such as sequential and triggered star formation, where feedback from early-forming stars or gas collisions may compress gas and trigger new stellar generations \citep{Tan00, Inutsuka15, Posch23}. On larger scales, the locations of these clouds at formation collectively trace the sites of dense gas in our Galaxy over time. An improved census of recent star formation will therefore reveal not just local star formation patterns, but also kpc-scale gas patterns that trace larger structures like spiral arms.

The data releases of the Gaia spacecraft are making the local star formation record more complete than ever before \citep{GaiaMission, GaiaDR322}. Recent \textit{Gaia}-based surveys have substantially deepened our record of nearby associations, ranging from large, heavily substructured stellar complexes to small, seemingly isolated associations \citep[e.g.][]{Kounkel19, Hunt23}. Many of the larger populations in the solar neighborhood have already been studied, often revealing extensive substructure and multi-generational star formation patterns \citep[e.g.,][]{CantatGaudin19,Kerr21,Briceno23, Ratzenbock23,SanchezSanjuan24, Kerr24}.  

However, associations containing less than a few hundred solar masses often lack dedicated studies of their structure, demographics, and ages, especially outside of the nearest 100 pc \citep{deZeeuw99, Gagne18BXIII}. As such, we know little about their typical dynamics and star formation patterns. We also know little about whether they emerge from the kpc-scale gas structures in the solar neighborhood or any clouds that may have preceded their assembly, or whether they form in truly isolated environments. If their formation is isolated, they may be better-suited for precise dynamical studies, having had minimal interactions with particularly massive nearby gas clouds that shift stellar motions. 

By identifying co-moving structures in a sample of photometrically young stars, SPYGLASS-IV \citep{Kerr23} has produced one of the deepest young association surveys to date, revealing 116 young ($<50$ Myr) populations within 1 kpc. Much of this sample is in this largely unknown demographic of low-mass associations, including 10 entirely new populations. These populations are defined by as few as 10 high-confidence photometrically young stars, and as such, they are among the lowest-mass associations ever discovered. They may therefore represent a unique demographic that traces gas structures distinct from the kpc-scale structures that dominate the solar neighborhood \citep[e.g.,][]{Lallement19, Alves20}, which could produce unusual initial mass functions (IMFs) and binarity fractions \citep[e.g.,][]{Moraux07,Conroy12, Grudic23}.  

In this paper, we perform the first comprehensive study of the ages and demographics of 15 low-mass stellar populations identified in \citetalias{Kerr23}, revealing new coherent substructures, and providing the high-level overview necessary to guide follow-up dynamical studies. In Section \ref{sec:data}, we introduce our \textit{Gaia}-based dataset alongside new and existing spectroscopic data. We analyze this data in Section \ref{sec:methods}, assessing stellar membership and identifying substructure. We then compute robust ages for all populations and substructures and measure their masses and demographics in Section \ref{sec:results}. We discuss those results in Section \ref{sec:discussion}, before summarizing in Section \ref{sec:conclusion}. 

\section{Data} \label{sec:data}

\subsection{Initial Population Selection} \label{sec:basesample}

Our initial membership samples are drawn from \citetalias{Kerr23}, which identifies photometrically young stars using a Bayesian framework, and then applies the HDBSCAN clustering algorithm \citep{McInnes2017} to detect overdensities of ten or more photometrically young \textit{Gaia} stars that mark the site of an association. The resulting catalog provides both the photometrically young stars with robust membership used to define each group, in addition to phase space neighbors, which include early-type stars that are no longer on the pre-main sequence. We select target associations based on a combination of proximity, location, and size, requiring a distance within 350 pc, on-sky position accessible to telescopes at subtropical northern latitudes ($\delta > -30$), and a small population defined by $N \le 50$ photometrically young stars. We also require that each population has an age solution $\tau<40$ Myr from \citetalias{Kerr23} to exclude older populations where dynamical ages are less reliable and youth indicators like Lithium are less informative. This produces an initial sample of 19 populations: SCYA-2, SCYA-3, SCYA-26, UPK 88 (SCYA-30), SCYA-35, SCYA-43, Theia 232 (SCYA-54), SCYA-58, Taurus-Orion 1 (SCYA-64), Canis Major North (SCYA-65), Theia 72 (SCYA-66), SCYA-70, SCYA-72, Ophiuchus Southeast (SCYA-75), Aquila East (SCYA-78), SCYA-95, SCYA-97, SCYA-104,  and Theia 78 (SCYA-112). We remove SCYA-95, as it shows substantial field contamination that greatly increases the observing time required to spectroscopically survey its members. Finally, we add Theia 98 (SCYA-79), which passes all restrictions except distance, at $d = 436$ pc, but has a position and velocity that may suggest dynamical connections to the Cep-Her Complex, alongside Theia 232, making its inclusion useful to assess that possibility in future dynamical studies. 

\subsection{Supplemental Data} \label{sec:suppdat}

High-resolution spectroscopy is essential for studying young stellar populations and their star formation histories, providing both radial velocities with uncertainties less than 1 km s$^{-1}$ and equivalent widths for lines that inform youth and stellar age. These observations are therefore required for dynamical and lithium depletion ages, while also confirming the motion of each association in 3D space. In the case of small populations with limited available targets like the 19 we consider here, the availability of observations is a limiting factor for their characterization, as particularly small populations may host few stars where radial velocities are available alongside strong youth indicators. We therefore gather spectroscopic observations of bright members throughout all populations in our sample, and use the resulting database to identify and characterize populations with adequate coverage.

\subsubsection{Literature Values}

The small associations we investigate have not been previously targeted by any dedicated spectroscopic observations. However, several large spectroscopic surveys cover candidate members of our target associations, reporting radial velocities and Li and H$\alpha$ EWs. To identify literature measurements of RVs and line EWs in our target associations, we cross-match our datasets with an updated form of the spectroscopic survey database from \citetalias{Kerr24}, which combines the source compilations from \citet{Anderson12}, \citet{Luhman22}, and \citet{Zerjal23} with Vizier \citep{Vizier00} and Simbad \citep{Wenger00} searches across several associations in the \citetalias{Kerr23} sample. Gaia DR3 \citep{GaiaDR3RVs_Katz23}, GALAH DR4 \citep{Buder24}, the Gaia-ESO Survey \citep{Randich22}, and LAMOST \citep{LAMOSTDR8VAC, LAMOSTSCRV} are the largest and most recent of these surveys, although some more localized datasets also cover candidate members of these associations.

\subsubsection{New Spectroscopy} 

While literature observations exist in our target associations, few candidates have high-quality radial velocities alongside Li and H$\alpha$ equivalent widths. As such, characterizing these associations requires new spectroscopy. We therefore conducted a long-term survey of these associations with the McDonald Observatory's 2.7m Harlan J. Smith Telescope (HJST), with observation dates ranging from summer 2020 to fall 2024. These observations used the Tull Coud\'e Echelle Spectrograph, which provides non-continuous coverage from 3400~\AA~to 10900~\AA~at $R=60000$ \citep{Tull95}. Some early observations used a configuration that prioritized maximizing signal at the expense of the H$\alpha$ line, however most observations use a configuration that covers both the H$\alpha$ and the Li-6708\AA~lines.

We processed and reduced our spectra from HJST using a publicly available pipeline designed for the Tull spectrograph\footnote{\url{https://github.com/dkrolikowski/tull_coude_reduction}}. This pipeline also provides radial velocity measurements, which use spectral line broadening functions from the \texttt{saphires} package \citep{Tofflemire19}. We extract Li and H$\alpha$ equivalent widths by fitting a Gaussian profile the lines, following \citet{Kerr22a}. This method does not consider veiling, or the systematic underestimation of EWs due to excess emission from accretion \citep{Saad25}, but we expect accretion to be uncommon at the ages of these populations. Our other literature sources and the EAGLES package that we use for lithium depletion ages in Section \ref{sec:liages} \citep{Jeffries23} also do not consider veiling, so this decision ensures a consistent analysis.

\begin{deluxetable*}{ccccccccccccccccccccccccc}
\tablecolumns{25}
\tablewidth{0pt}
\tabletypesize{\scriptsize}
\tablecaption{Data for all stars analyzed in this paper, \added{including the RA and Dec coordinates, \citet{BailerJones21} distances, and apparent $g$ magnitudes available through \textit{Gaia}, in addition to }new and literature spectroscopic data, membership and quality flags, and three measures of membership probability.}
\label{tab:spectres}
\tablehead{
\colhead{Gaia ID} &
\colhead{Assn.} &
\colhead{SG} &
\colhead{RA} &
\colhead{Dec} &
\colhead{d} &
\colhead{g} &
\colhead{M} &
\multicolumn{3}{c}{RV (km s$^{-1}$)\tablenotemark{a}} &
\multicolumn{3}{c}{EW$_{Li}$ (\AA)\tablenotemark{a}\tablenotemark{b}} &
\multicolumn{3}{c}{EW$_{H\alpha}$ (\AA)\tablenotemark{a}} &
\multicolumn{4}{c}{Flags \tablenotemark{c}} &
\multicolumn{3}{c}{Membership \tablenotemark{d}} \\
\colhead{} &
\colhead{} &
\colhead{} &
\colhead{(deg)} &
\colhead{(deg)} &
\colhead{(pc)} &
\colhead{} &
\colhead{(M$_{\odot}$)} &
\colhead{val} &
\colhead{err} &
\colhead{src} &
\colhead{val} &
\colhead{err} &
\colhead{src} &
\colhead{val} &
\colhead{err} &
\colhead{src}&
\colhead{$V$} &  
\colhead{Li} & 
\colhead{H$\alpha$} & 
\colhead{F} &  
\colhead{$P_{spat}$} &
\colhead{$P_{sp}$} &
\colhead{$P_{fin}$} \\
}
\startdata
2799017095743054080 & AndS & -1 & 5.6904 & 19.9496 & 163.7 & 19.12 & 0.14 &  &  &  &  &  &  &  &  &  & 0 & 0 & 0 & 0 & 0.067 & 0.0 & 0.0 \\
2799882346970157696 & AndS & -1 & 6.8938 & 22.0631 & 164.8 & 17.39 & 0.24 &  &  &  &  &  &  &  &  &  & 0 & 0 & 0 & 0 & 0.144 & 0.022 & 0.022 \\
2800152620672963968 & AndS & -1 & 4.66 & 21.2949 & 176.0 & 18.28 & 0.21 &  &  &  &  &  &  &  &  &  & 0 & 0 & 0 & 8 & 0.081 &  &  \\
2800952442366616576 & AndS & -1 & 6.385 & 23.8265 & 184.6 & 17.94 & 0.18 &  &  &  &  &  &  &  &  &  & 0 & 0 & 0 & 0 & 0.07 & 0.404 & 0.404 \\
2800965361628641792 & AndS & -1 & 4.9067 & 22.9071 & 171.0 & 18.99 & 0.14 &  &  &  &  &  &  &  &  &  & 0 & 0 & 0 & 0 & 0.43 &  &  \\
2806653242652754048 & AndS & -1 & 9.3791 & 24.7374 & 185.5 & 19.25 & 0.14 &  &  &  &  &  &  &  &  &  & 0 & 0 & 0 & 0 & 0.07 & 0.0 & 0.0 \\
2806734602218348928 & AndS & -1 & 9.8256 & 24.822 & 172.3 & 15.91 & 0.43 &  &  &  &  &  &  &  &  &  & 0 & 0 & 0 & 0 & 0.189 & 0.042 & 0.042 \\
2807197668412073472 & AndS & -1 & 6.9167 & 24.6447 & 158.0 & 17.23 & 0.22 &  &  &  &  &  &  &  &  &  & 0 & 0 & 0 & 0 & 0.091 & 0.273 & 0.273 \\
2807636133033582464 & AndS & -1 & 9.5937 & 26.075 & 174.0 & 17.94 & 0.2 & 19.86 & 6.65 & BOSS &  &  &  & 7.64 & 1.69 & BOSS & 0 & 0 & 0 & 0 & 0.29 & 0.206 & 0.206 \\
2807673898681458304 & AndS & -1 & 8.7561 & 26.0849 & 169.4 & 17.55 & 0.23 &  &  &  &  &  &  &  &  &  & 0 & 0 & 0 & 0 & 0.262 & 0.023 & 0.023 \\
2807845005882860672 & AndS & -1 & 7.7592 & 26.3164 & 161.5 & 18.63 & 0.14 &  &  &  &  &  &  &  &  &  & 0 & 0 & 0 & 0 & 0.479 &  &  \\
2808165208580376320 & AndS & -1 & 11.9928 & 25.7936 & 178.4 & 15.95 & 0.43 &  &  &  &  &  &  &  &  &  & 0 & 0 & 0 & 0 & 0.073 & 0.053 & 0.053 \\
\enddata
\tablenotetext{a}{The source of the measurement. The codes used are as follows: McDonald Observatory 2.7m Harlan J. Smith Telescope (HJST), Gaia DR3 (GDR3), SDSS APOGEE spectrograph \citep{APOGEE} (APOG), SDSS BOSS spectrograph \citep{BOSS} (BOSS), DESI DR1 \citep{DESIDR1} (DESI), GALAH DR4 \citep{Buder24} (GAL4), LAMOST DR9 MRS spectra \citep{LAMOSTSCRV} (LAM9M), LAMOST DR8 LRS parameters of A, F, G, and K stars (LAM8L), LAMOST DR8 Catalog of M Stars (LAM8LMCAT), LAMOST DR8 Value Added Catalog \citep{LAMOSTDR8VAC} (LAM8LV), \citet{Famaey05} (F5),  \citet{Torres06} (T6), \citet{Gontcharov06} (G6), \citet{Grieves18} (G18), \citet{Zerjal21} (Z21)}
\tablenotetext{b}{Reported values include the correction for Fe I line deblending described in Section \ref{sec:specyouth}}
\tablenotetext{c}{Flags include: $V$, velocity membership assessment, where 1 is a kinematic member, 0 is inconclusive, and -1 is a non-member; Li, Lithium membership assessment, where 1 is a member, 0 is inconclusive, and -1 is a non-member; H$\alpha$, H$\alpha$ membership assessment, where 0 is inconclusive, and -1 is a non-member; and $F$, General flag, where 1 indicates a star with a  resolved companion within 10,000 au in the plane of the sky, 2 indicates a bad broadening function solution, 4 indicates a bimodal line profile likely indicative of spectroscopic binarity, 8 indicates an RUWE$>$1.2, indicating likely unresolved binarity, and 16 indicates that the RV recorded was ambiguously attributed to two components of a binary pair. General flags are when multiple are true; for example, flag 6 indicates both flags 2 and 4.} 
\tablenotetext{d}{Membership probabilities provided are: $P_{spat}$, the spatial membership probability, previously referred to as $P_{mem}$ in \citetalias{Kerr23}, compares number of nearby young and old stars; $P_{sp}$, the spatial-photometric membership probability, defined by correcting the prior on $P_{Age<50Myr}$ from \citetalias{Kerr23} to align with $P_{spat}$; and $P_{fin}$, the final membership probability, which includes radial velocity and spectroscopic youth indicators}. 
\vspace*{0.1in}
\end{deluxetable*}

Early HJST observations focused on brighter members ($m_G < 12$) with no restrictions, while later observations covered stars with $m_G \la 15$, de-prioritizing objects with either Gaia radial velocities inconsistent with the population's mean radial velocity if known, or evidence for unresolved binarity. The presence of a companion introduces a velocity dispersion internal to the system as well as a flux contribution that can warp Lithium equivalent widths, and fitting the bulk system motion requires multi-epoch observations. Binaries are therefore slow to survey and less useful for the age analyses we perform in this paper (see Sec. \ref{sec:binaries}), but still important for the demographic analysis in Section \ref{sec:demographics}. SCYA-26, SCYA-35, SCYA-58, and SCYA-72 have few enough bright members that their RV peaks are unclear in the Gaia data, reducing the efficiency of our survey work there. We observed some of the brighter candidates in these populations but have yet to find strong youth indicators, so a deeper follow-up survey will be necessary to characterize them. We therefore exclude these populations from this work, leaving 15 well-covered associations.

Finally, we supplement our dataset with RVs and spectral line measurements from SDSS-V \citep{Kollmeier25}, which operates 2.5m telescopes at the Apache Point and Las Campanas observatories \citep{DuPontTelescope, SDSSTelescope}. These instruments provide high-quality radial velocities through the $R\sim22,500$ near-infrared APOGEE spectrograph \citep{APOGEE}, as well as Li-6708\AA~and H$\alpha$ spectral line measurements from the BOSS spectrograph, which provides $R\sim2000$ with good throughput for 3650-9500\AA~\citep{BOSS}. The observations from HJST dominate our dataset, although APOGEE in particular contributes many new spectra through the Milky Way Mapper (MWM) survey of young stellar objects \citep{Kounkel23}. Some of the data we present here are available through SDSS DR19 \citep{SDSSDR19} and other previous data releases \citep[e.g.,][]{APOGEEDR17}, while we have early access to the rest through the SDSS-V collaboration. 

\subsubsection{Data Combination}

Following \citetalias{Kerr24}, we use the lowest-uncertainty measurement in cases where multiple sources are available for the same star. We de-prioritize Gaia DR3, RAVE DR6, and LAMOST LRS radial velocities, as these have been shown to have discrepancies from ground based RV measurements of young stars of order 1 km s$^{-1}$ that are not acknowledged by the reported uncertainties \citep{Kounkel23, Kerr24}. Non-deprioritized observations are always selected over deprioritized sources if both have sub-km s$^{-1}$ RV measurements. Finally, using the Gaia DR3-2MASS best neighbor cross-match, we add 2MASS K-band photometry to our sample, which will be useful for upcoming Lithium depletion boundary studies \citep{Skrutskie06}. We compile our complete dataset for all 15 target populations in Table \ref{tab:spectres}.

\section{Methods \& Analysis} \label{sec:methods}

\subsection{Membership} \label{sec:membership}

Our membership assessment in these populations largely follows the probabilistic approach used in \citetalias{Kerr24}, where some steps discussed here are explained in more detail. This approach combines existing spatial and photometric membership assessments with spectroscopic membership indicators like radial velocities and Li EWs to produce aggregate membership probabilities, which we use to separate probable members from background contamination. 

\subsubsection{Initial Selection}

Our final selection of young associations with adequate spectroscopic coverage contains the following 15 populations: SCYA-2, SCYA-3, UPK 88 (SCYA-30), SCYA-43, Taurus-Orion 1 (TOR1, SCYA-64), Canis Major North (CMaN, SCYA-65), Theia 72 (SCYA-66), SCYA-70, Ophiuchus Southeast (OphSE, SCYA-75), Aquila East (AqE, SCYA-78), SCYA-97, SCYA-104,  Theia 78 (SCYA-112), Theia 98 (SCYA-79), and Theia 232 (SCYA-54). For ease of reference, we assign new names to populations with distinct on-sky positions that lack common names. Following the tradition of naming populations after their parent constellations, we refer to SCYA-2 as Leo Central (LeoC), SCYA-3 as Leo East (LeoE), SCYA-30 as Vulpecula East (VulE), SCYA-43 as Cassiopeia East (CasE), SCYA-70 as Scutum North (ScuN), SCYA-97 as Andromeda South (AndS), and SCYA-104 as Aries South (AriS). 

Many of these populations have a pre-main sequence that separates cleanly from the field. In these cases we use the full \citetalias{Kerr23} sample without restriction. However, other populations have stellar distributions that are more intertwined with the field, and this is marked by low values of $P_{spatial}$, a metric from \citetalias{Kerr23} that measures the local fraction of young stars compared to old stars. In these contaminated populations, stars with low $P_{spatial}$ can dominate the color-magnitude diagram (CMD) such that overluminous field stars like binaries fill the gap between the field and pre-main sequence. In these cases, we remove stars with low $P_{spatial}$ to suppress this potentially 
dominant source of contamination. We therefore apply restrictions to Aries South ($P_{spatial}>0.1$) and Taurus-Orion 1  ($P_{spatial}>0.15$), as well as SCYA-54 and SCYA-79, where we use the $P_{spatial}>0.2$ cut that \citetalias{Kerr24} used in the adjacent and similarly-aged Cep-Her complex. We do not remove stars that were identified as founding photometrically young members of the association by HDBSCAN in \citetalias{Kerr23} regardless of their $P_{spatial}$. The remaining sample is used for analysis, and comprises 4244 stars across 15 populations.  

\subsubsection{Spatial-Photometric Membership Probabilities}

\citetalias{Kerr23} provides photometric youth probabilities, $P_{Age<50 Myr}$, in addition to spatial membership probabilities, $P_{spatial}$, which estimate the local membership fraction using the ratio between the number of near-certain young and old stars. $P_{spatial}$ is available for all stars in the candidate member lists in \citetalias{Kerr23}, while $P_{Age<50 Myr}$ is available for a subset of stars that pass the photometric and astrometric quality cuts used in that publication. \citetalias{Kerr23} calculates $P_{Age<50 Myr}$ assuming the demographics of the field, which is taken to have a constant star formation rate over the last 11.2 Gyr \citep{Binney00}. This implies that 0.4\% of stars are younger than 50 Myr, however associations contain young stars at a much higher rate than the field, and that rate is measured by $P_{spatial}$. We therefore produce spatial-photometric membership probabilities, $P_{sp}$, which readjust the initial youth probability prior for a given value of $P_{spatial}$, following the routine presented in \citetalias{Kerr24} and also employed in \citetalias{Kerr25}.

Contaminants in CasE necessitate adjustments to our probability assessments, as the field sequence there is elevated slightly above the typical field sequence seen in other populations. This may be caused by the presence of one or more older open clusters that share the parameter space and are younger than the rest of the field. The most probable contaminant population is HSC 1040 \citep[see][]{Hunt23}, which overlaps with CasE in spatial coordinates and is within 3 mas/yr in proper motion space. To avoid potential contamination from these sources, we apply a manual cut, which we present in Section \ref{sec:finalmembership}, and set $P_{sp}$ for all sources below it to zero. Due to the youth of CasE, most stars excluded by this cut are more than a magnitude below the association's PMS, so the risk of false rejections due to this cut is low.

\subsubsection{Velocity Membership} \label{sec:velmem}

\citetalias{Kerr23} selected association members in 5-D space-transverse velocity coordinates, and thus, stars with consistent positions and transverse velocities but inconsistent radial velocities may be marked as candidate members. We therefore use the new and literature radial velocities to assess membership. 

Our RV coverage is often incomplete and contaminated with field stars, so using the distribution of those measurements to set our membership selection risks significant over or under-estimates of the true size of the distribution. We therefore use the 2-D transverse velocity distributions to predict the overall velocity distribution. To do this, we compute a median transverse velocity in $l$ and $b$ across stars with $P_{sp}>0.5$, and compute median UVW galactic cartesian velocities using stars with $P_{sp}>0.5$ and $\sigma_{RV}<1.5$ km s$^{-1}$. We use RA/Dec rather than $l$/$b$ for LeoE and LeoC due to their proximity to the north galactic pole. Assuming that the maximum extent of the population in 3-D space is the same as the clustering-defined maximum extent in 2-D space, we take the difference between the median transverse velocity and the transverse velocity of the most outlying candidate member as the radius within which stars are considered members. This produces a relatively generous membership determination that will produce some false positives, which we account for in Section \ref{sec:falseposneg}.

Stars within this radius of the UVW median are marked as velocity members ($V=1$), and stars outside this radius are marked as non-members ($V=-1$). We leave stars with no RV or an RV uncertainty greater than half of that search radius as inconclusive kinematic members ($V=0$), as they lack sufficient observational precision for membership assessment. 

\subsubsection{Velocity False Positives and Negatives} \label{sec:falseposneg}

Following \citetalias{Kerr24}, we estimate false positive and false negative rates for velocity membership. These quantities can have a substantial influence on the overall membership probability of candidate members, especially on the main sequence, where field contamination is most common. 

\begin{figure*}
\centering
\begin{tabular}{ccccc}
    \includegraphics[height=3.1cm]{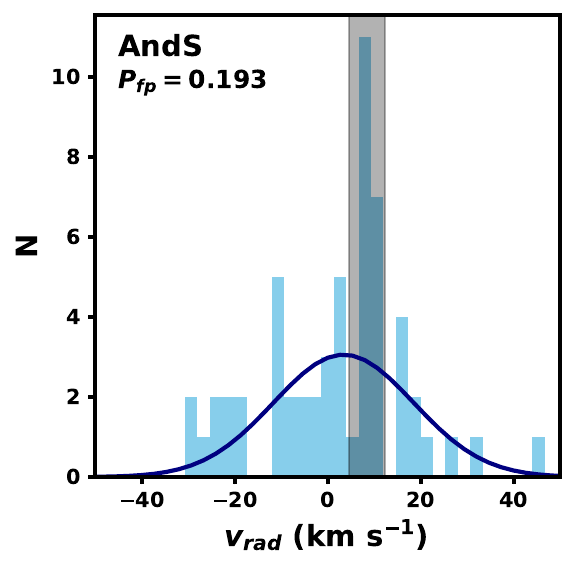}
    & \includegraphics[height=3.1cm]{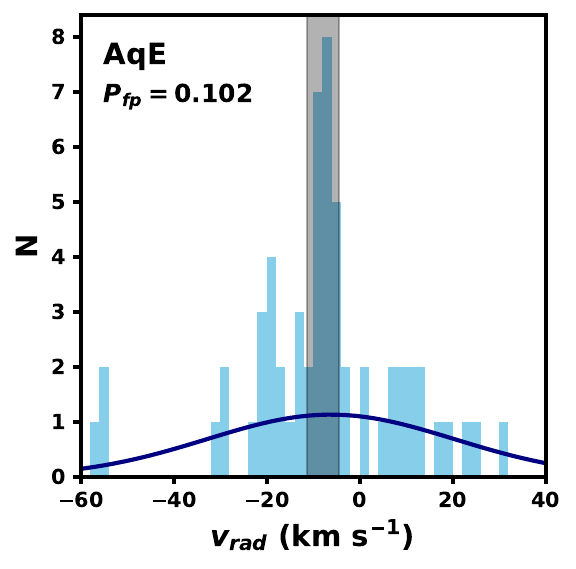}
    & \includegraphics[height=3.1cm]{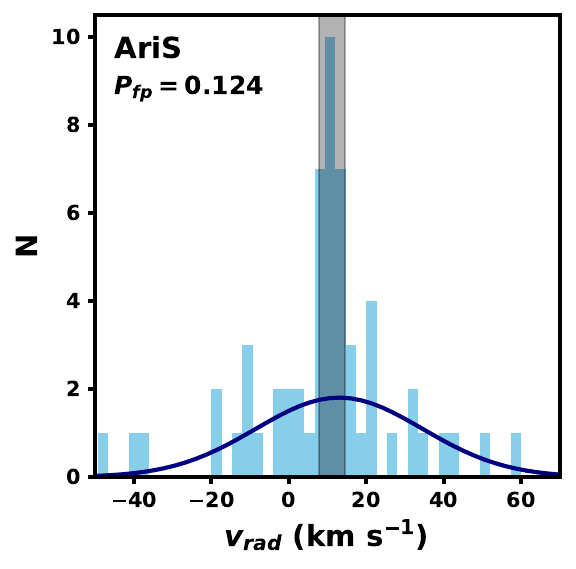}
    & \includegraphics[height=3.1cm]{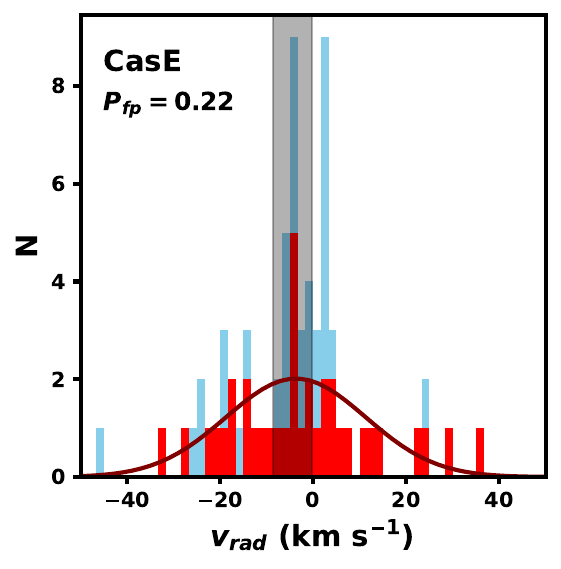}
    &\includegraphics[height=3.1cm]
    {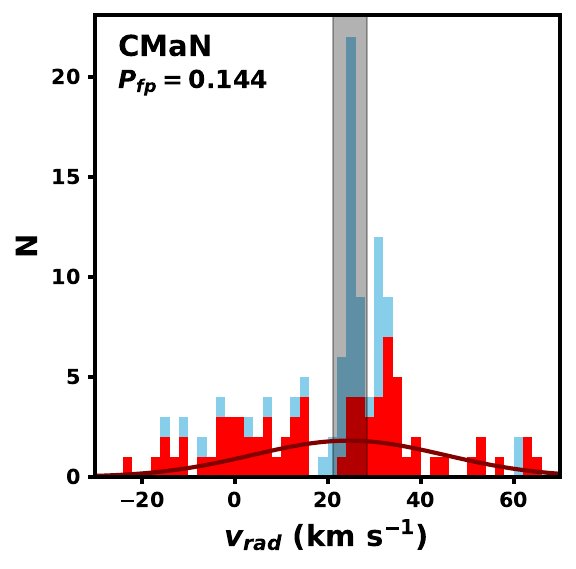}\\[-4pt]
    \includegraphics[height=3.1cm]{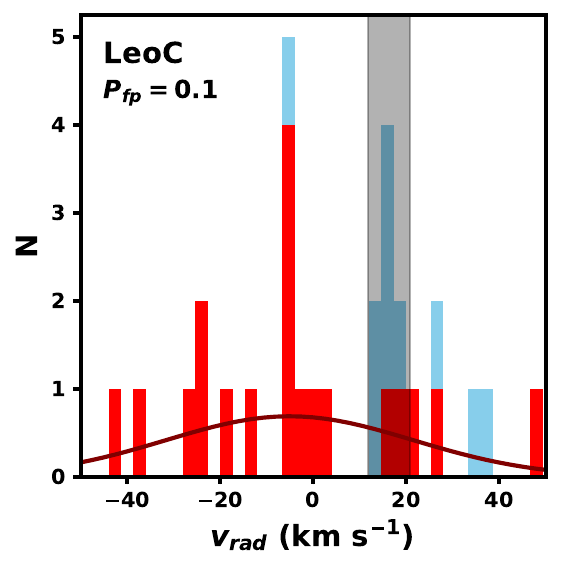}
    & \includegraphics[height=3.1cm]{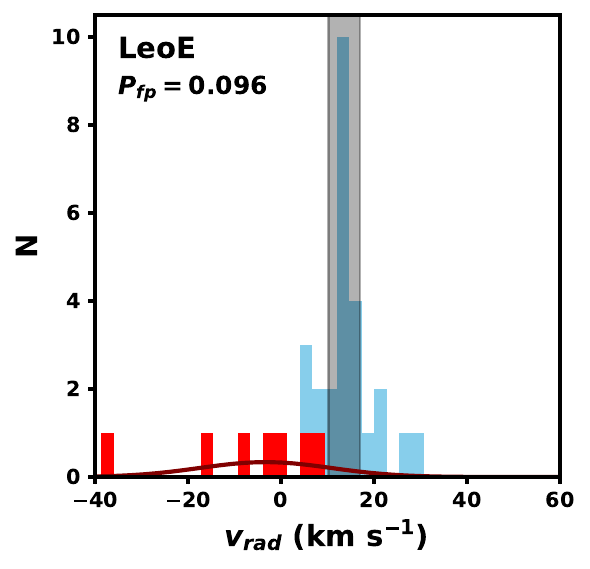}
    &\includegraphics[height=3.1cm]{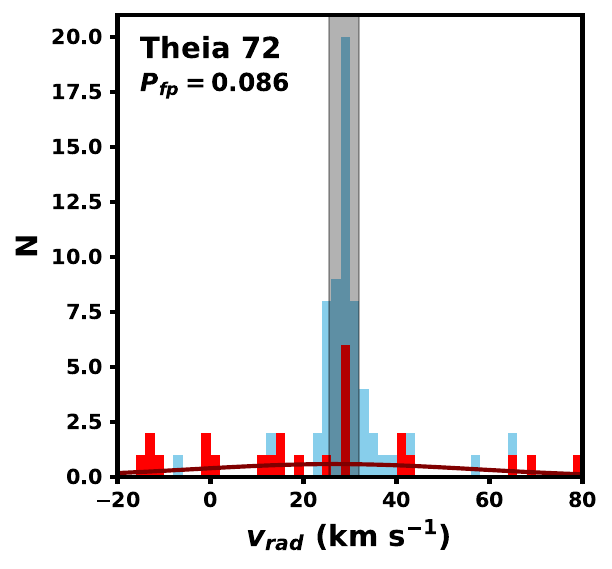}
    & \includegraphics[height=3.1cm]{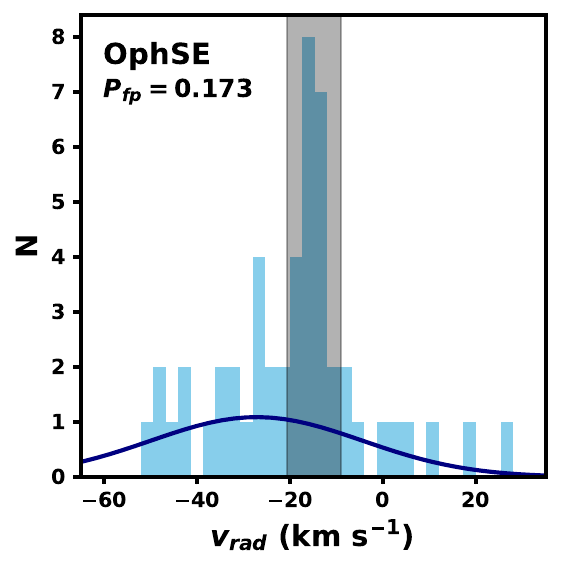}
    & \includegraphics[height=3.1cm]{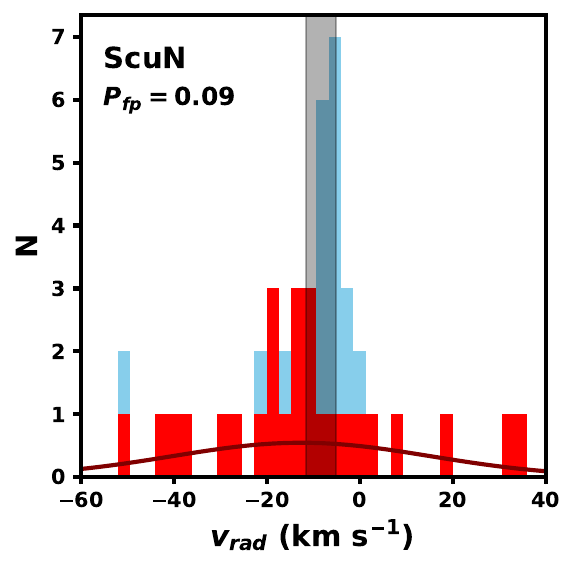} \\[-4pt]
    \includegraphics[height=3.1cm] {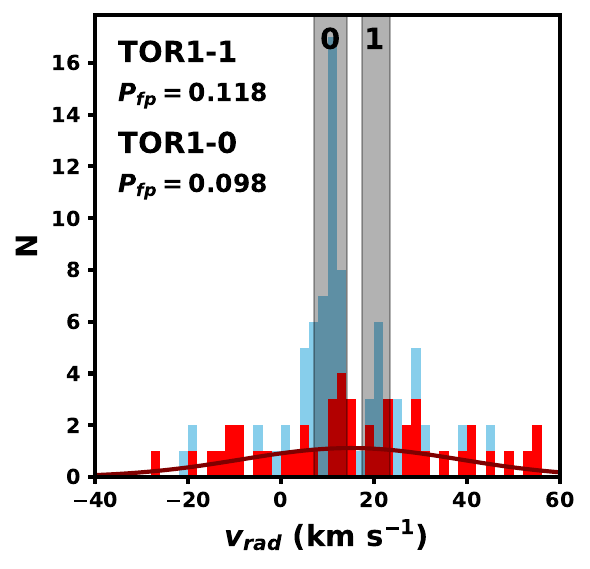}
    & \includegraphics[height=3.1cm]{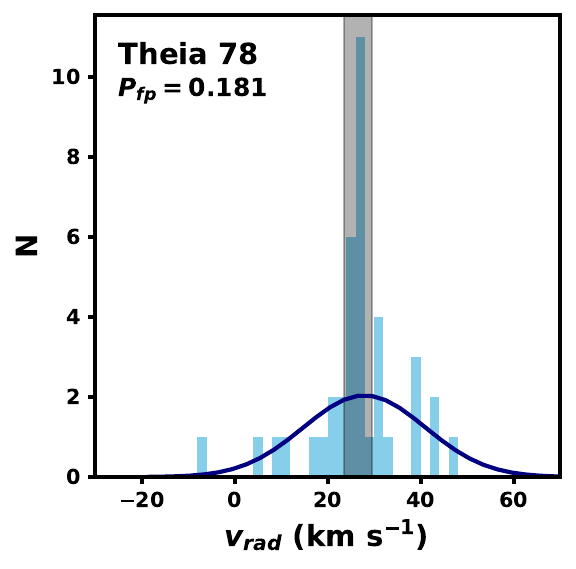}
    & \includegraphics[height=3.1cm]{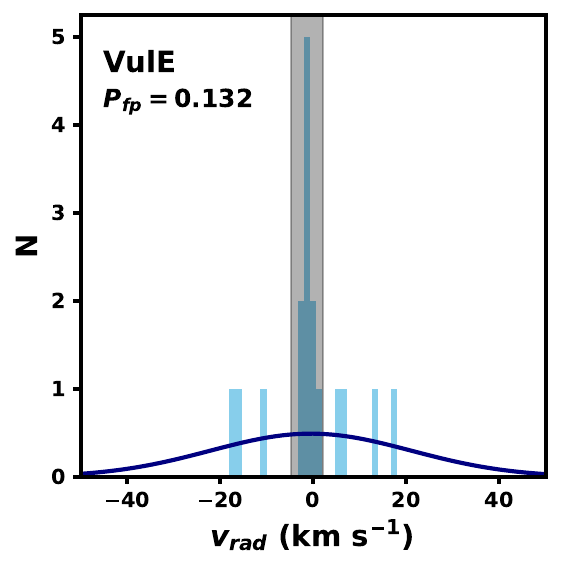}
    & \includegraphics[height=3.1cm]{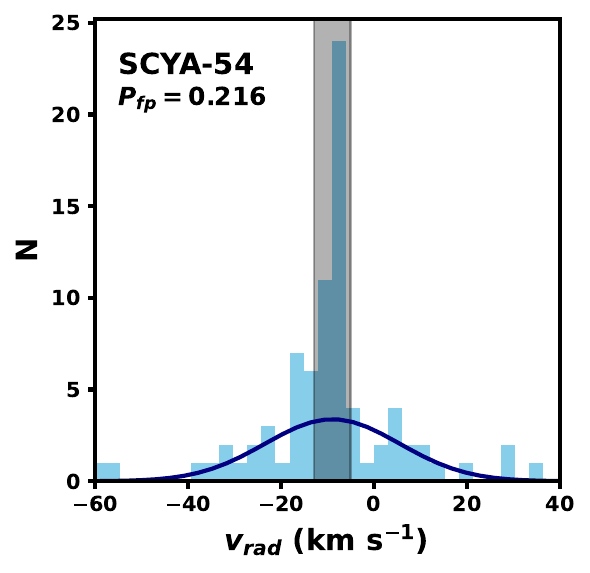}
    & \includegraphics[height=3.1cm]{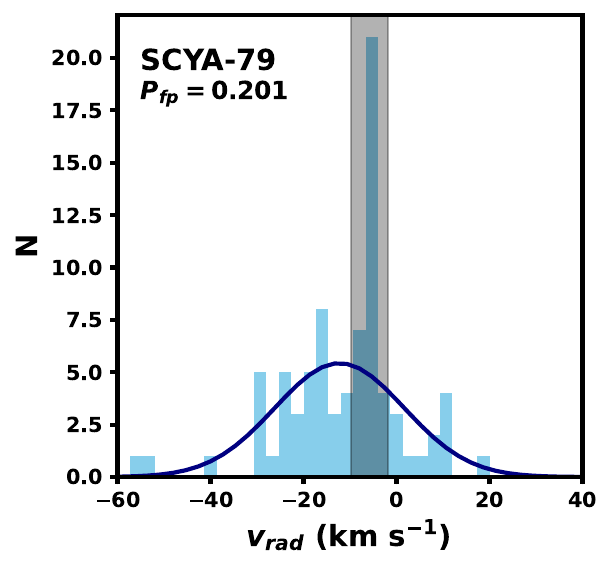}\\[-4pt]
\end{tabular}
\caption{Fits to the background contamination for all 15 associations included in this paper, which provide false positive rates, or the rate at which field stars are incorrectly assigned to the association. For populations where a Gaussian fit was used to fit the background, the fit is shown in dark blue, while the RV histogram of all stars is shown in light blue. The shaded region is masked, as that is the RV range occupied by the association. For populations with features that make a fit difficult, we take the mean and standard deviation of a set of probable non-members, which are indicated by the red bins. A curve representing the field model in these cases is shown in dark red. In TOR1, we show $P_{fp}$ results separately for the two components discussed in Section \ref{sec:tor1}.}
\label{fig:pfp_grid}
\end{figure*}

We compute the false positive rate by producing a histogram of the RVs with $\sigma_{RV} < 5$ km s$^{-1}$, masking bins in the range occupied by association members as per our RV selection, and fitting a Gaussian to the resulting distribution. Some populations in our sample are tenuous enough that a narrow association component is not favored by RV fitting, and in others, modeling the background using something like a KDE struggles to simultaneously model both the dense population and sparse field. Our approach therefore allows us to fit the field alone, without needing to model the much narrower peak associated with the association.

In populations with substructure that makes excluding the association more challenging (e.g., TOR1, see Sec. \ref{sec:tor1}), or with a field distribution that is either notably non-Gaussian (e.g. CasE) or particularly tenuous (e.g., LeoC), we instead compute an RV, mean, and standard deviation among probable non-members with $P_{sp} < 0.5$, or $P_{sp} < 0.2$ if the more lenient cut does not remove most of the spike associated with the population. We then use a Gaussian with those parameters to model the background. The false positive rate is provided by the fraction of field stars, as set by the field model, that would fall in the range of RVs where stars are marked as association members by our RV membership assessment. We show our fits to the field RV distributions in Figure \ref{fig:pfp_grid}. The values of this false positive rate range from 9\% in Theia 72 and ScuN to 22\% in CasE and SCYA-54. Establishing a non-zero false-positive rate is important to allow the removal of otherwise non-credible RV members, although the factor of two variation between these associations has a relatively limited effect, typically translating to a $\sim10$\% change in final membership probability in Section \ref{sec:finalmembership}.

For the false negative rate, $P_{fn}$, we use the 5\% value adopted in \citetalias{Kerr24}. This percentage describes the fraction of association members that are expected to have RVs inconsistent with the population, most likely due to velocities induced by the presence of a companion. The small populations discussed in this paper make it difficult to measure this value independently for each association, however this 5\% value is consistent with the result in \citetalias{Kerr24} and with other cases where false negatives have been discussed \citep[e.g.,][]{Shkolnik17}. 

\subsubsection{Spectroscopic Youth Indicators} \label{sec:specyouth}

For stars with spectroscopic measurements, we include the H$\alpha$ and Lithium 6708 \AA~lines as membership indicators (see Sec. \ref{sec:suppdat}). Lithium depletes rapidly after formation in fully convective late-type stars with sufficiently hot cores, going from Li-rich to essentially Li-free in 15-20 Myr for mid-M stars, while depleting more gradually in earlier-type stars over timescales ranging from tens of Myr for early M and late K stars to a few hundred Myr \citep{Jeffries23}. Lithium is therefore a strong indicator of youth, especially in later type stars. The $H\alpha$ line, when seen in emission, traces processes like accretion that continue to be active for the first few tens of Myr of the lives of later-type stars. H$\alpha$ is generally not a reliable indicator of youth due to the presence of active H$\alpha$ emitters among older field stars, however late-type young stars consistently show at least some H$\alpha$ emission from chromospheric activity \citep{Stauffer97}. This was verified by \citet{Kraus14}, demonstrating that among a set of stars selected without regard for H$\alpha$ in Tuc-Hor, none fall below the limit set by \citet{Stauffer97}. A lack of H$\alpha$ emission can therefore reliably flag non-members of young associations. This indicator helps to reduce contamination in the early M-dwarf regime where stars older than $\sim10$ Myr are essentially free of a 6708 \AA~Li line, but H$\alpha$ emission is still expected.

\begin{figure*}
\centering
\begin{tabular}{ccccc}
    \includegraphics[width=3.3cm]{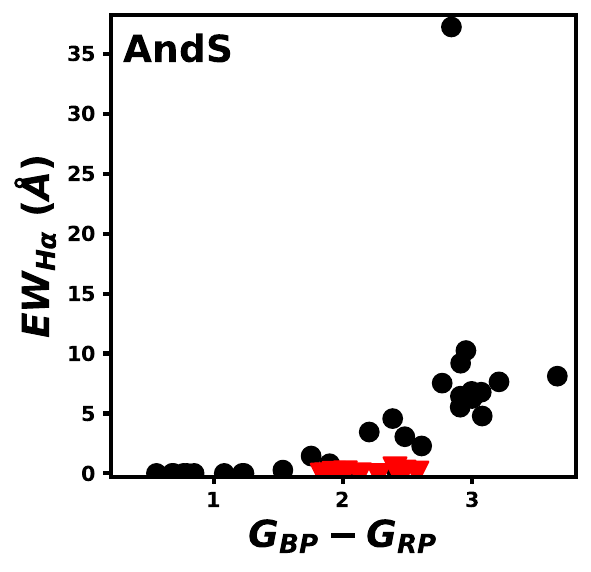}
    & \includegraphics[width=3.4cm]{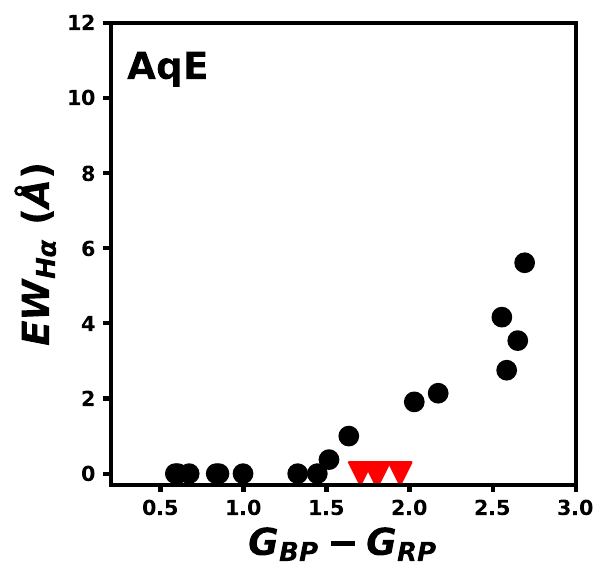}
    & \includegraphics[width=3.45cm]{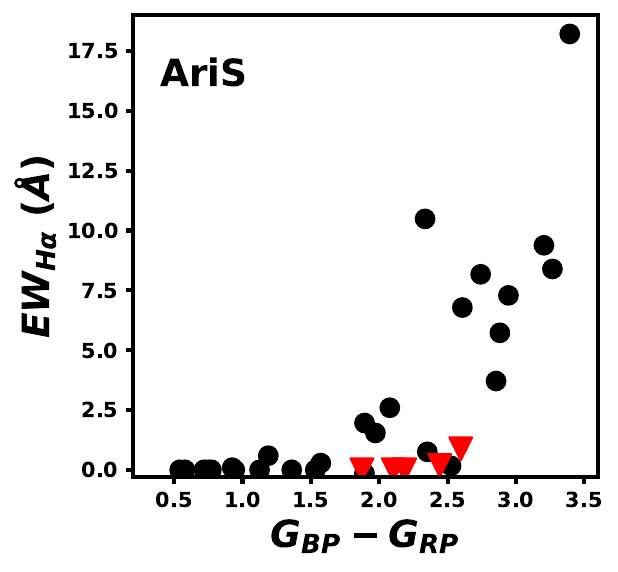}
    & \includegraphics[width=3.3cm]{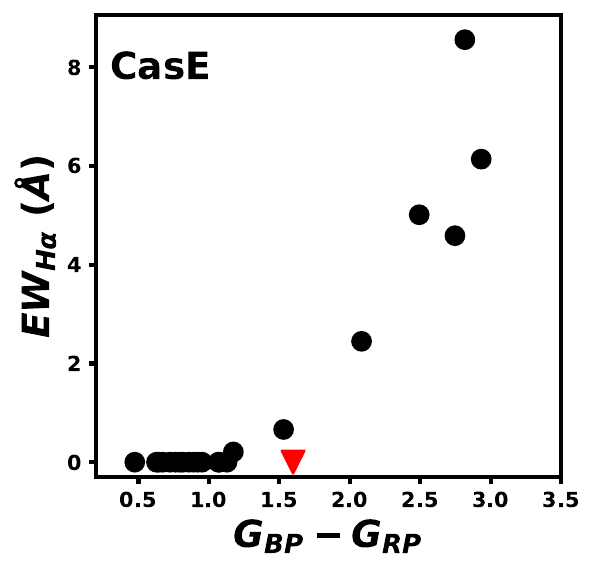}
    & \includegraphics[width=3.3cm]{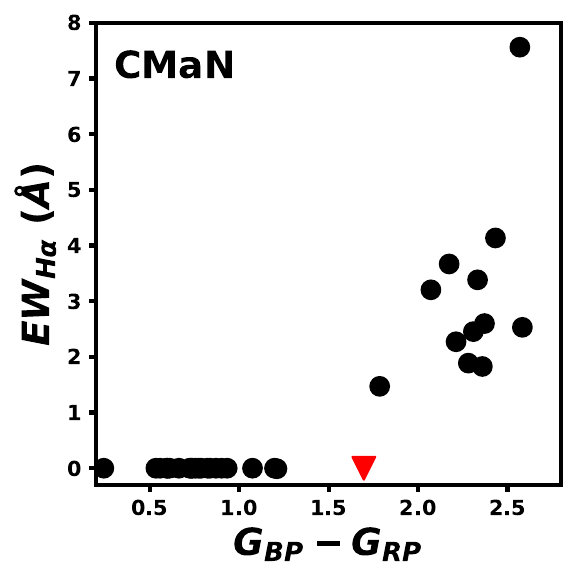}\\[-4pt]
    \includegraphics[width=3.3cm]{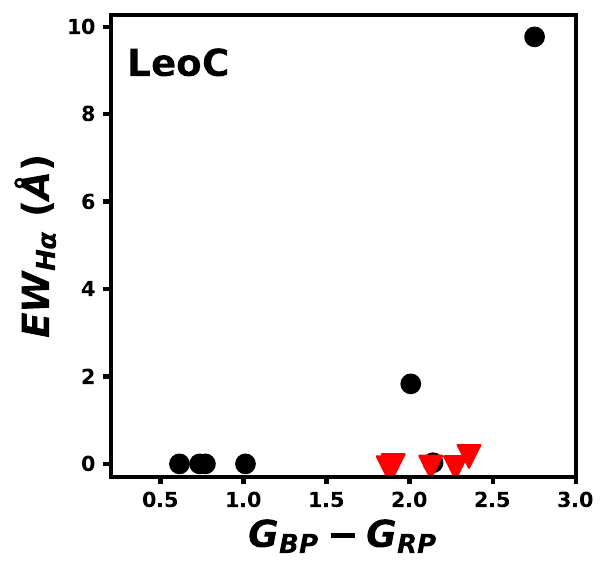}
    & \includegraphics[width=3.3cm]{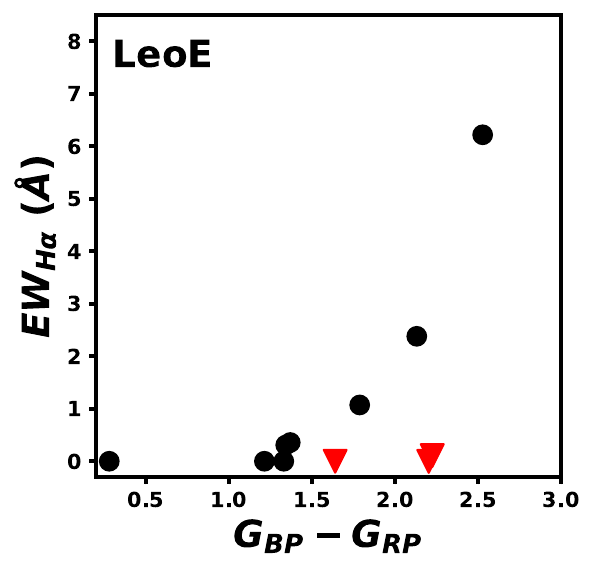}
    & \includegraphics[width=3.3cm]{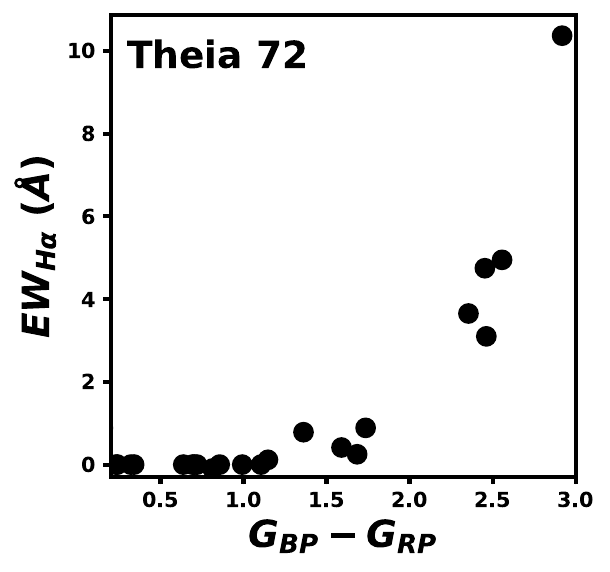}
    & \includegraphics[width=3.3cm]{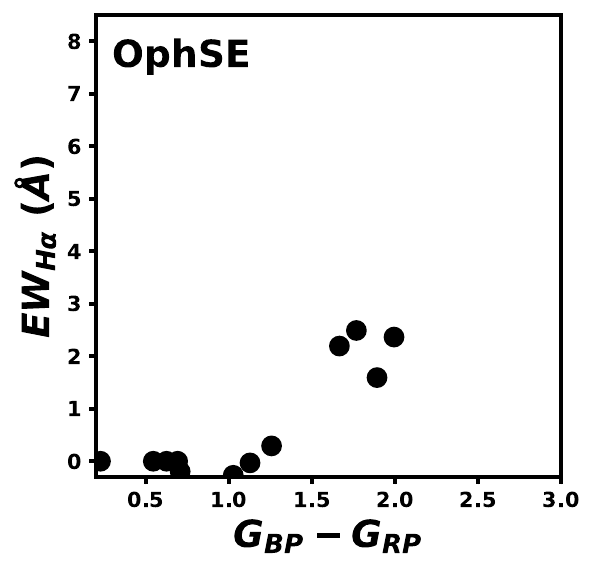}
    & \includegraphics[width=3.3cm]{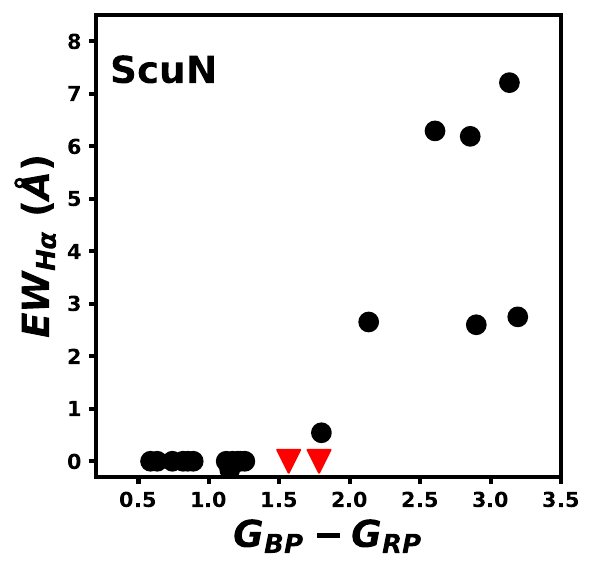}\\[-4pt]
    \includegraphics[width=3.3cm]{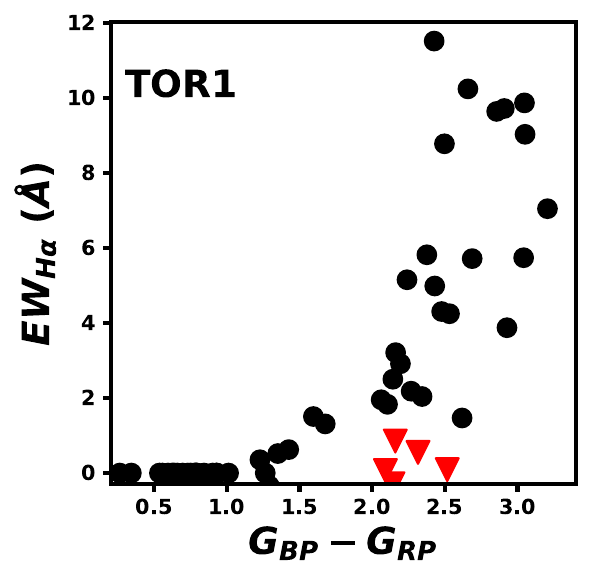}
    & \includegraphics[width=3.3cm]{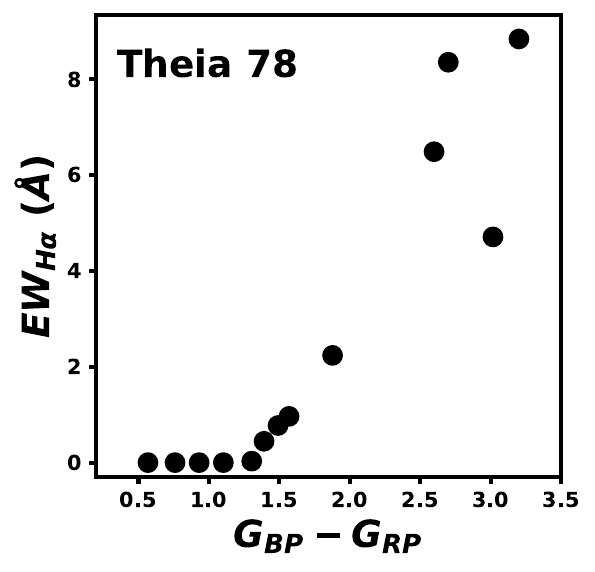}
    & \includegraphics[width=3.3cm]{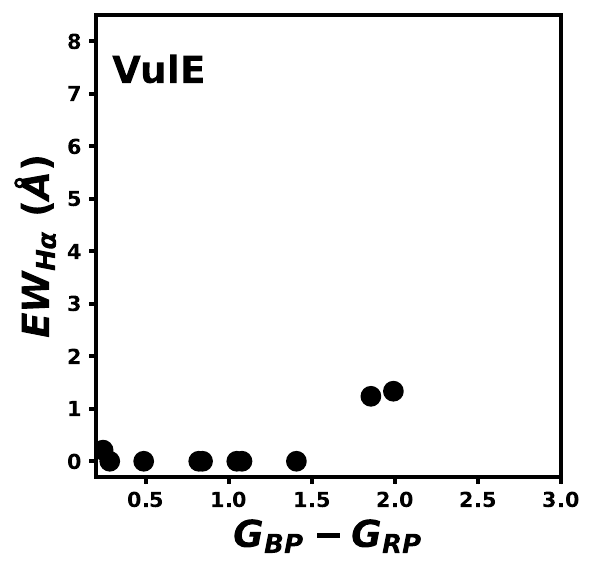}
    & \includegraphics[width=3.3cm]{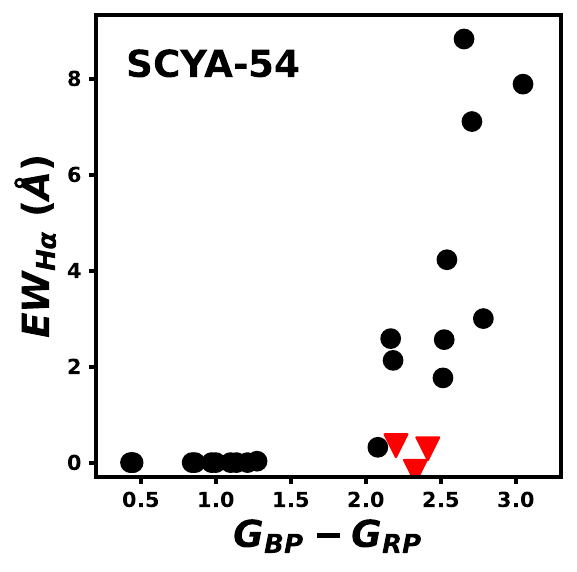}
    & \includegraphics[width=3.3cm]{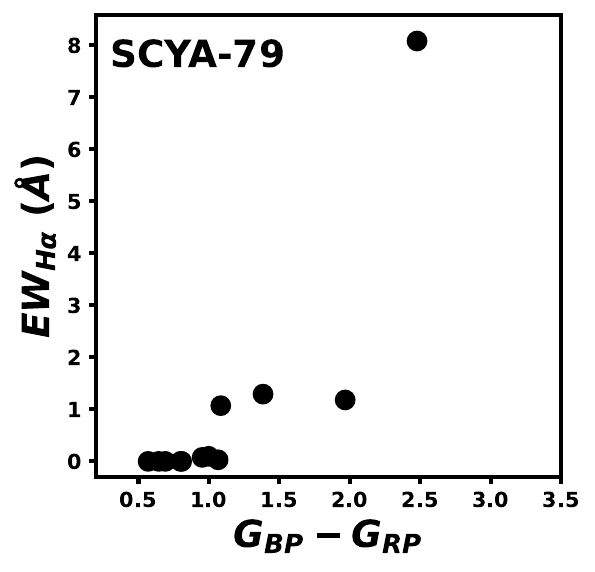}\\[-4pt]
\end{tabular}
\caption{H$\alpha$ equivalent width against color for all 15 associations covered by this paper. Stars rejected as members by H$\alpha$ are marked as red triangles, and all other stars are marked with black dots.}
\label{fig:hamem}
\end{figure*}

To assess Li membership, we use the EAGLES package, which computes age posteriors for stars of a given Li EW and $T_{eff}$ value. The latter can be acquired from $G_{BP}-G_{RP}$ using a conversion presented in \citet{Jeffries23}, which was shown to produce age values that differ from those of directly-measured values of $T_{eff}$ by only 0.01 dex on average. Li EWs in our sample typically do not account for the presence of the Fe I line at 6707.44 \AA~, but \citet{Soderblom93} provides a corrective factor as a function of $T_{eff}$ that accounts for this often-unmodeled line. We apply this correction to all Li EWs prior to running EAGLES, except for the GALAH DR4 values, which model the Fe I line \citep[see][]{Wang24}. EAGLES provides models of the expected Li EW for $3000<T_{eff}<6500$, so our ability to assess youth with EAGLES is limited to that range \citep{Jeffries23}. Following \citetalias{Kerr24}, we mark stars with EAGLES analytical age posteriors $P_{age<80 Myr}<0.005$ as non-members, which is a strict selection that only removes stars inconsistent with membership at ages near the SPYGLASS limit of detectability or with substantial contributions from a Li-free unresolved companion. We additionally require $RUWE<1.2$ for an assessment of non-membership to exempt stars with evidence of an unresolved binary \citep{RUWELindegren18, Bryson20}, which reduces false negatives caused by a companion masking the Li line depth. We then mark stars with $\tau_{mean}<50$ Myr as Li-verified members, a cut that reflects the typical age range of SPYGLASS associations. The remaining stars are marked as having ambiguous Li membership. Of the stars marked as Lithium members, 9 of 100 across all groups are marked as RV non-members, which is within the 90\% binomial confidence interval of the 5\% velocity false negative rate chosen in Section \ref{sec:falseposneg}. One of the 3 velocity non-members in TOR-1 later gets reclassified as a velocity member of a dynamically distinct subgroup in Section \ref{sec:tor1}, bringing these values into even closer agreement. 

\begin{figure*}
\centering
\begin{tabular}{ccccc}
    \includegraphics[width=3.3cm]{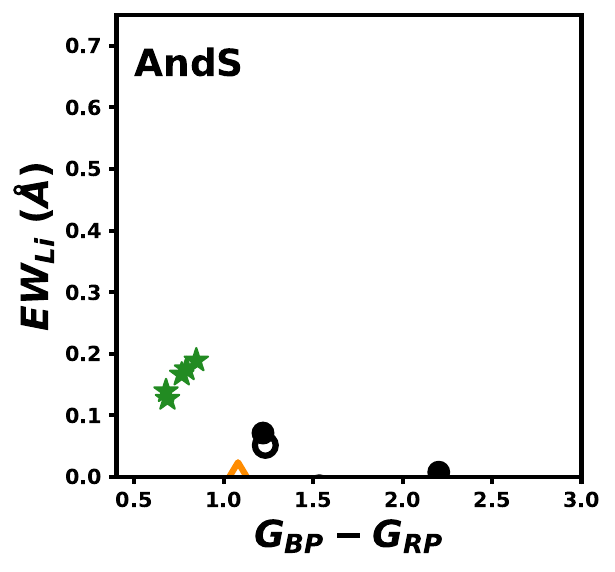}
    & \includegraphics[width=3.3cm]{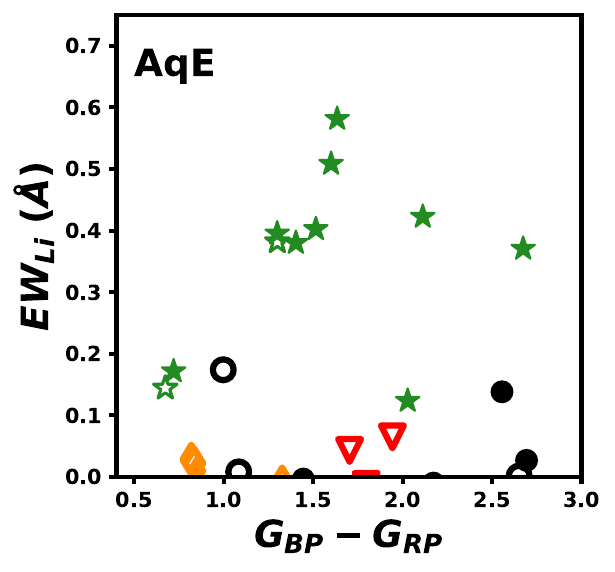}
    & \includegraphics[width=3.3cm]{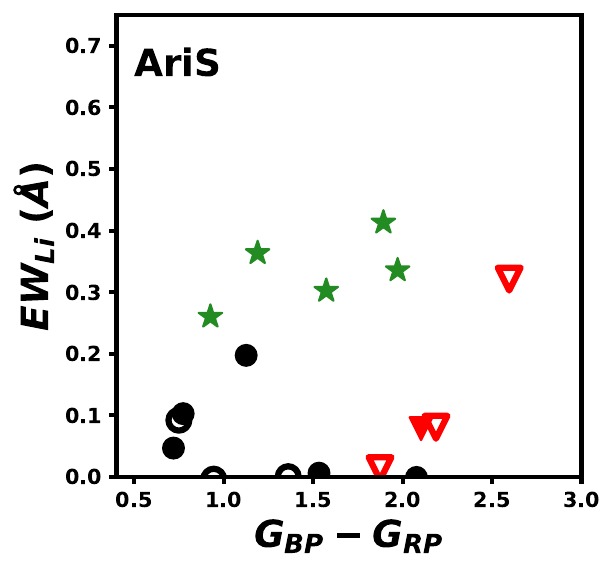}
    & \includegraphics[width=3.3cm]{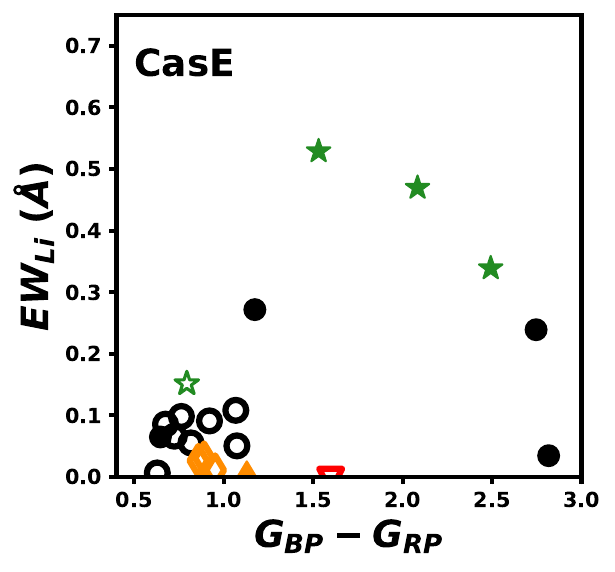}
    & \includegraphics[width=3.3cm]{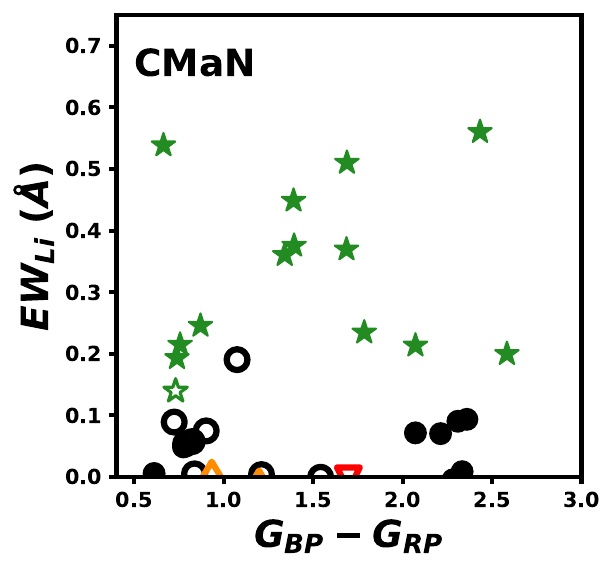}\\[-4pt]
    \includegraphics[width=3.3cm]{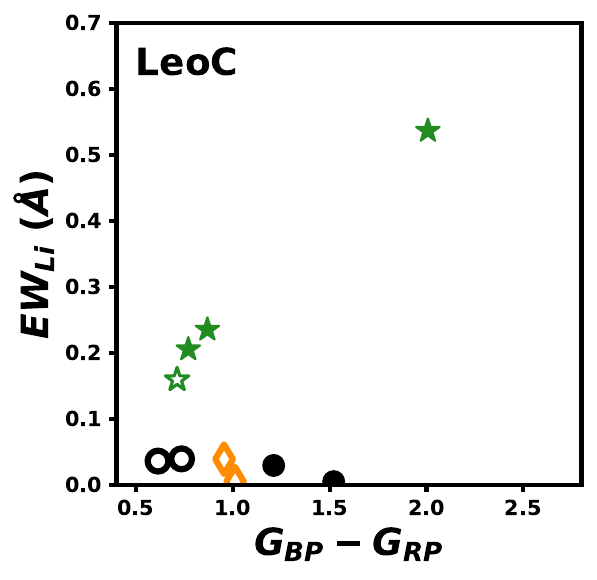}
    & \includegraphics[width=3.3cm]{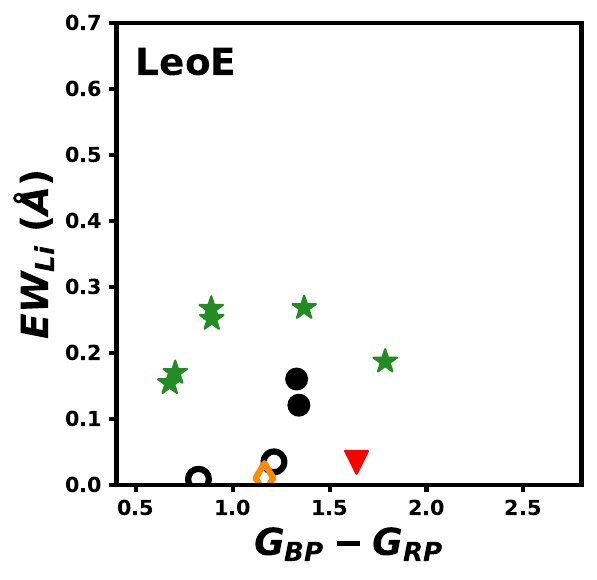}
    & \includegraphics[width=3.3cm]{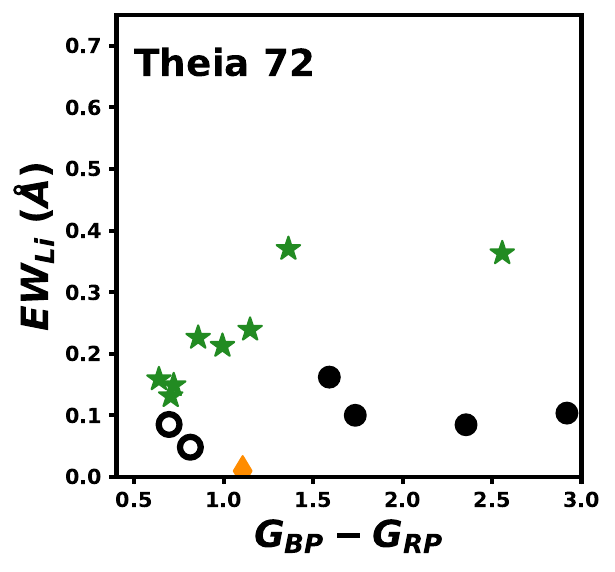}
    & \includegraphics[width=3.3cm]{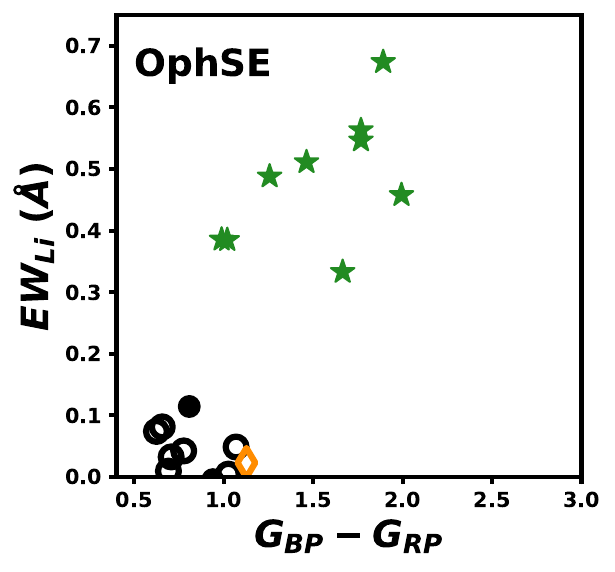}
    & \includegraphics[width=3.3cm]{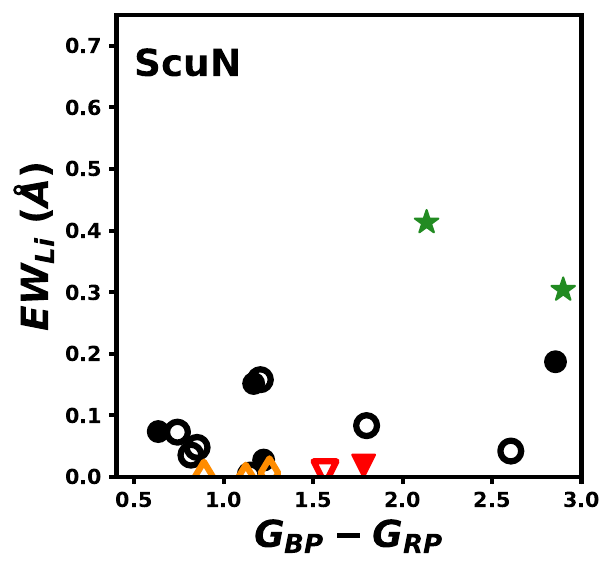}\\[-4pt]
    \includegraphics[width=3.3cm]{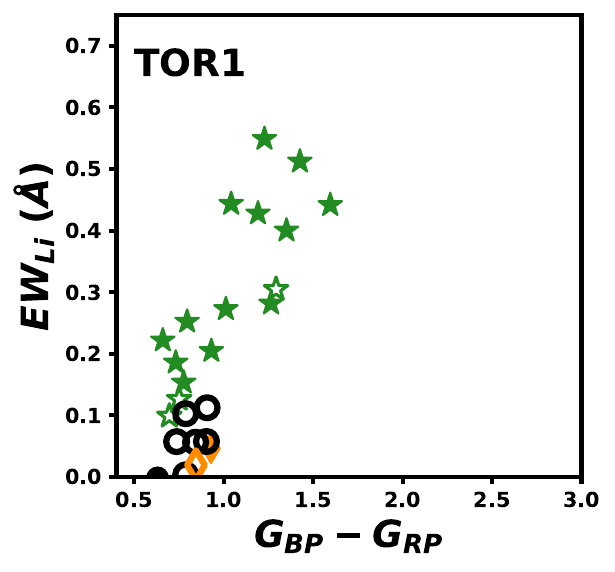}
    & \includegraphics[width=3.3cm]{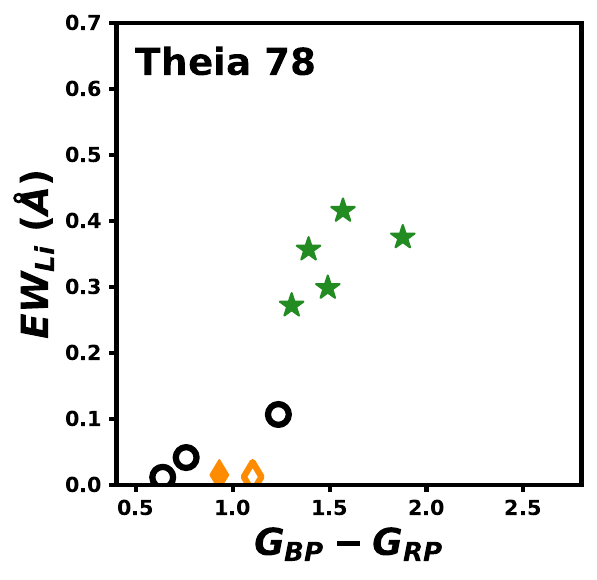}
    & \includegraphics[width=3.3cm]{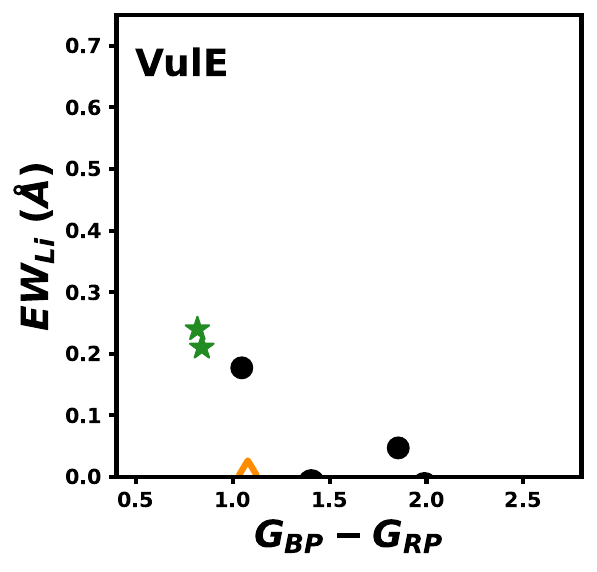}
    & \includegraphics[width=3.3cm]{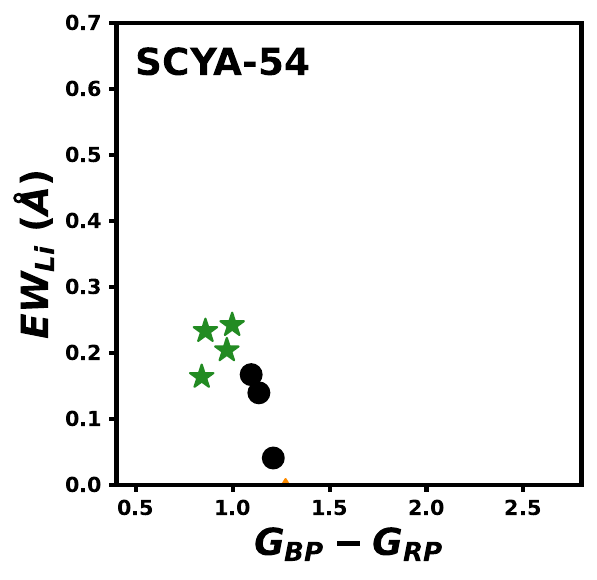}
    & \includegraphics[width=3.3cm]{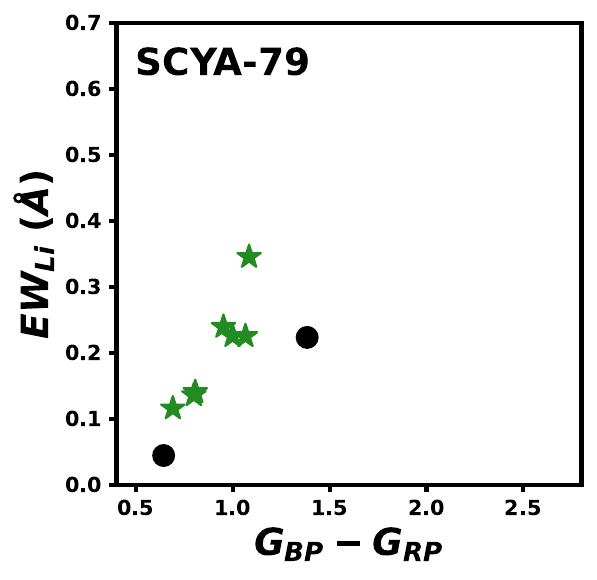}\\[-4pt]
\end{tabular}
\caption{Li-6708\AA~equivalent width against color for all 15 associations covered by this paper. Stars marked as lithium members are shown with green star icons, while stars that are marked as lithium non-members are marked with orange diamonds. Black circles indicate stars with inconclusive Lithium membership, while red triangles indicate the stars without a conclusive Li assessment that are rejected due to their lack of H$\alpha$ emission in Figure \ref{fig:hamem}. Stars with open markers are velocity non-members.}
\label{fig:limem}
\end{figure*}

To assess H$\alpha$ membership, we use the lower limits of the H$\alpha$ distribution for IC 2602 and IC 2391 presented in \citet{Stauffer97}. Following \citet{Kraus14}, we mark stars below this empirical limit as H$\alpha$ non-members, unless they show evidence for unresolved binarity ($RUWE>1.2$). All other stars are given an inconclusive H$\alpha$ membership assessment. We present the spectroscopic membership selection for all 15 associations in Figures \ref{fig:hamem} and \ref{fig:limem}. In Fig. \ref{fig:hamem}, we show the H$\alpha$ membership assessment. We then show the results of the Lithium membership assessment in Figure \ref{fig:limem}, separately marking stars identified as non-members by other membership metrics. 

\subsubsection{Final Membership Probabilities} \label{sec:finalmembership}

Finally, we combine the results across all membership indicators to produce aggregate membership probabilities for all stars in the sample. First, we adjust the spatial-photometric membership probabilities using the velocity membership assessment. Velocity is not a conclusive membership indicator due to the non-zero false positive and negative rates discussed in Section \ref{sec:falseposneg}, and final membership rates can therefore be computed using Bayes' Theorem as the posterior probability $P(mem|V = x)$, where $V$ is the velocity membership assessment, following \citetalias{Kerr24}:

\begin{equation} \label{eq:pmem|mem}
P(mem|V = 1) = \frac{(1-P_{fn}) P_{sp}}{P(V=1)},
\end{equation}

\noindent where $P_{fn}$ is the false negative rate (5\%, see Section \ref{sec:falseposneg}). The marginal probability $P(V=1)$ can be written as:

\begin{equation}
P(V = 1) = (1-P_{fn})P_{sp} +P_{fp}(1-P_{sp}),
\end{equation}

\noindent where $P_{fp}$ is the false positive rate for the population derived in Section \ref{sec:falseposneg}. The analogous formula for velocity non-members $P(mem|V=-1)$ is provided by:

\begin{equation} \label{eq:pmem|nmem}
P(mem|V = -1) = \frac{P_{fn} P_{sp}}{P_{fn} P_{sp} + (1 - P_{fp})(1 - P_{sp})}.
\end{equation}

For a given star with a given velocity membership assessment $V$, we evaluate the corresponding formula for its membership probability given its value of $V$, with the result providing its combined membership probability. For stars with inconclusive membership probabilities ($V=0$), the membership probability remains $P_{sp}$. A more detailed motivation for these formulae is provided in \citetalias{Kerr24}.

\begin{figure*}
\centering
\begin{tabular}{ccccc}
    \includegraphics[width=3.3cm]{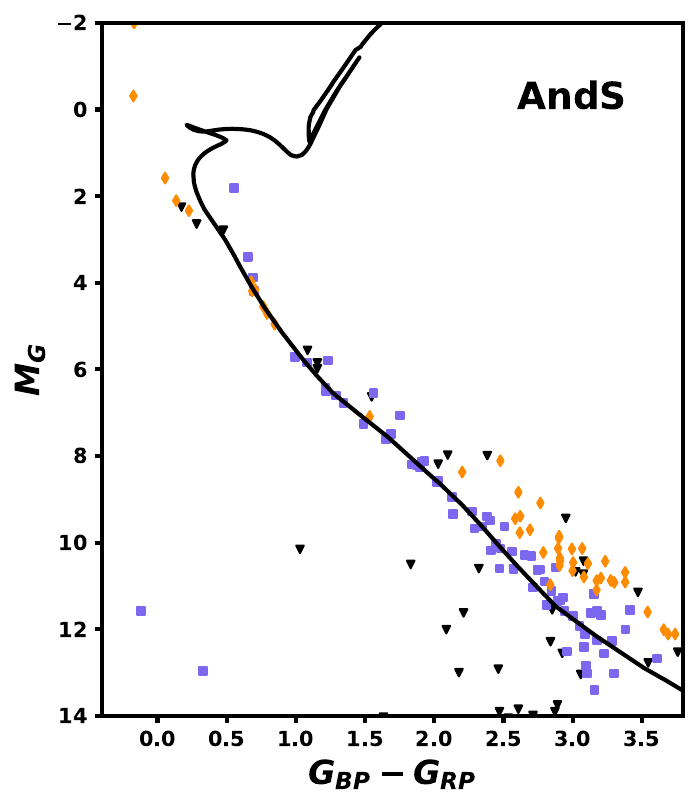}
    & \includegraphics[width=3.3cm]{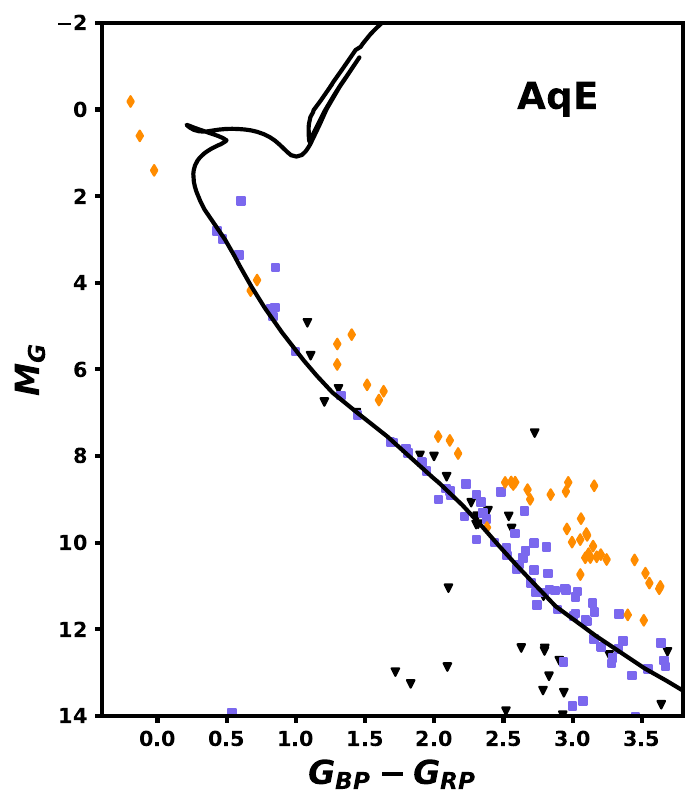}
    & \includegraphics[width=3.3cm]{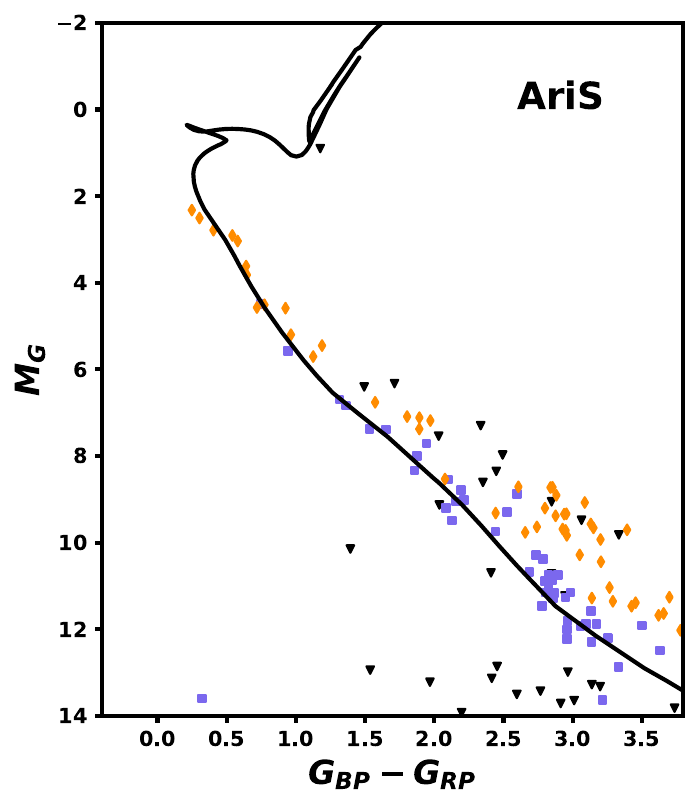}
    & \includegraphics[width=3.3cm]{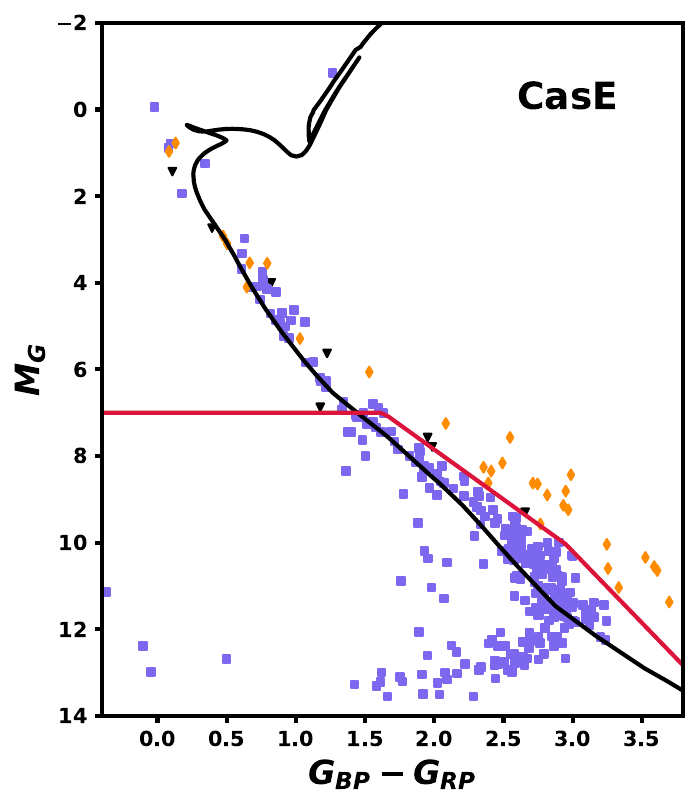}
    & \includegraphics[width=3.3cm]{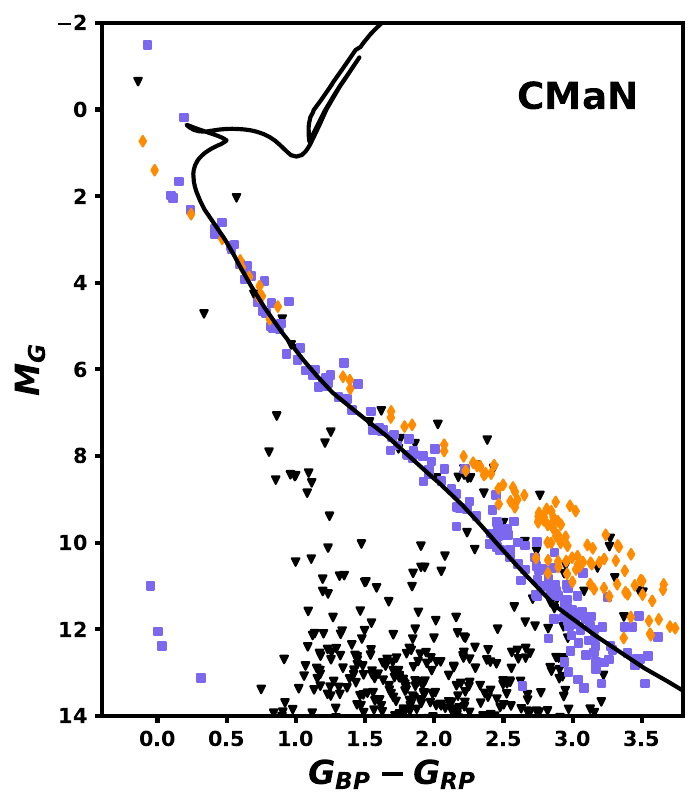}\\[-4pt]
    \includegraphics[width=3.3cm]{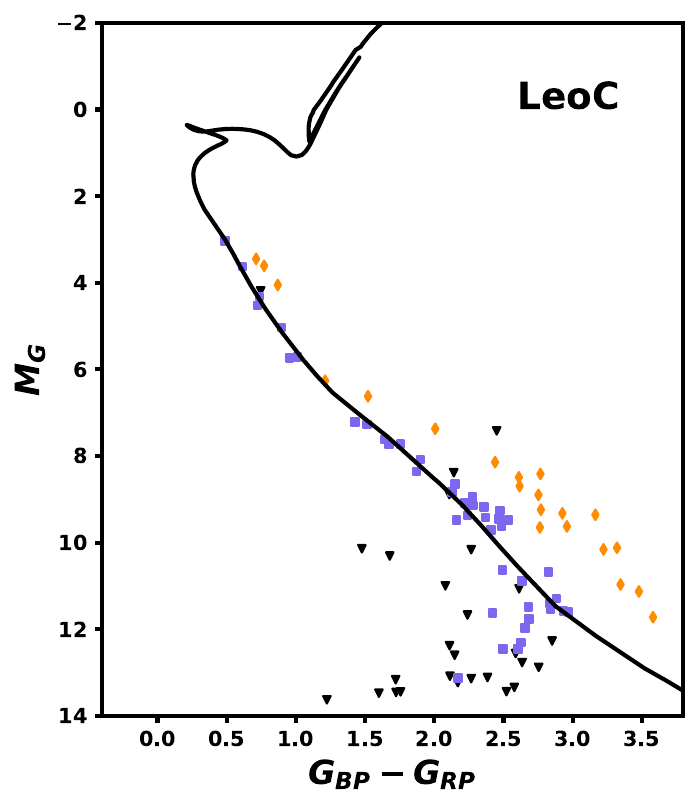}
    & \includegraphics[width=3.3cm]{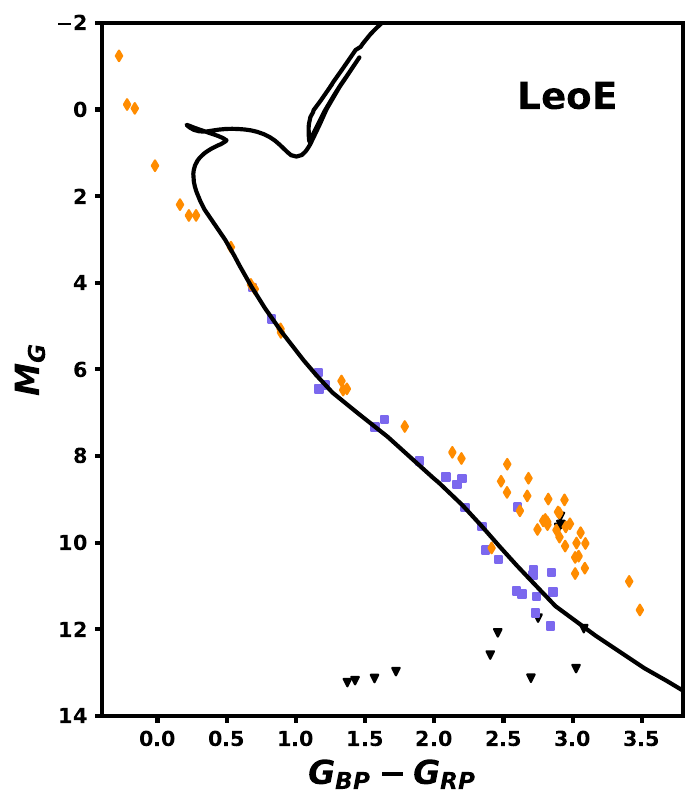}
    & \includegraphics[width=3.3cm]{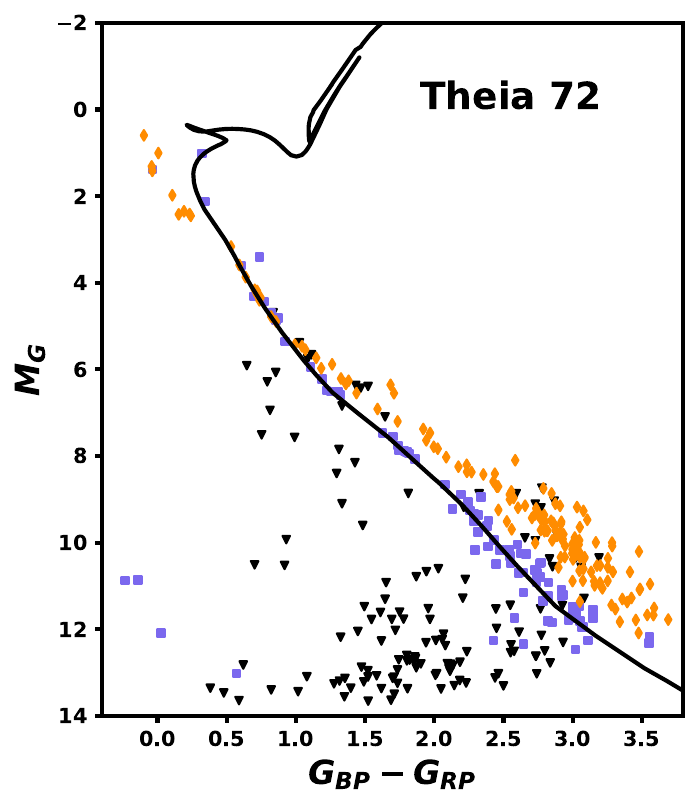}
    & \includegraphics[width=3.3cm]{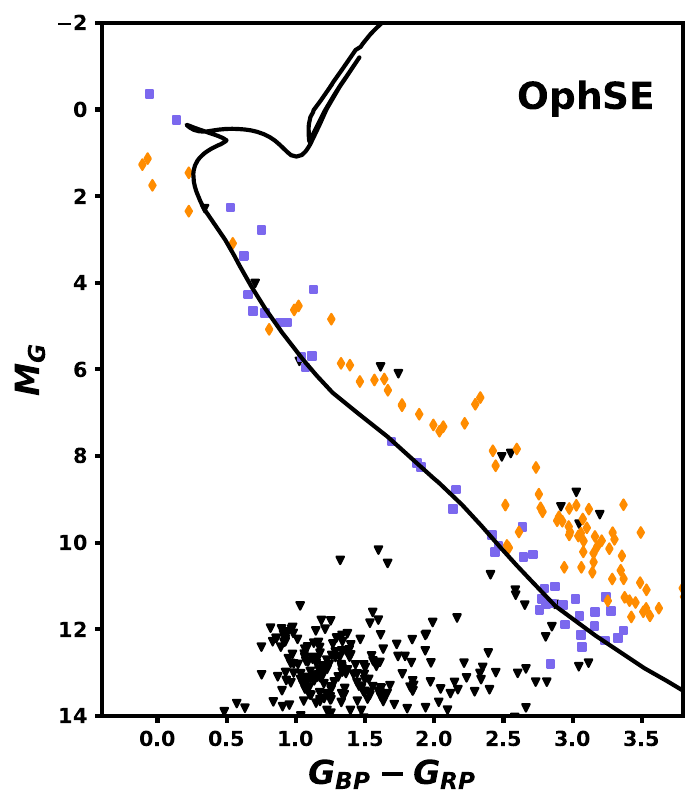}
    & \includegraphics[width=3.3cm]{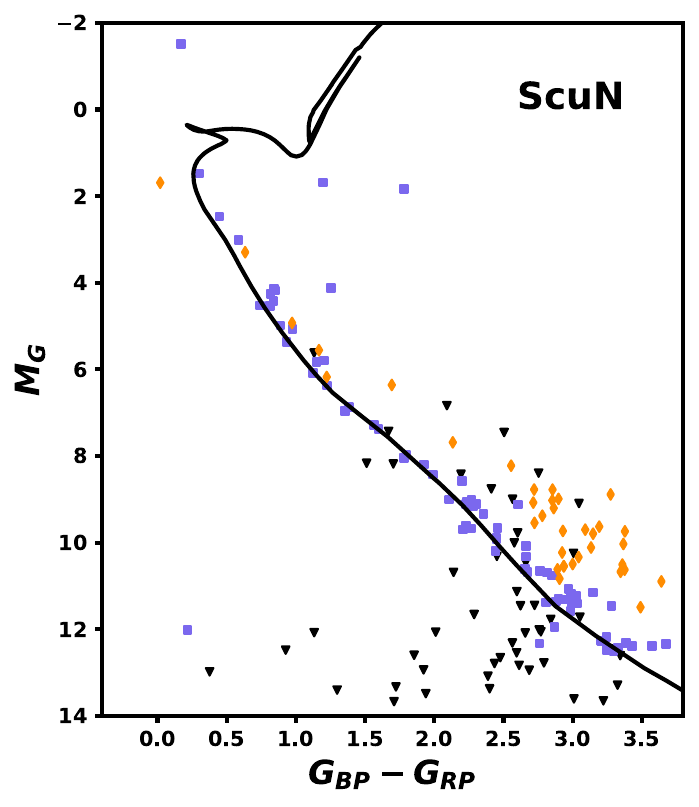}\\[-4pt]
    \includegraphics[width=3.3cm]{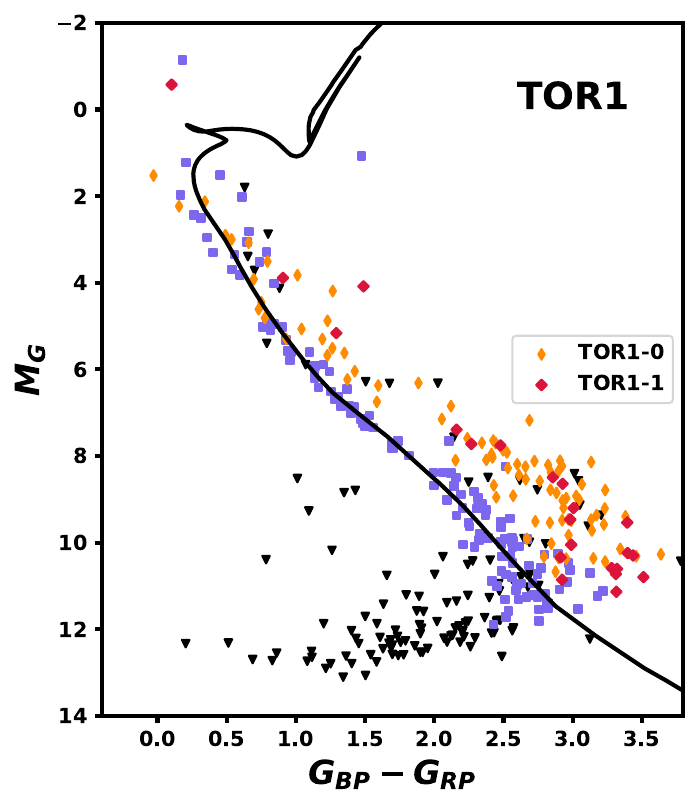}
    & \includegraphics[width=3.3cm]{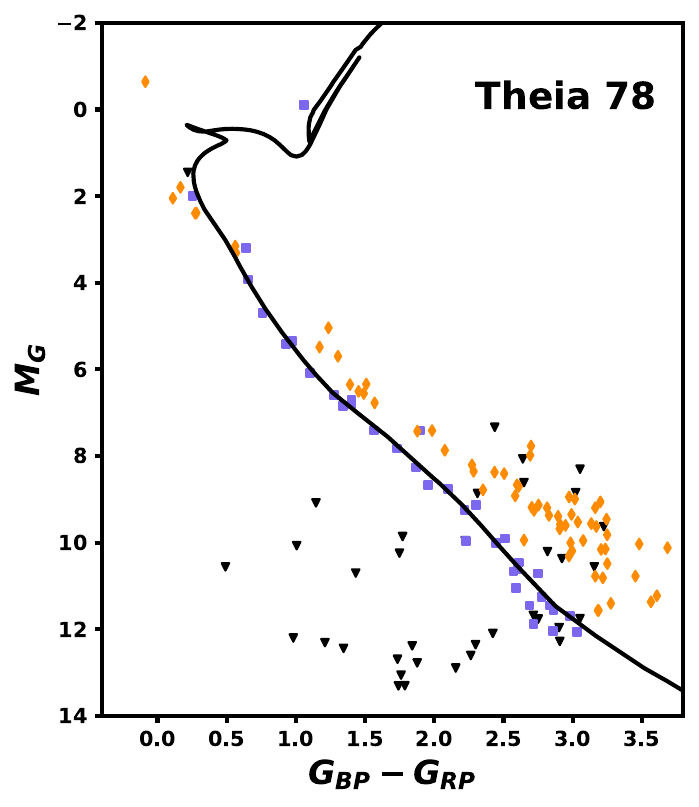}
    & \includegraphics[width=3.3cm]{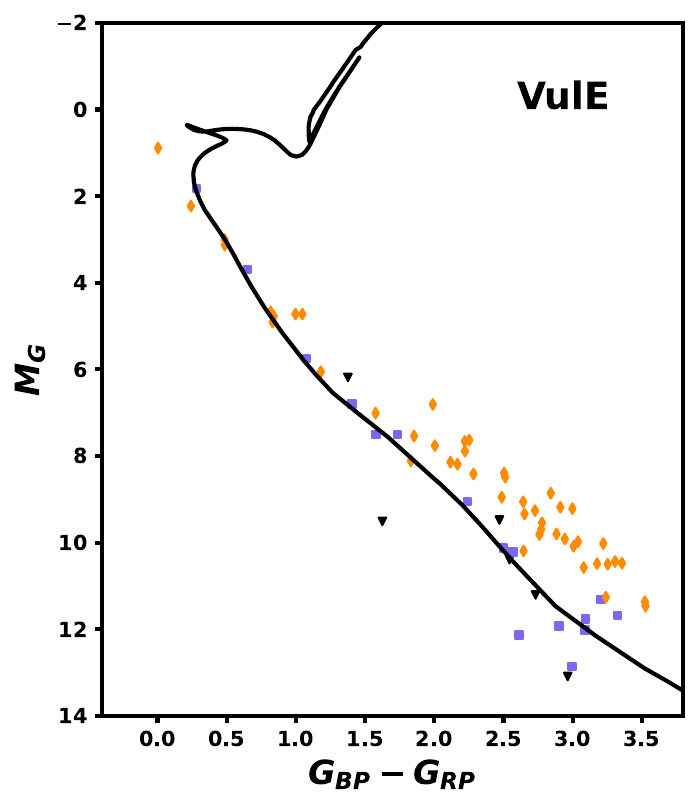}
    & \includegraphics[width=3.3cm]{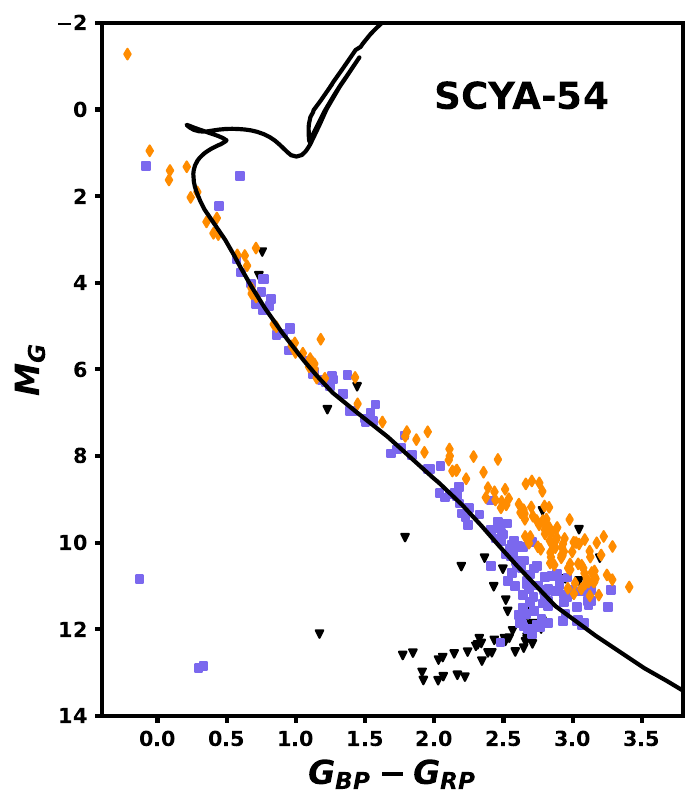}
    & \includegraphics[width=3.3cm]{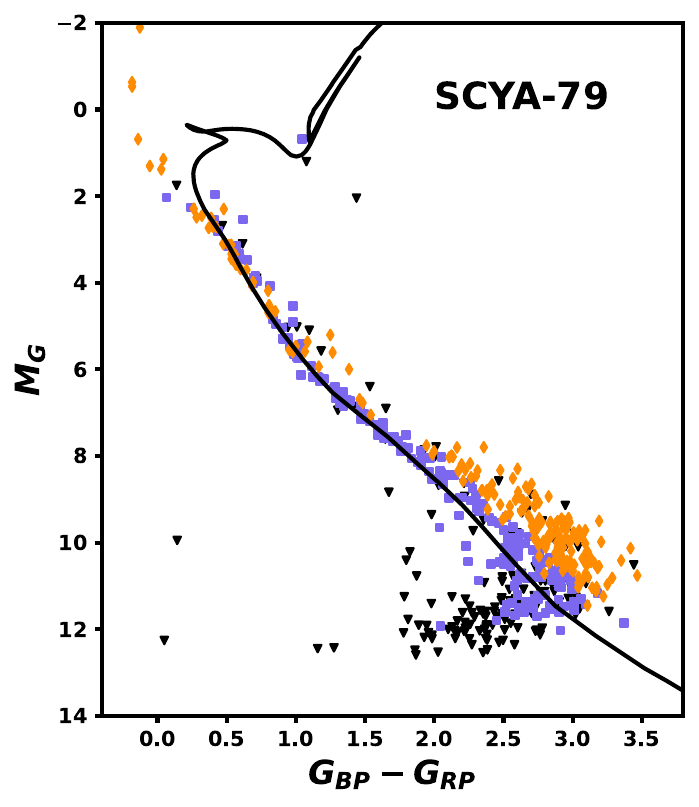}\\[-4pt]
\end{tabular}
\caption{CMDs for all 15 associations, adjusted for distance and reddening. Orange diamonds mark probable members with $P_{fin} > 0.5$, purple squares indicate stars with $P_{fin} < 0.5$, and black triangles mark stars that lack $P_{fin}$ because they fail one or more of the quality cuts required for a $P_{Age<50 Myr}$ result. The black curve indicates a typical field sequence, represented by a PARSEC 1 Gyr isochrone. In the panel for CasE, we include a cut to avoid contamination from an elevated field sequence in red, which avoids older populations that share its parameter space. In the TOR1 panel, we use different markers for probable members of the two components discussed in Section \ref{sec:tor1}, as noted in the legend.}
\label{fig:cmds}
\end{figure*}

Finally, we introduce the Lithium and H$\alpha$ membership assessments. We consider these assessments to be conclusive, and we therefore mark Lithium members as having a membership probability of 1, and Lithium non-members to have a membership probability of 0. We also mark H$\alpha$ non-members as having a membership probability of 0. The corrections for spectroscopic membership and velocity membership ($V$) produce final membership probabilities $P_{fin}$ which we use throughout the rest of this paper. We limit the structural and dynamical analyses later in this paper to stars with $P_{fin} > 0.5$, although our demographic analysis also considers stars with lower values of $P_{fin}$. 

We show CMDs for all 15 associations covered by this publication in Figure \ref{fig:cmds}, with photometry corrected for \citet{BailerJones21} distance and for reddening using the \citet{Lallement19} maps. There we also show the cut used for removing field contaminants in CasE. All of the associations, with the exception of SCYA-79 (see Section \ref{sec:agesynth}), show a clear separation between the stars in the field sequence (near the black isochrone), and the young populations defined by $P_{fin} > 0.5$.

\subsection{Substructure} \label{sec:substructure}

Small population sizes limit the potential for, and detectability of, substructure. However, substructure may nonetheless be present in our sample, especially in larger populations. The presence of multiple subgroups with divergent trajectories can affect dynamical age results, as the most compact configuration of a substructured population may be in the moment of closest approach between subgroups, not in the moment a coherent cloud becomes unbound. We must therefore search for substructure to identify populations with distinct ages, trajectories, and formation sites.

We use the HDBSCAN clustering algorithm to search each of the 15 populations in our sample for substructure. Like in \citetalias{Kerr22a} and \citetalias{Kerr24}, we cluster in 5D space-transverse velocity anomaly coordinates, $(X,Y,Z,c*\Delta$v$_{T,l},c*\Delta$v$_{T,b})$, where the transverse velocity anomaly is defined as the transverse velocity for the star, minus the projected transverse velocity of the median UVW 3D velocity vector at its location. We use $c=6$ pc km$^{-1}$, which is a constant that makes the scales of the spatial and velocity coordinates comparable. We cluster this coordinate space in HDBSCAN's leaf mode, which identifies the smallest scale overdensities present in the population, rather the EOM mode, which identifies structures due to their presence over the largest range of scales. We set {\tt min$\_$samples} and {\tt min$\_$cluster$\_$size} to 7, following the parameter choices used in \citetalias{Kerr22a}, which focused on an association at a similar distance to many of these populations, and with a scale and complexity not much larger than them. 


HDBSCAN assumes the presence of a background, however our samples are much more pure, so we therefore assign outlying stars to the closest parent population. Following previous SPYGLASS publications, we employ 5D $(X,Y,Z,c*\Delta$v$_{T,l},c*\Delta$v$_{T,b})$ space-velocity coordinates with $c=12$ pc km$^{-1}$ as the distance metric for assigning outlying members. This metric slightly emphasizes velocity, which is useful in these groups, as stars that remain unclustered after our HDBSCAN clustering tend to be more scattered in spatial coordinates than velocity. We assign unclustered stars to the subgroup with the nearest mean space-velocity position, measured using the distance metric above. 

This clustering reveals subgroups in 4 of the 15 populations studied in this publication: TOR1, Theia 72, SCYA-54, and SCYA-79. This is consistent with expectations, as the populations that host substructure are four of the largest in our sample, and a visual inspection verifies that the populations classified as non-substructured are relatively small, simple populations with only one major stellar concentration within. This does not exclude the possibility of further substructure, however, as we cannot detect structures smaller than our minimum cluster size. 

\begin{figure*}
\centering
\includegraphics[height=6cm]{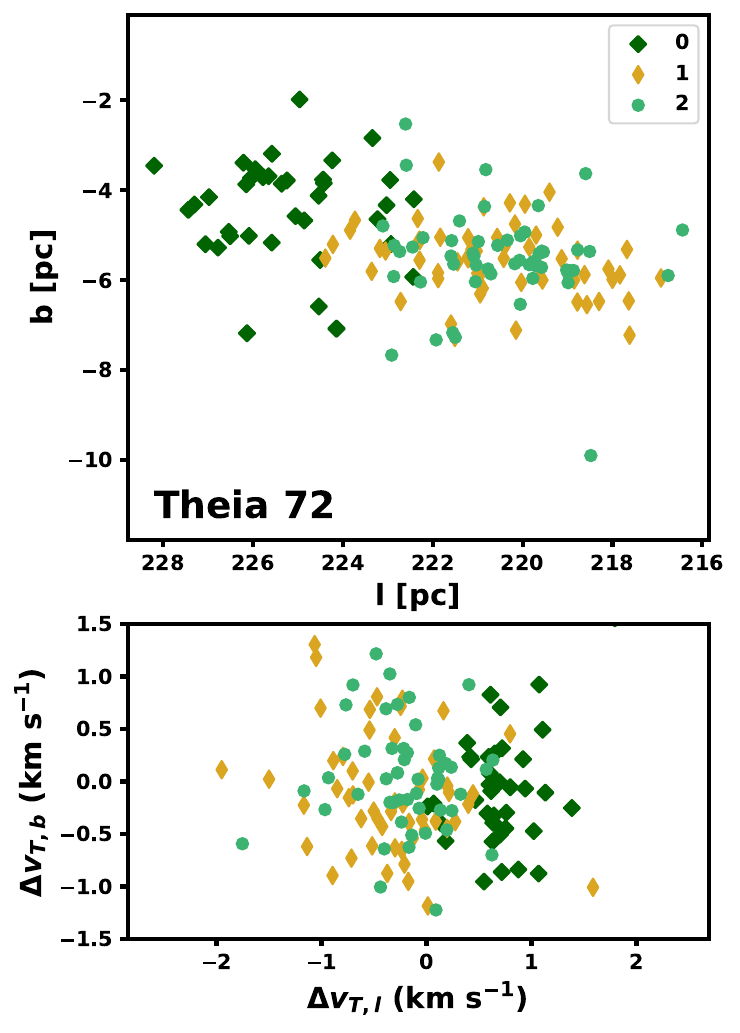}
\includegraphics[height=6cm]{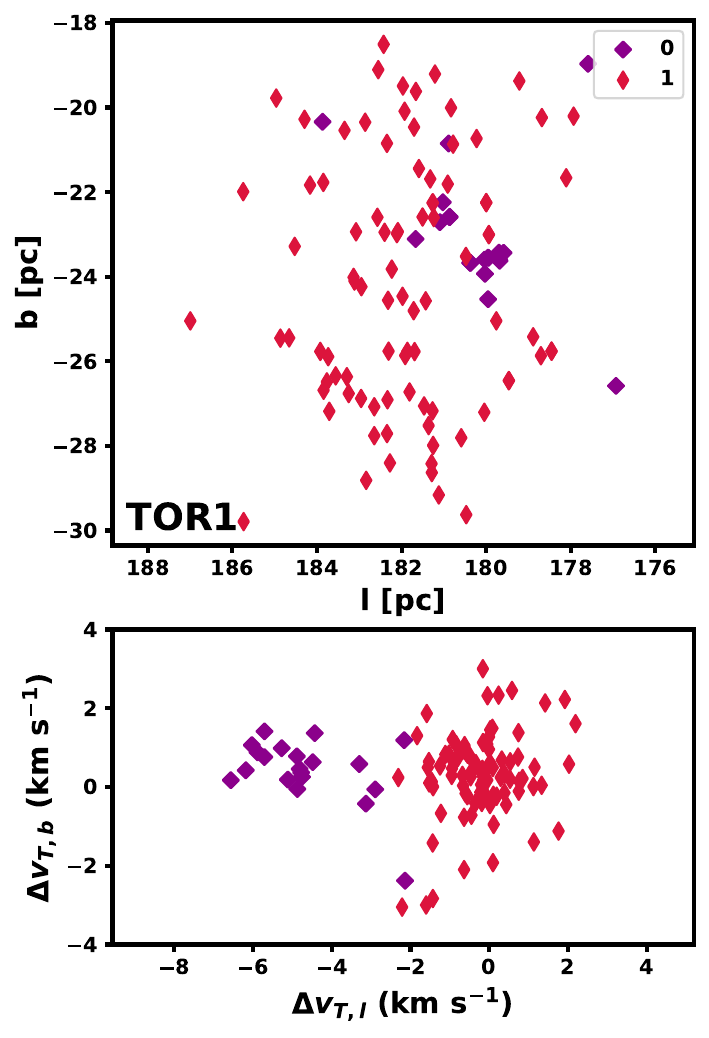}
\includegraphics[height=6cm]{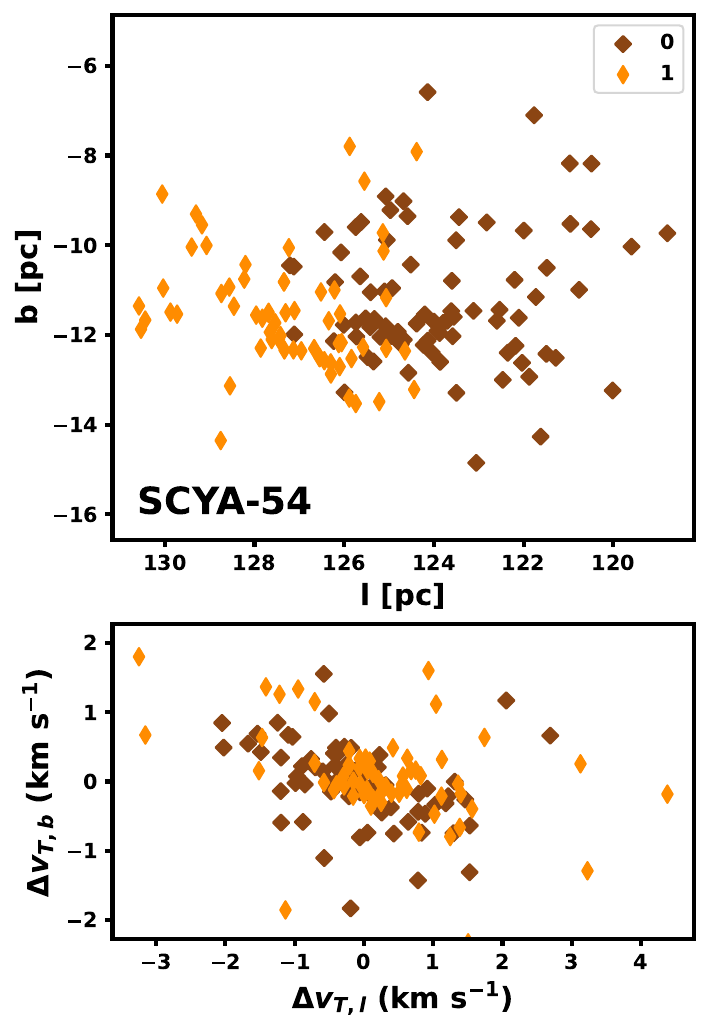}
\includegraphics[height=6cm]{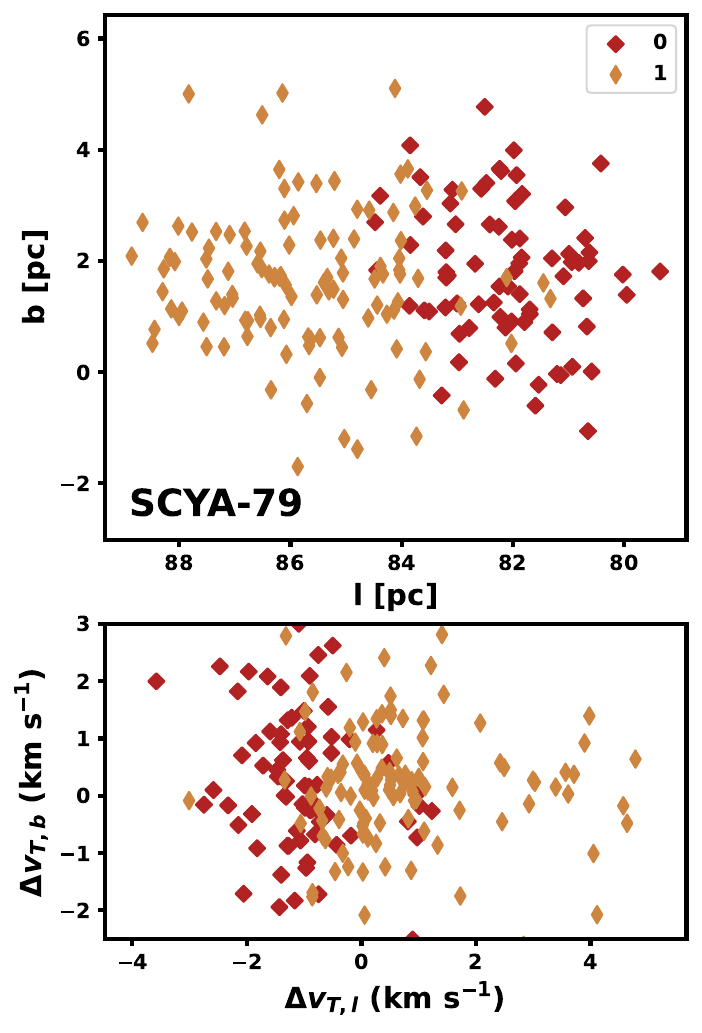}
\caption{Substructure in young associations where we detect it, shown in $l$ and $b$ galactic sky coordinates and $\Delta v_{T, l}$ and $\Delta v_{T, b}$ transverse velocity anomalies. We only show stars with $P_{fin}>0.5$. The two components in TOR1 also differ by $\sim10$ km s$^{-1}$ in radial velocity, indicating that they are dynamically distinct populations. We rename TOR1-1 to TOR1A, and TOR1-0 to TOR1B.}
\label{fig:substructures}
\end{figure*}

In three of the four subclustered associations (Theia 72, SCYA-54, and SCYA-79), substructure is subtle and driven primarily by local overdensities in a extended stellar spatial distribution. Future work can test whether these overdensities reflect distinct formation times and sites. In TOR1, the substructure is distinguished by two highly distinct velocity components. Figure \ref{fig:substructures} shows the substructures we reveal through this analysis.

\subsubsection{Taurus-Orion 1} \label{sec:tor1}

The subgroup we mark as TOR1-0 in Figure \ref{fig:substructures} has transverse velocity anomalies that differ from the larger TOR1-1 clump by about 6 km s$^{-1}$. We find that this discrepancy is reflected in the radial velocities, where TOR1-0 has $v_R \sim 19.8$ km s$^{-1}$, compared to the average value of $v_R \sim 10.8$ km s$^{-1}$ in the population as a whole and $v_R \sim 10.5$ km s$^{-1}$ in TOR1-1. A velocity difference of order 10 km s$^{-1}$ quickly propagates to a large positional difference of 100 pc after only 10 Myr, so this may indicate that the two components of TOR1 originate in completely different environments. We therefore use HDBSCAN's definition of the two subgroups presented above to split the population in two, and analyze the two halves separately throughout the rest of this paper. We hereafter refer to the larger TOR1-1 clump as TOR1A, and rename the smaller TOR1-0 clump to TOR1B. 

We update RV membership following the procedures in Section \ref{sec:velmem} for both populations given their new definitions, which results in one Li member/velocity non-member in Figure \ref{fig:limem} being redefined as a velocity member of TOR1B. The two components lie in a similar environment, and we therefore continue to use the original background fit for TOR1 shown in Figure \ref{fig:pfp_grid} for both components. However, the two subgroups have different RV ranges, changing their false positive rates. We show the RV ranges of both subgroups, along with the corresponding $P_{fp}$ for both subgroups in TOR1's panel in Figure \ref{fig:pfp_grid}. Using the updated membership assessments and $P_{fp}$ values, we update $P_{fin}$ for all members in both components. In Figure \ref{fig:cmds}, we present a CMD that distinguishes between probable members of each component, showing that TOR1A and TOR1B both have clear pre-main sequences with similar ages.

\section{Results} \label{sec:results}

In this section, we calculate the ages and population demographics of our associations. Our ages combine three independent methods, which we compare to converge on the best available age fits. Our demographics correct the known membership for completeness, contamination, and binaries to measure the number of members and mass produced by these star-forming events. Together, these values provide critical information on the scales and histories of these associations, which contextualize their formation environments. 

\subsection{Ages} \label{sec:ages}

\subsubsection{Isochronal Ages} \label{sec:isoages}

Isochronal ages are the most widely available method of age computation, requiring only absolute photometry, which is available for nearly 2 billion stars in \textit{Gaia} \citep[e.g.,][]{Soderblom14}. While disagreements between different isochronal models result in systematic uncertainties, isochronal ages are quite reliable in relative terms, as small differences in the height of the pre-main sequence can reliably indicate age differences of order a few Myr. 

\begin{figure*}
\centering
\setlength{\tabcolsep}{0pt}
\begin{tabular}{cccccc}
    \includegraphics[height=3.3cm]{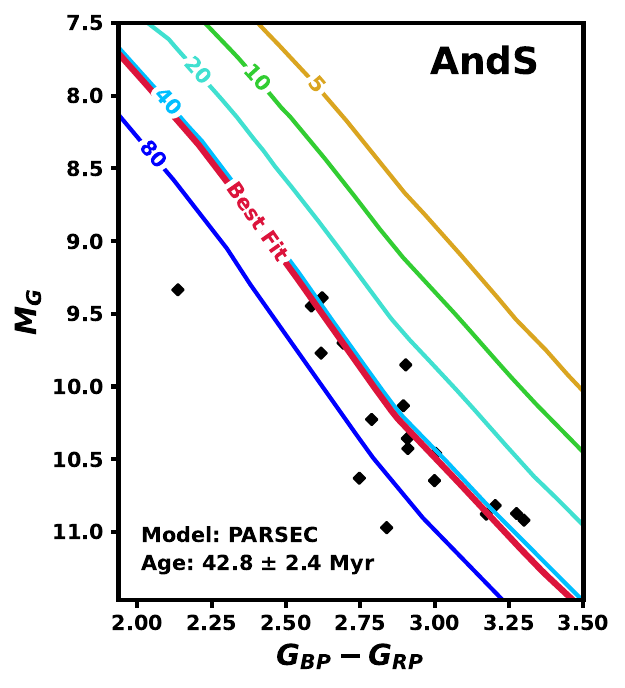}
    & \includegraphics[height=3.3cm]{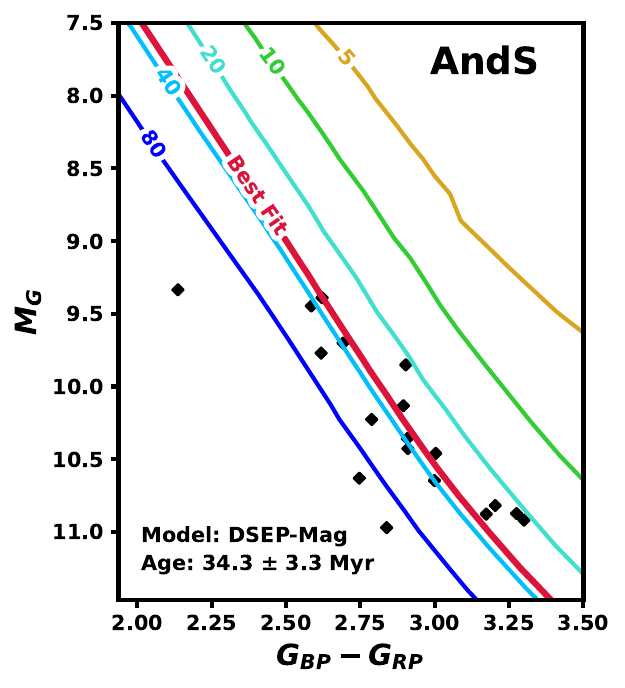}
    & \includegraphics[height=3.3cm]{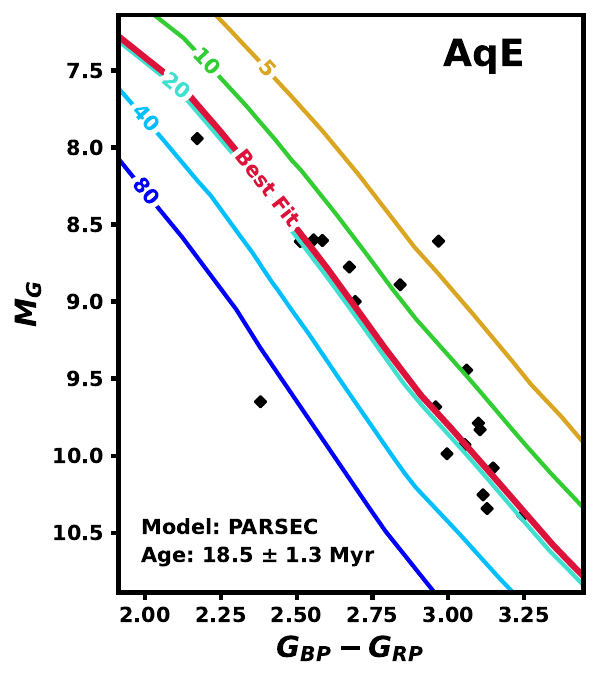}
    & \includegraphics[height=3.3cm]{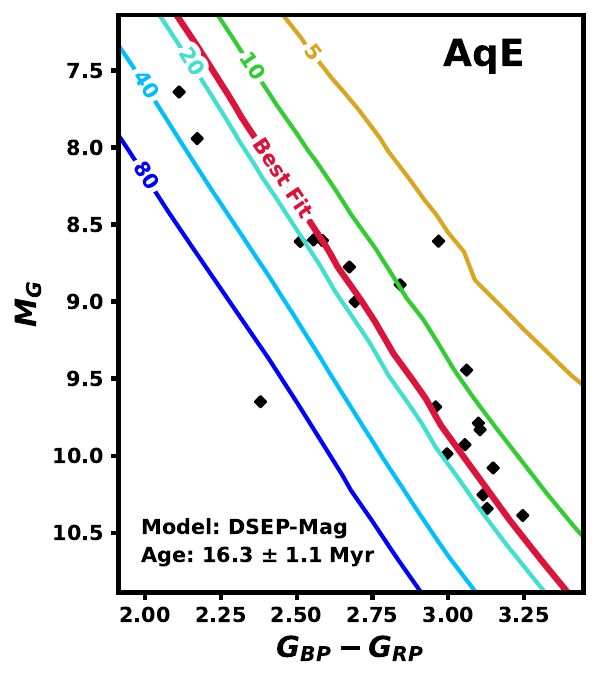}
    & \includegraphics[height=3.3cm]{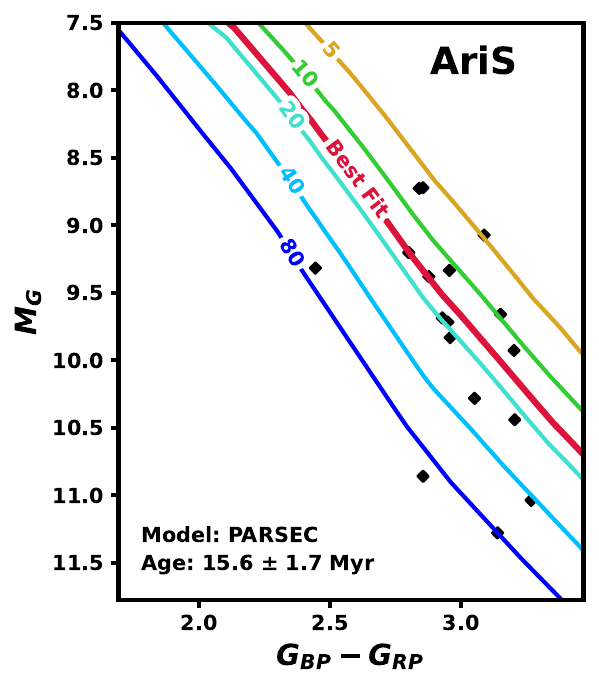}
    & \includegraphics[height=3.3cm]{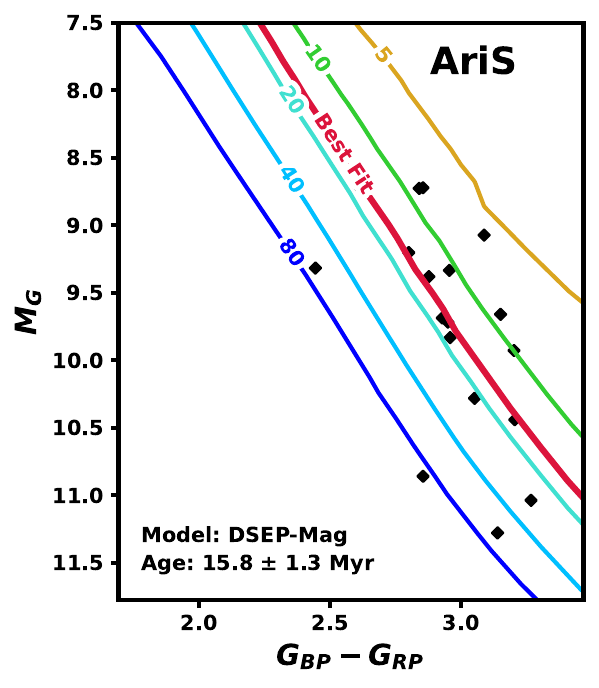}\\[-4pt]
    \includegraphics[height=3.3cm]{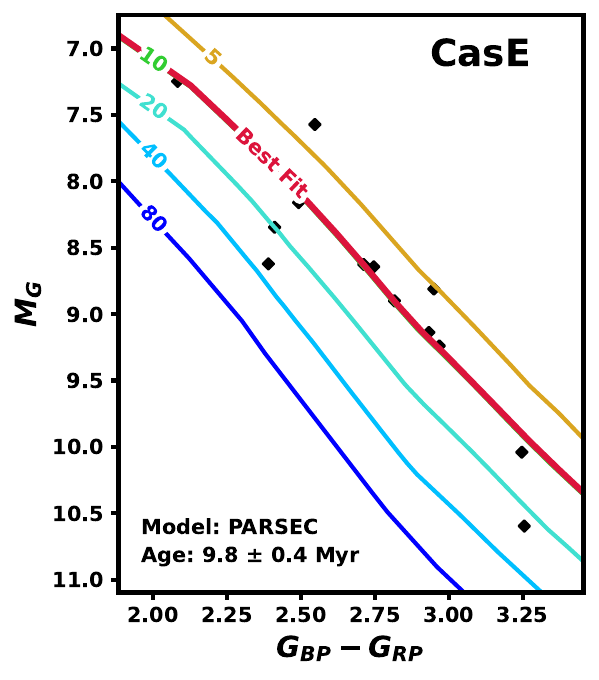}
    & \includegraphics[height=3.3cm]{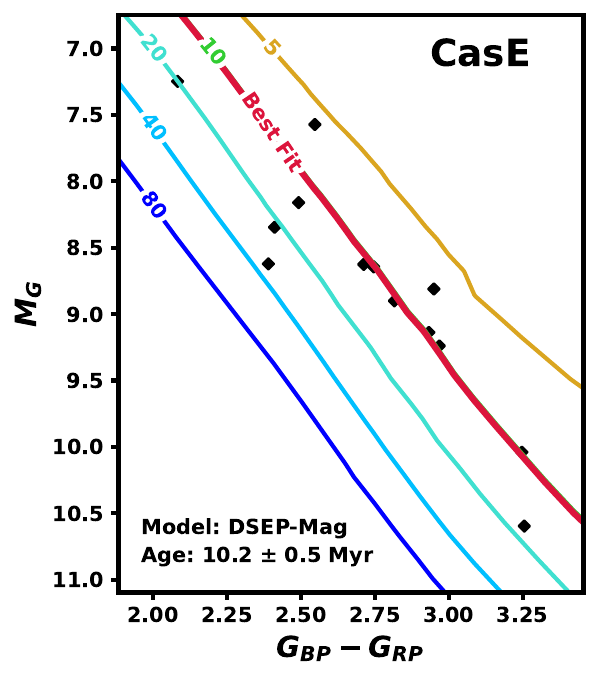}
    & \includegraphics[height=3.3cm]{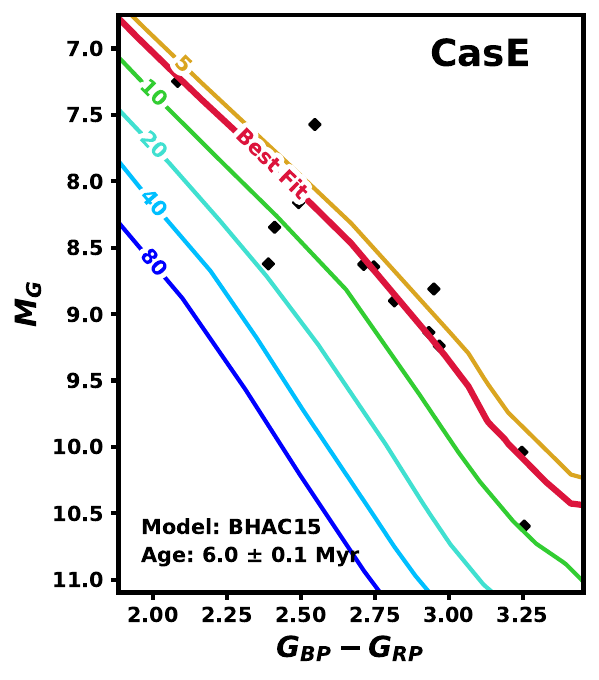}
    & \includegraphics[height=3.3cm]{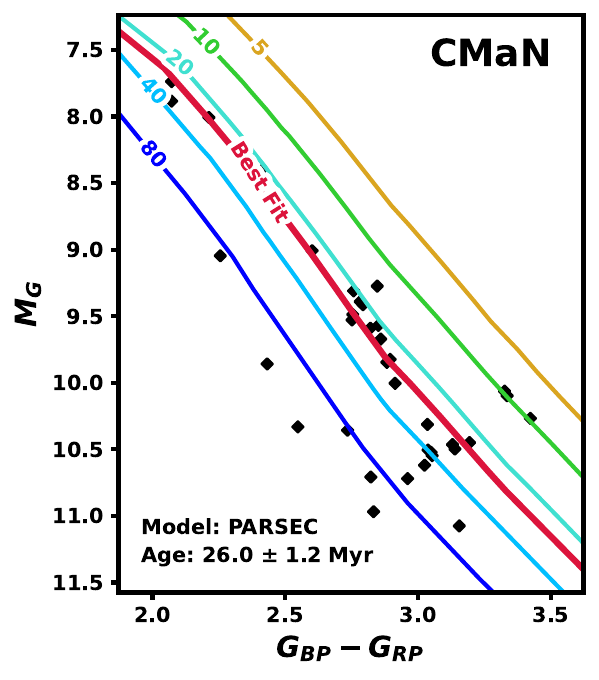}
    & \includegraphics[height=3.3cm]{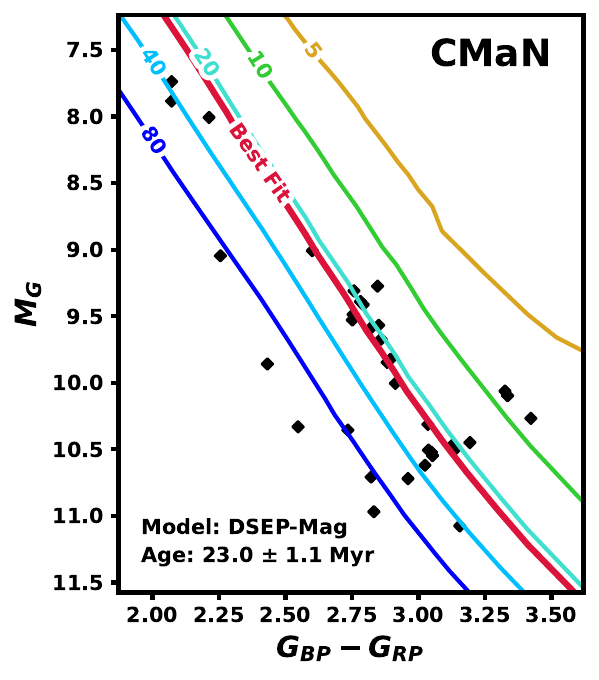}
    & \includegraphics[height=3.3cm]{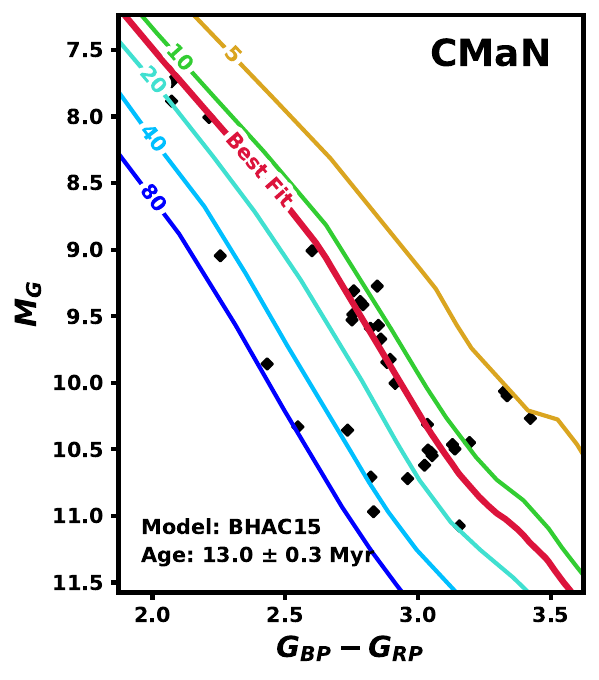}\\[-4pt]
    \includegraphics[height=3.3cm]{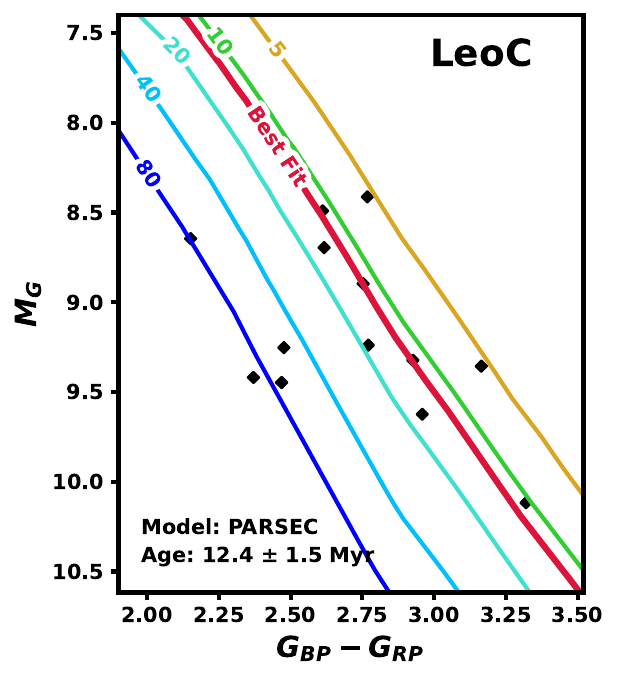}
    & \includegraphics[height=3.3cm]{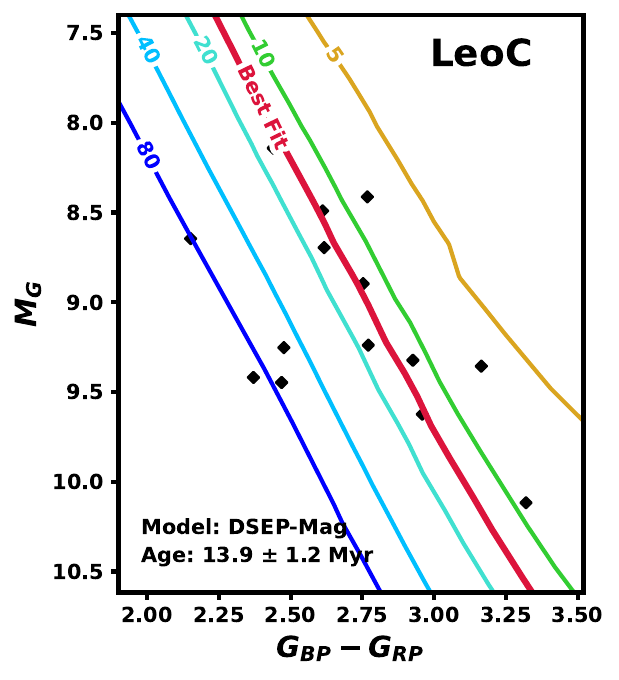}
    & \includegraphics[height=3.3cm]{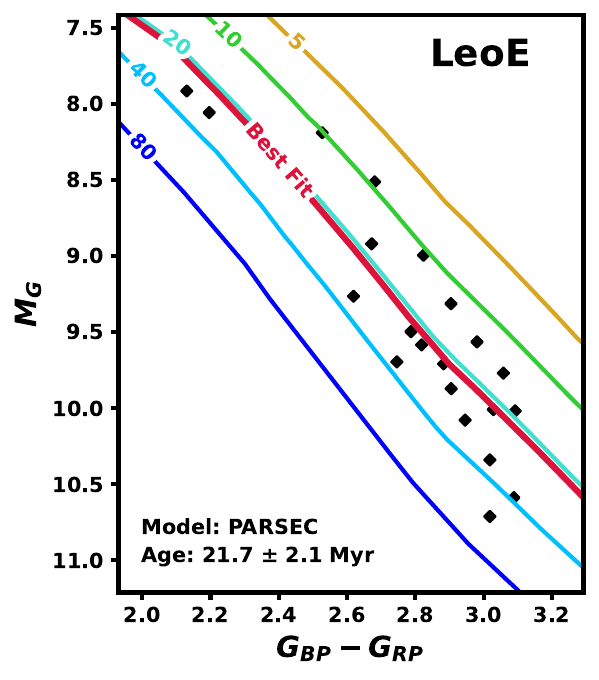}
    & \includegraphics[height=3.3cm]{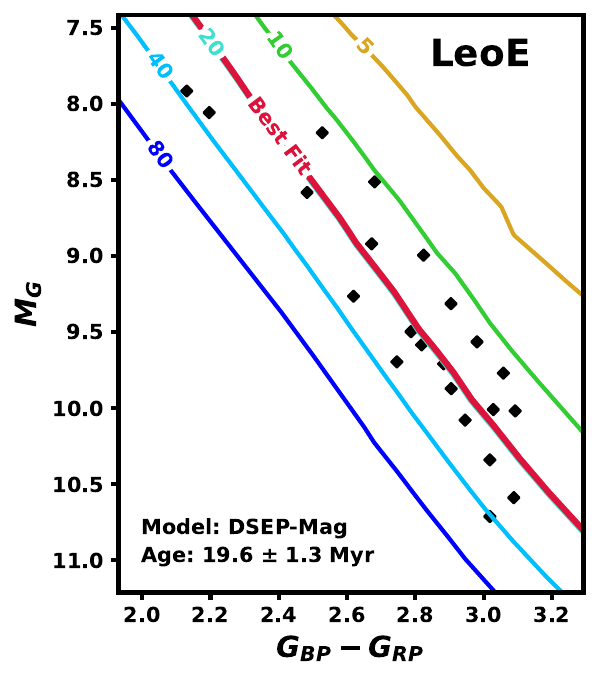}
    & \includegraphics[height=3.3cm]{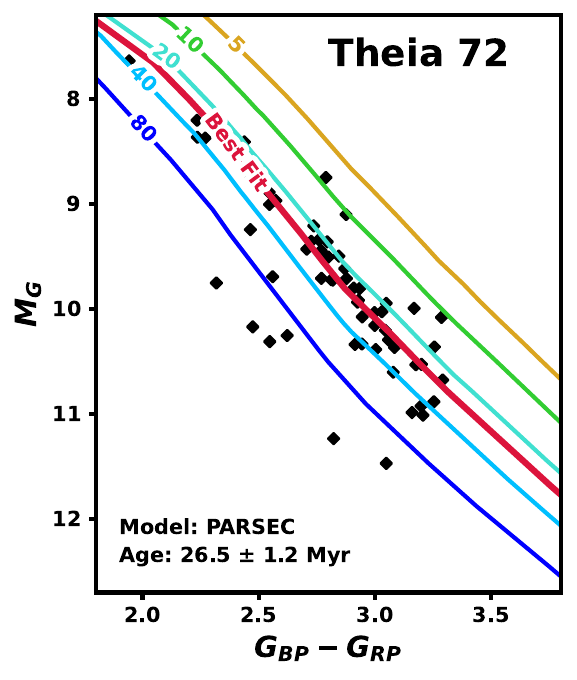}
    & \includegraphics[height=3.3cm]{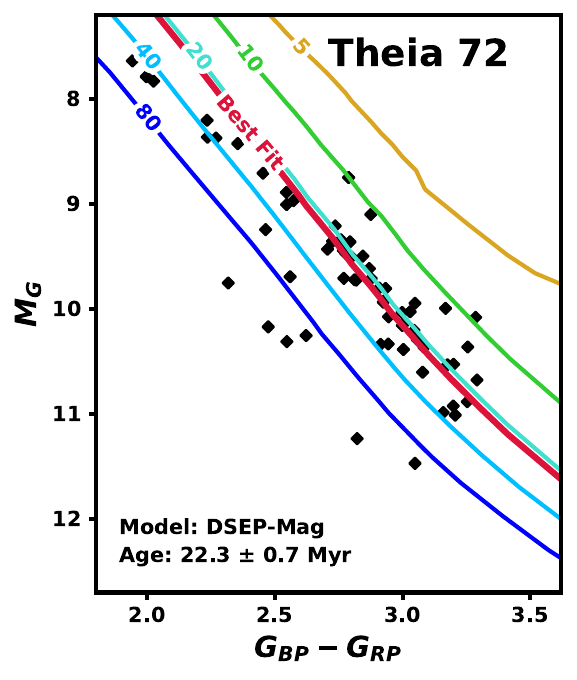}\\[-4pt]
    \includegraphics[height=3.3cm]{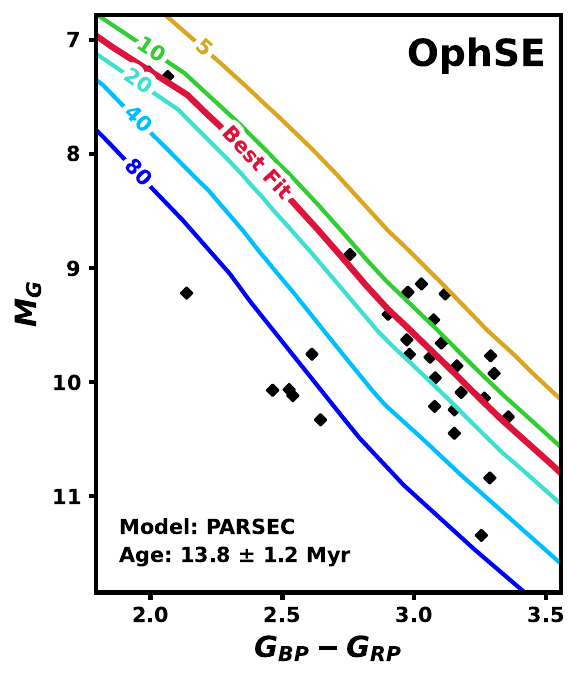}
    & \includegraphics[height=3.3cm]{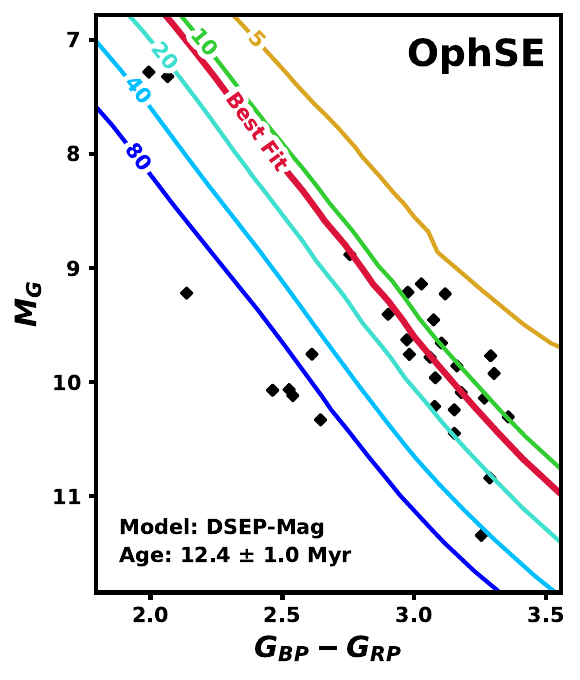}
    & \includegraphics[height=3.3cm]{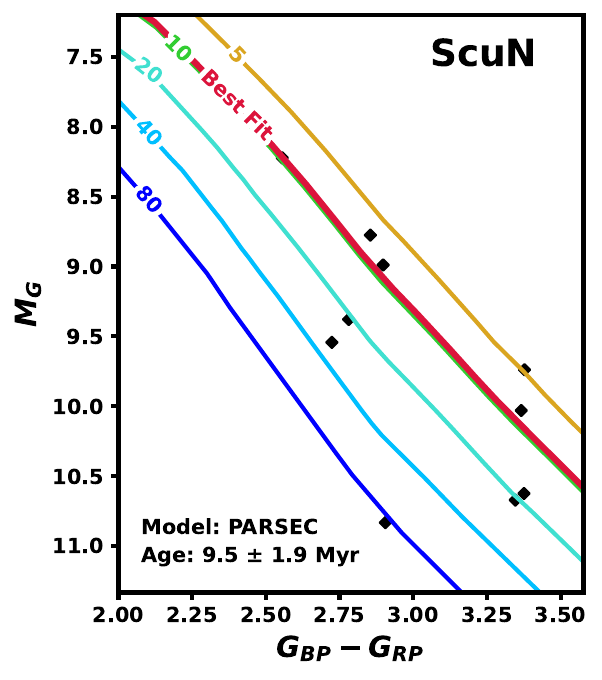}
    & \includegraphics[height=3.3cm]{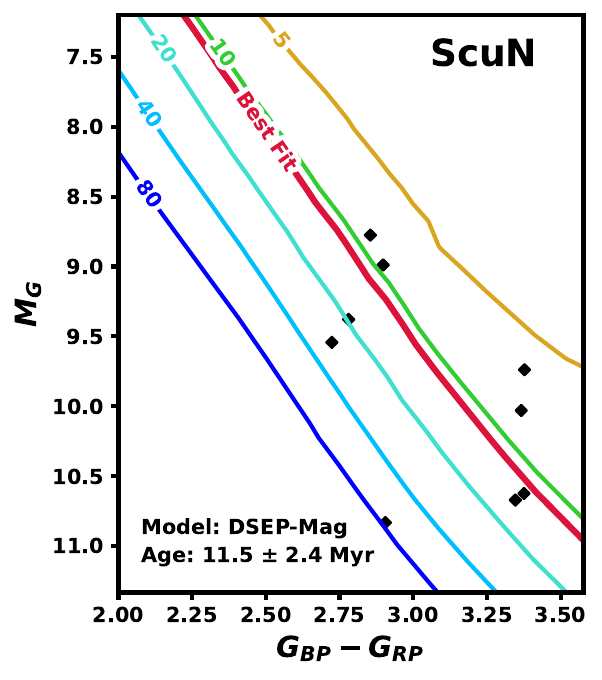}
    & \includegraphics[height=3.3cm]
    {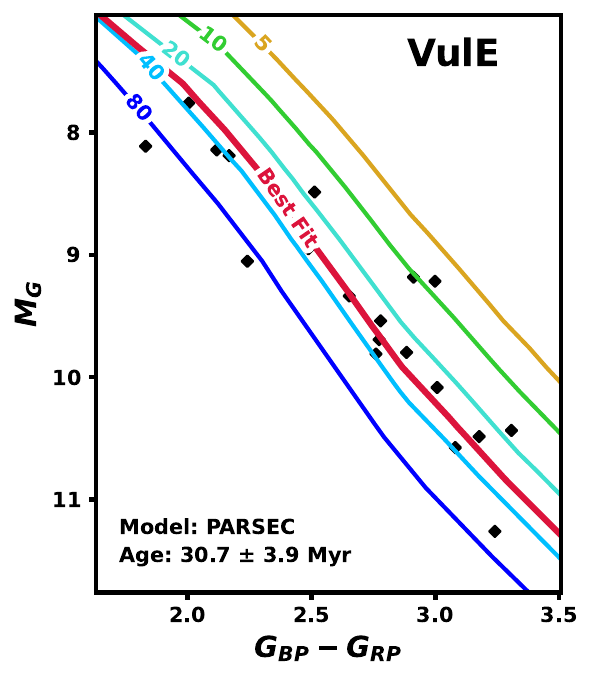}
    & \includegraphics[height=3.3cm]{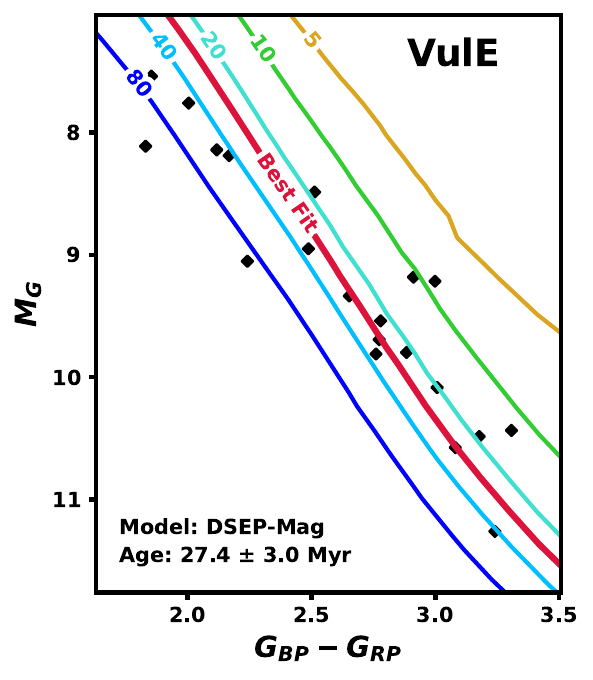}\\[-4pt]
    \includegraphics[height=3.3cm]{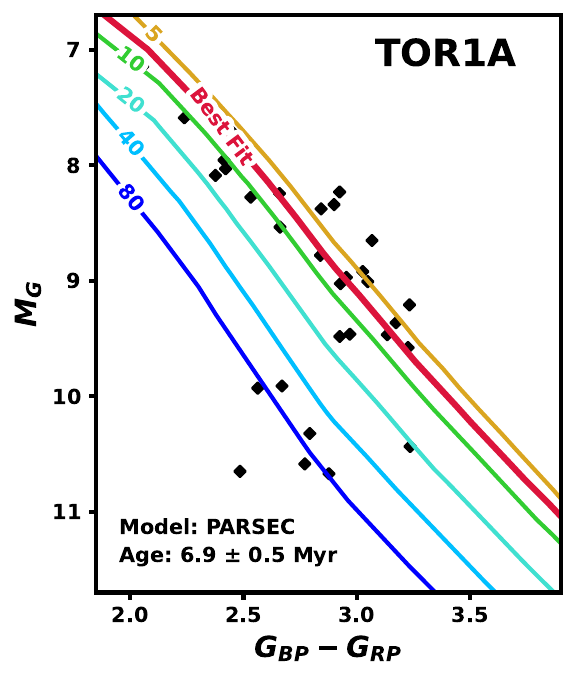}
    & \includegraphics[height=3.3cm]{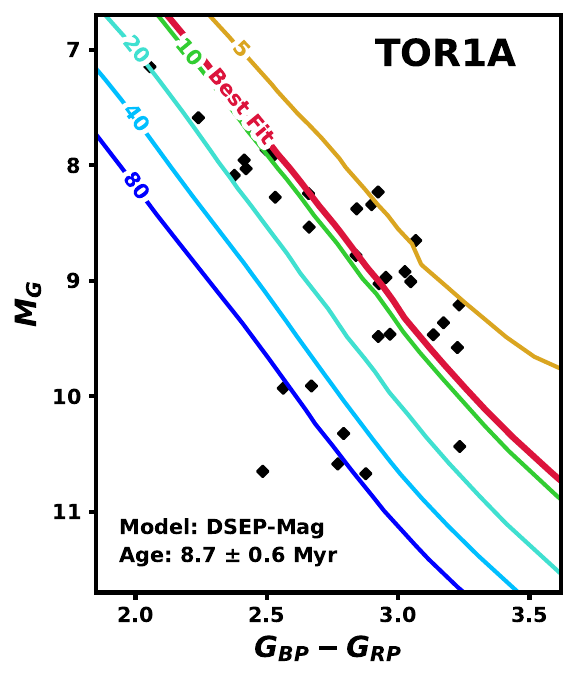}
    & \includegraphics[height=3.3cm]{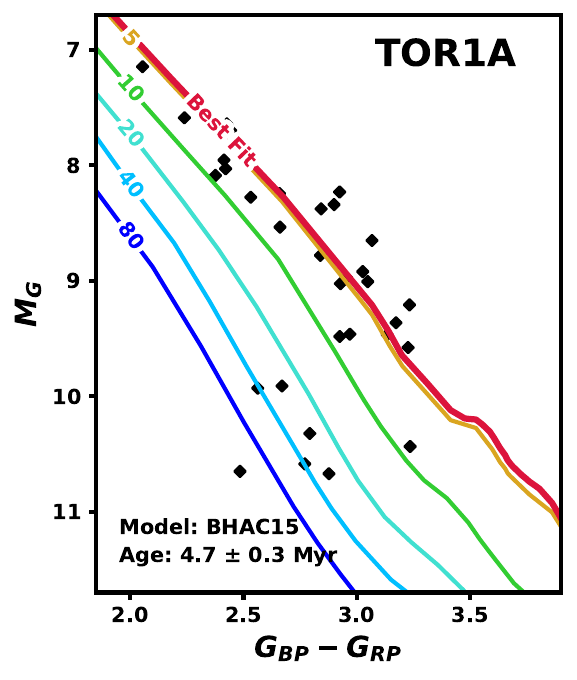}
    & \includegraphics[height=3.3cm]{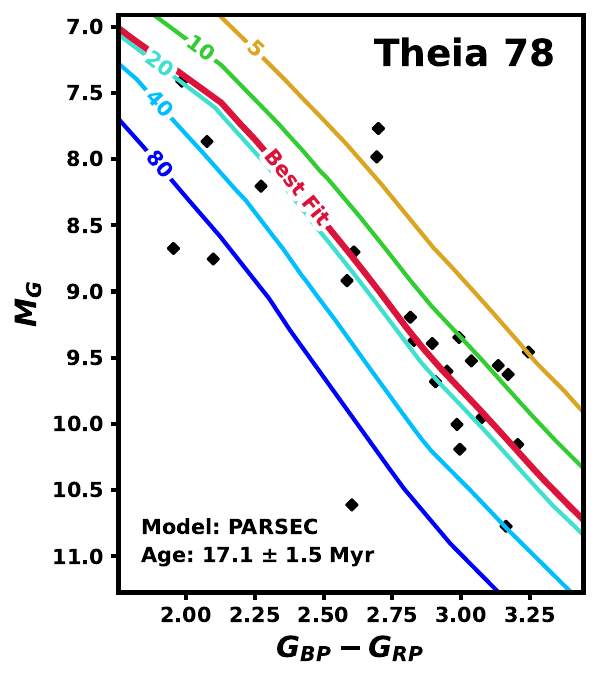}
    & \includegraphics[height=3.3cm]{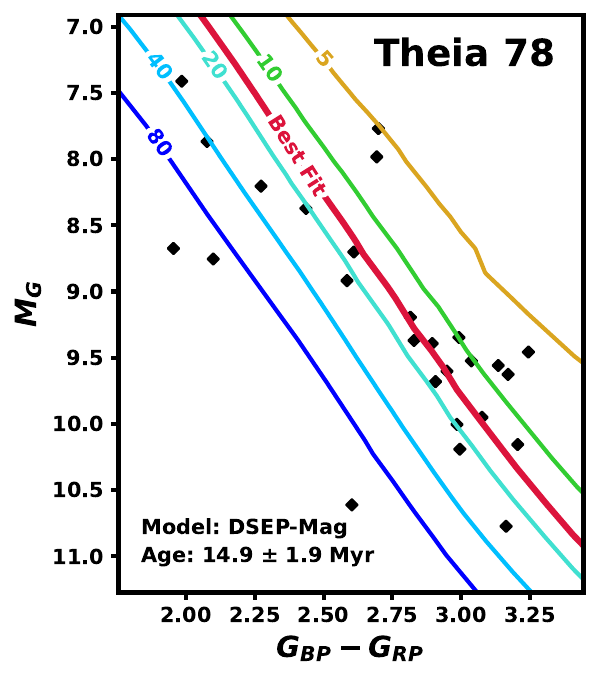}
    &\includegraphics[height=3.3cm]{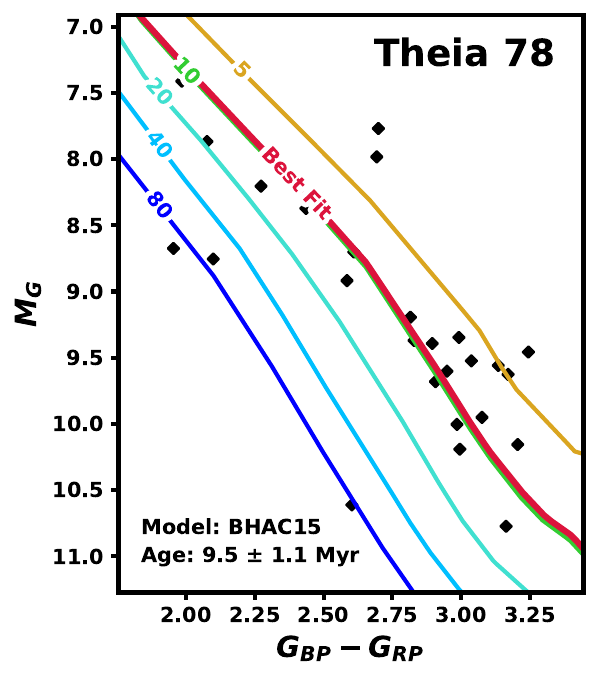}\\[-4pt]
    \includegraphics[height=3.3cm]{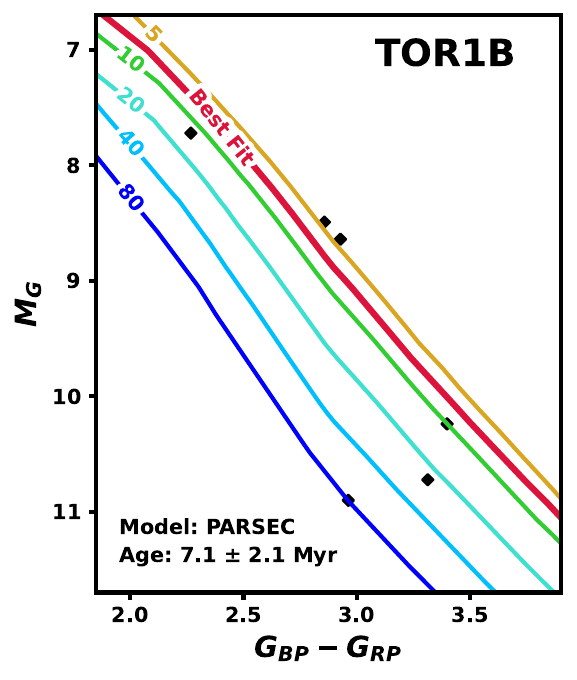}
    & \includegraphics[height=3.3cm]{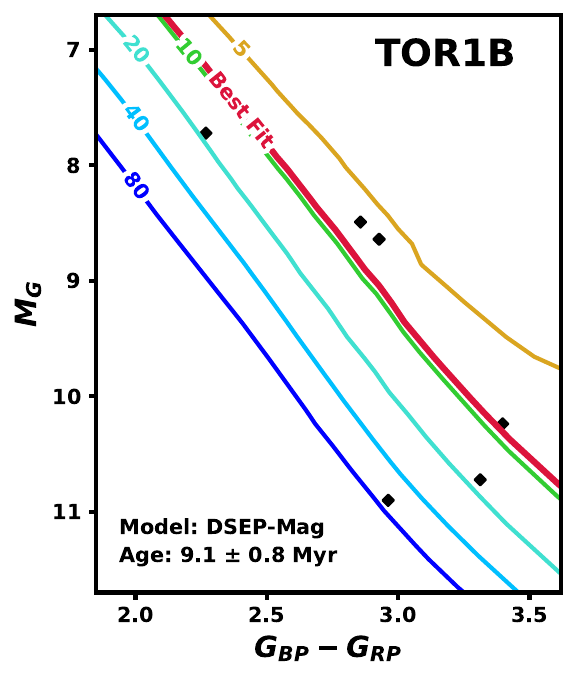}
    & \includegraphics[height=3.3cm]{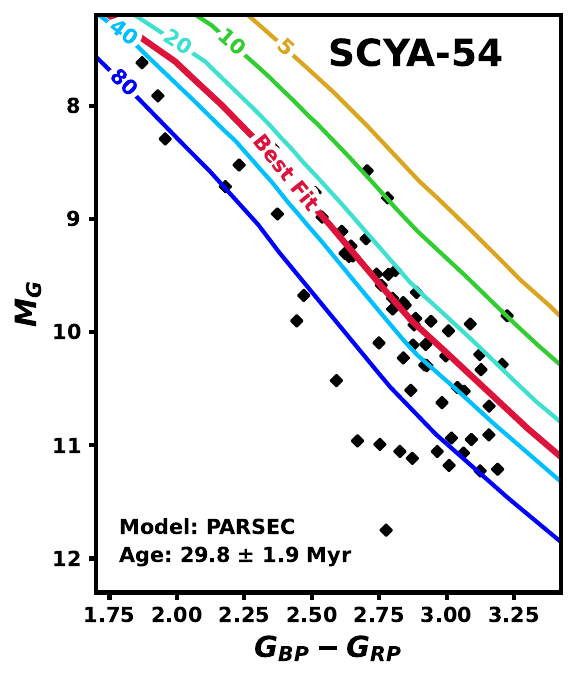}
    & \includegraphics[height=3.3cm]{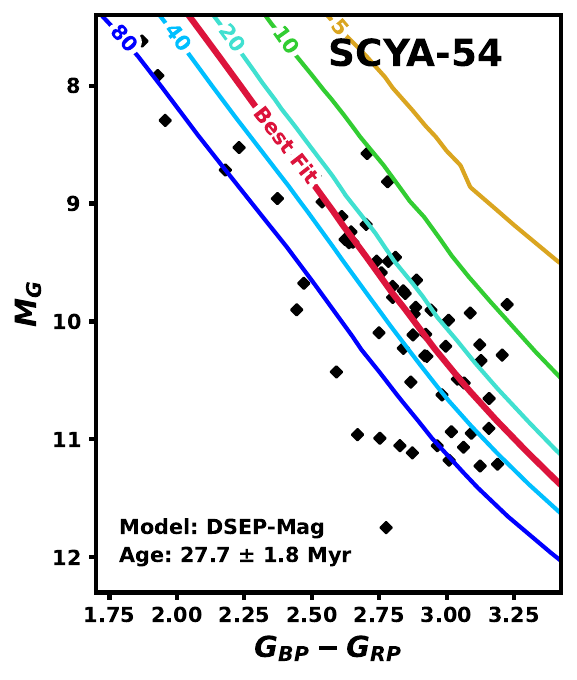}
    & \includegraphics[height=3.3cm]{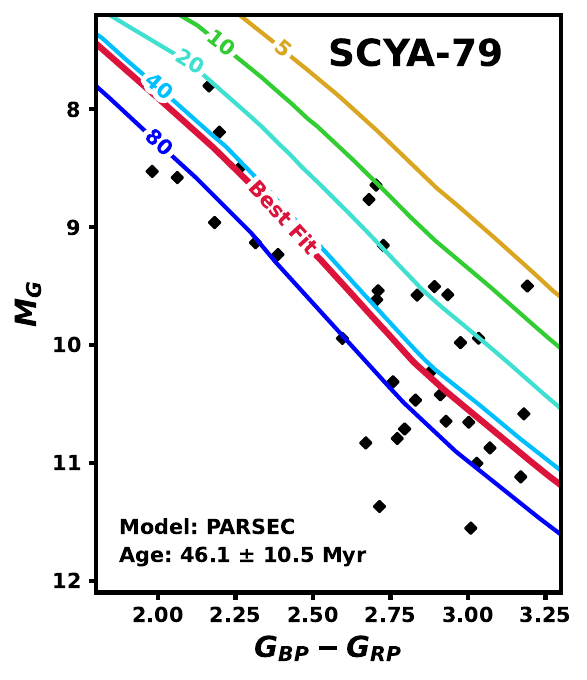}
    & \includegraphics[height=3.3cm]{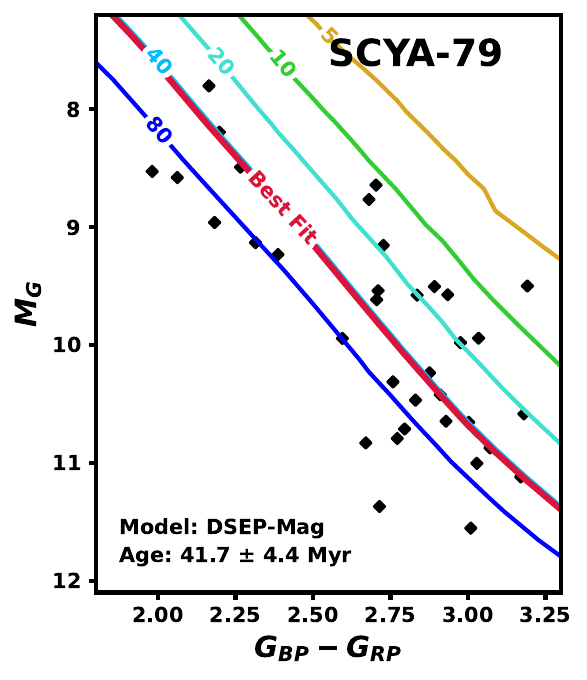}
\end{tabular}
\caption{Isochronal fits in $G_{BP}-G_{RP}$ vs $M_G$ absolute color-magnitude space, using PARSEC, BHAC15, and DSEP-Magnetic isochrones. The population, best fit age, and model are annotated in each panel. Stars included used in the fit are marked with black diamonds. We plot the best-fit isochrone alongside 5, 10, 20, 40, and 80 Myr isochrones (top to bottom) for reference. We show PARSEC and DSEP-Magnetic fits for all populations, as well as BHAC15 ages for select populations. The full set, including all models in addition to age fits for subgroups, is available in the online version.}
\label{fig:isochronefits}
\end{figure*}

To compute isochronal ages, we follow a generalized form of the fitting routine outlined in \citetalias{Kerr24}, designed for use with several different isochronal models. We restrict our stellar fitting to $1.8<G_{BP}-G_{RP}<4$, covering much of the pre-main sequence, and minimize the influence of binaries by removing stars with $RUWE>1.2$, following \citet{Bryson20}. We further restrict the sample to $RUWE<1.1$ in cases where this additional restriction does not reduce the number of available stars available below 15, providing a maximally pure sample of single stars in well-populated groups where further restriction does not adversely affect sample size. We also remove stars with mass $M < 0.2 M_{\odot}$ (see Sec. \ref{sec:mass}). Stars in our samples below this cutoff rarely follow the same isochrone as more massive members, and the offset differs by model. Their inclusion results in age solutions that grow older with the number of stars in this mass range, which correlates with distance. \citet{Wang25} acknowledged this pattern and corrected the PARSEC isochrones to remove it, although here we simply remove these low-mass stars for distance-unbiased results across all models. 

Finally, we impose two cuts on membership probability. To avoid excessive field contamination, we require $P_{fin}>0.1$, which removes clear non-members, without removing potential members that lie low on the pre-main sequence. This cut consistently removes most, but not all visually identifiable field stars, and only removes plausible pre-main sequence stars in cases of spectroscopic or velocity non-membership. Finally, we cut on $P_{spatial}$ to remove candidates in a field-contaminated regime farther from the region's space-velocity locus. For most populations, we impose $P_{spatial}>0.2$, but further restrict to $P_{spatial} > 0.3$ for TOR1B and to $P_{spatial} > 0.5$ for CMaN, TOR1A, Theia 72, ScuN, SCYA-54, and SCYA-79, ensuring that the pre-main sequence dominates the sample without the excessive removal of credible members. $P_{spatial}$ has no dependence on photometry, so restricting it does not risk biasing the age result unless unresolved age substructure is present. 

For each population, we take 10000 sub-sampled samples of the stars that survive the cuts above, selecting half of those stars at random to a minimum sample size of 5 with selection probability weighted by $P_{spatial}$. We then apply least-squares fitting with 2-$\sigma$ outlier removal to each sample, selecting a best fit isochrone model based on the $G_{BP}-G_{RP}$ and $M_G$ absolute photometry of the selected stars, corrected for \textit{Gaia} geometric distance \citep{BailerJones21} and extinction from \citet{Lallement19}. This routine provides a reliable uncertainty measurement while suppressing contamination from any remaining binaries or stars with excess emission from accretion \citep[e.g.,][]{Muzerolle98}. The final age solution and uncertainty are provided by the 2-$\sigma$ clipped mean and standard deviation, respectively, across all samples. We apply this fitting routine to all populations and subgroups included in this paper, using the BHAC15 \citep{BHAC15}, DSEP-Magnetic \citep{Feiden16}, and PARSEC \citep{PARSECChen15} isochronal models. The PARSEC and BHAC15 isochrones are chosen to cover sets of isochrones that produce similar age solutions in \citet{Herczeg15}, while the DSEP-Magnetic isochrones are chosen due to their success in reproducing young association ages computed using other methods, with those results accomplished by introducing strong magnetic fields to the model \citep[e.g.,][]{Kerr22a}. 

In Figure \ref{fig:isochronefits}, we show PARSEC and DSEP-Magnetic age fits to each of the 16 populations, along with some select BHAC15 isochrone fits. The online only version of this figure provides all isochronal age fits using all models, including subgroup-specific ages where relevant. We summarize the age solutions for each population and subgroup across all isochronal models in Table \ref{tab:ages}.

\subsubsection{Dynamical Ages} \label{sec:dynagecalc}

Dynamical ages are measured by tracing a population back in time through the galactic potential and measuring the time of most compact configuration, which marks the start of stellar dispersal after formation. These measurements typically assume minimal self-gravity (see Sec. \ref{sec:virialstates}), and that on the short timescales between the formation of these stars and today, external gravitational perturbations have a limited effect. \citet{Couture23} reviews the strengths and weaknesses of various dynamical age calculation methods, highlighting the importance of a pure sample of members and a reliable size metric. They note that the presence of observational uncertainties increases the velocity dispersion of a population, resulting in an overestimation of the rate of stellar dispersal. This produces an offset that, if uncorrected, results in age underestimation. We therefore compute dynamical ages that include a correction for this bias. 

We first correct our radial velocities for gravitational redshift, which is caused by general relativistic effects in the gravitational field of a star, and convective blueshift, which is caused by brightness differences of rising and falling convective cells producing line asymmetries that slightly blueshift the spectrum. As described in \citet{Couture23}, both effects change the velocity by a few tenths of a km s$^{-1}$, usually producing a net redshift. We correct for these effects for each star using the relations in \citet{Couture23}, which provide the expected shifts as a function of spectral class. We determine the spectral class for each star by first interpolating the $T_{eff}$ corresponding to its stellar mass for the parent association's best fit PARSEC isochrone.  We then convert from $T_{eff}$ to spectral class using the relationship from \citep{Herczeg14}, which covers F5 and later YSOs, and use the \citep{Pecaut13} relation for earlier type stars, which evolve quickly onto the main sequence. We do not apply a convective blueshift correction to APOGEE data, as this effect is most prominent at short wavelengths and often negligible in the near-infrared \citep{Meunier17}. We use these corrected RVs for dynamical traceback. 

To ensure that our sample contains only stars with reliable velocities, we restrict it to include only high-probability ($P_{fin}>0.8$) velocity-verified members without evidence for binarity (resolved or unresolved as indicated with $RUWE > 1.2$). We only include stars with $\sigma_{RV} < 1$ km s$^{-1}$, and exclude stars with only $\textit{Gaia}$ RVs, which have been shown to host systematic uncertainties at the 1 km s$^{-1}$ level among young stars \citep{Kounkel23}. We use the median length of all branches connecting stars as our distance metric, where branches are defined as a line connecting any two members, similar to the median mutual distance metric from \citetalias{Kerr22a}. In \citet{Couture25}'s study of Tuc-Hor, Carina, and Columba, the mean branch length provides the best combination of precision, accuracy, and dynamical age contrast, but we find that our use of the median instead of the mean produces more consistent results given the more limited datasets in our associations. 

\begin{figure*}
\centering
\setlength{\tabcolsep}{0pt}
\begin{tabular}{cccc}
    \includegraphics[height=3.3cm]{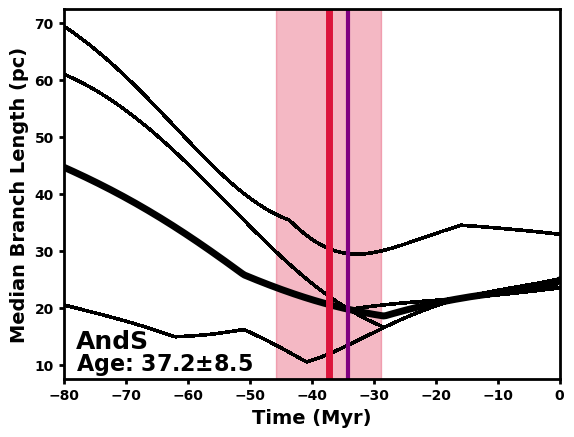}
    & \includegraphics[height=3.3cm]{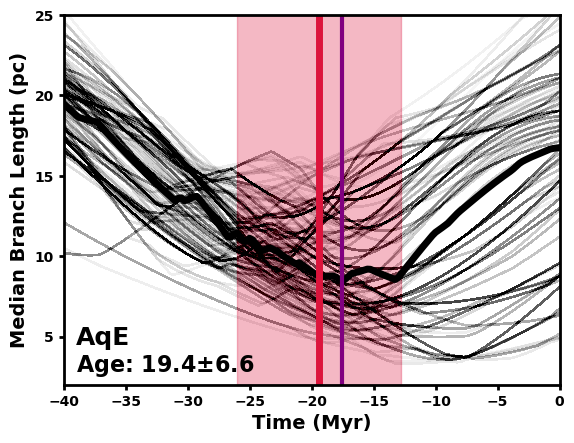}
    & \includegraphics[height=3.3cm]{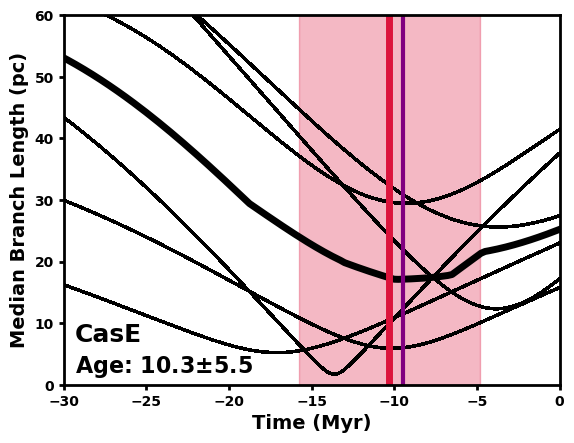}
    & \includegraphics[height=3.3cm]{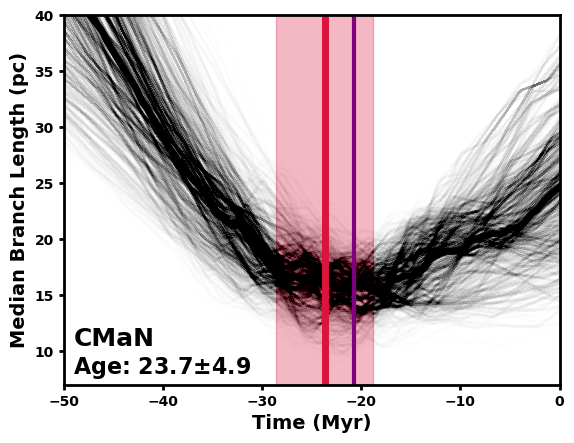}\\[-4pt]
    \includegraphics[height=3.3cm]{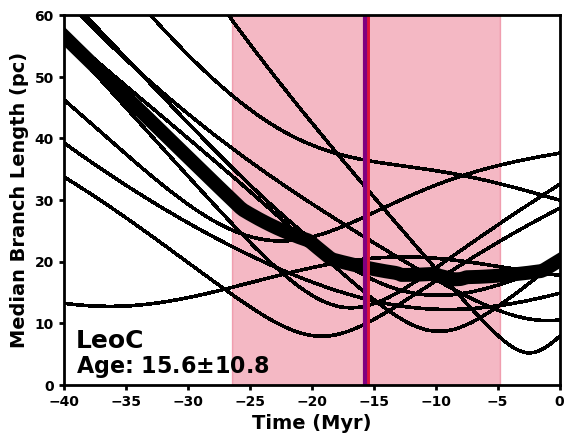}
    & \includegraphics[height=3.3cm]{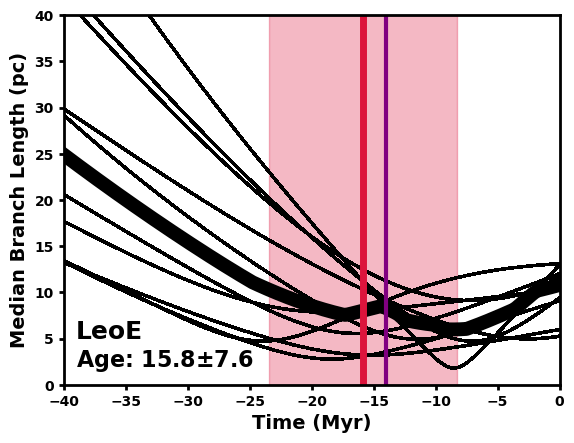}
    & \includegraphics[height=3.3cm]{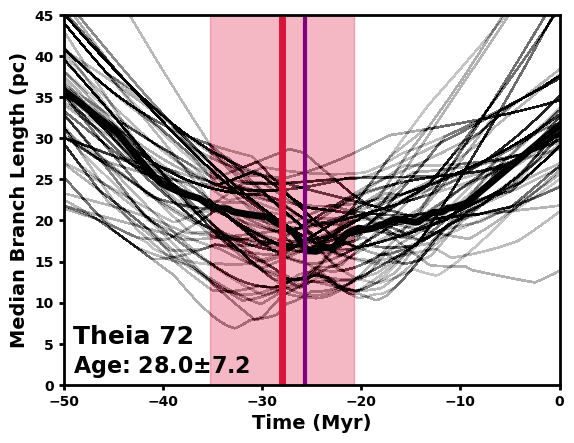}
    & \includegraphics[height=3.3cm]{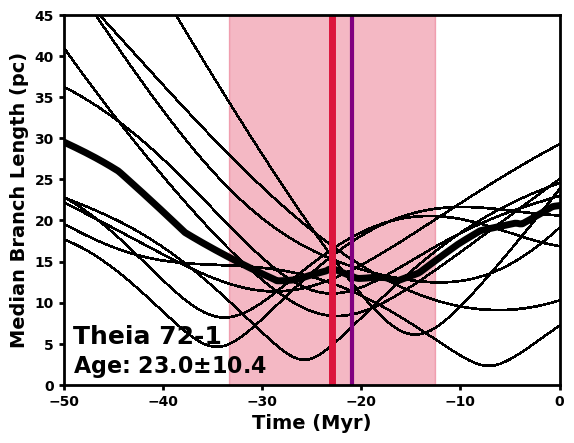}\\[-4pt]
    \includegraphics[height=3.3cm]{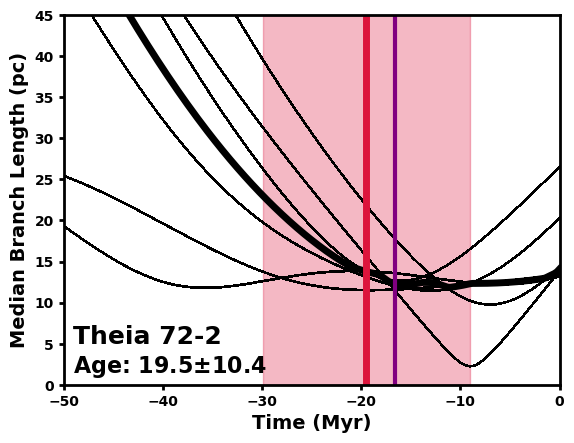}
    & \includegraphics[height=3.3cm]{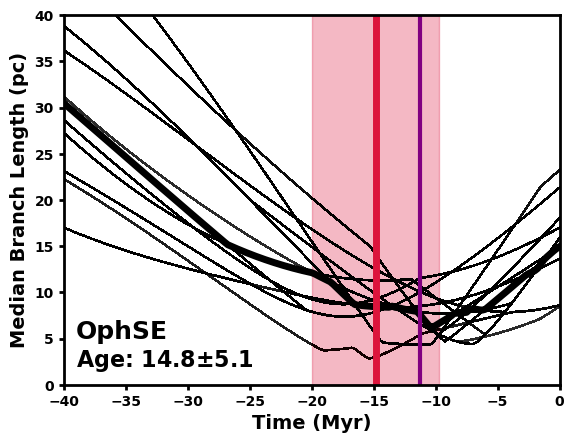}
    & \includegraphics[height=3.3cm]{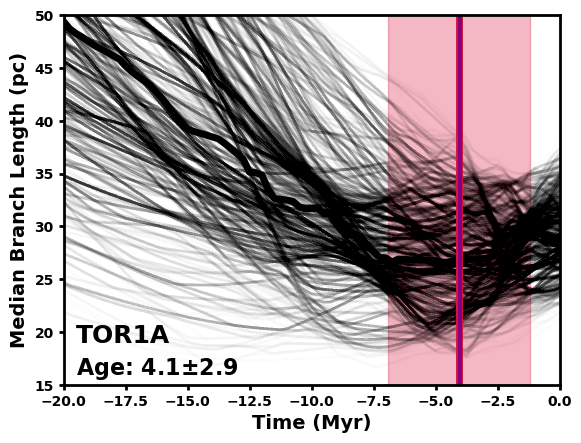}
    & \includegraphics[height=3.3cm]{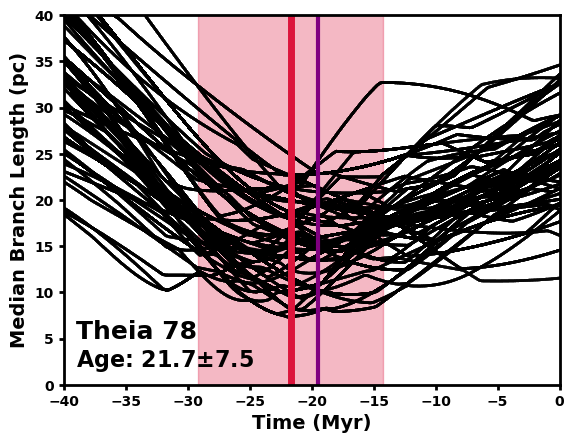}\\[-4pt]
    \includegraphics[height=3.3cm]{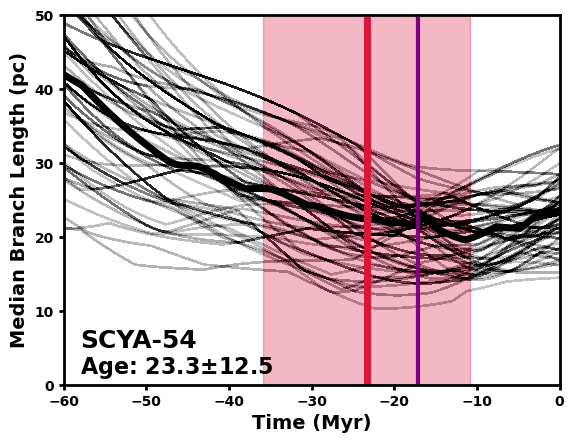}
    & \includegraphics[height=3.3cm]{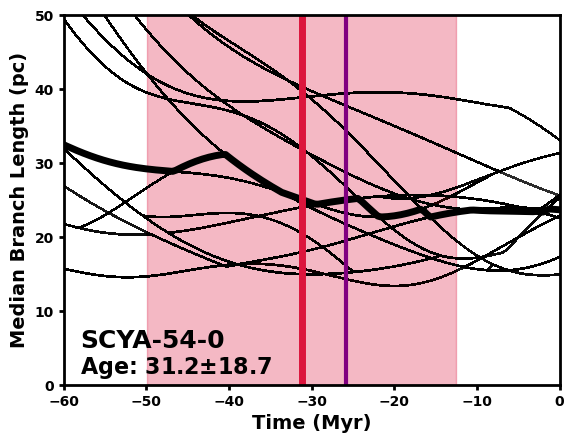}
    & \includegraphics[height=3.3cm]{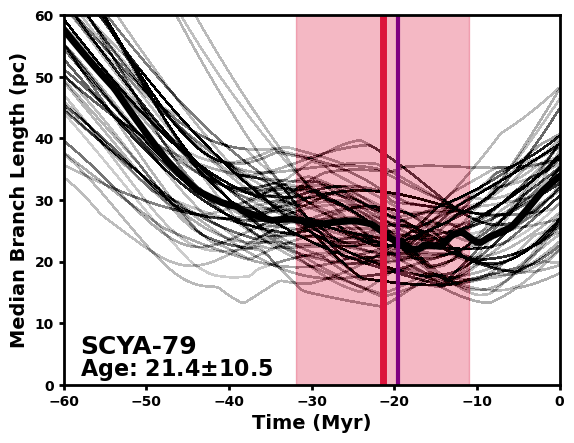}
    & \includegraphics[height=3.3cm]{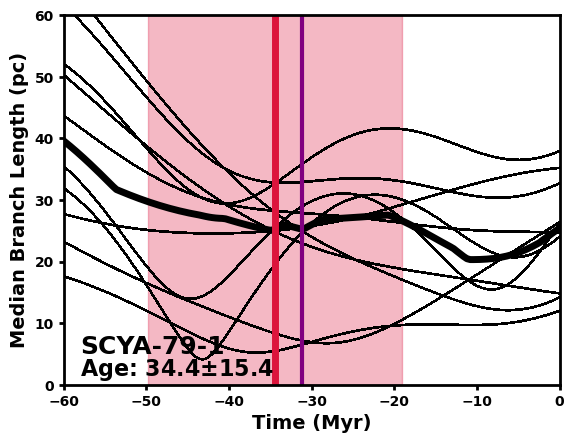}\\[-4pt]
\end{tabular}
\caption{Dynamical ages for all associations and sub-populations where a solution is possible. The black lines represent the median branch length over time for each monte carlo sample used in age calculation, and the thickest black line represents the median branch length curve of all members included in sampling. The thick vertical red line and shading represent the best fit dynamical age solution after debiasing and its 1-$\sigma$ confidence range, while the thinner purple vertical line provides the age solution before debiasing. The population and age solution are annotated.}
\label{fig:dynages}
\end{figure*}

Prior to computing dynamical ages, we further restrict the sample to eliminate outliers. We first trace stars back through the  \texttt{MWPotential2014} galactic potential using galpy \citep{Bovy15} and assuming the \citet{Hogg05} solar motion. We trace back 50 Myr for most populations and 80 Myr for the older populations of AndS, SCYA-54, and SCYA-79. The choice of model was shown in \citet{MiretRoig20} to have a minimal impact on dynamical age results for populations like those in our sample, so \texttt{MWPotential2014} is chosen primarily due to its computational efficiency. Following \citet{Couture23}, we eliminate any objects that deviate more than 3-$\sigma$ from the median velocity and position of the association at any point during that traceback history. Finally, we remove stars that reach their minimum average distance from other stars within 5 Myr of the present day for PARSEC ages $\tau > 20$Myr, within 1 Myr for populations with $10 <\tau < 20$Myr, and at 0 Myr for $\tau<10$Myr. This excludes objects with velocities inconsistent with dispersal from a common origin on a timescale similar to their presumed ages. These cuts usually remove less than 20\% of candidates, but these restrictions can be quite impactful in associations with fewer reliable RVs. 

To compute uncorrected dynamical ages, we take 5000 Monte Carlo samples containing half of the stars in each final restricted stellar sample to minimum of 3, and compute the time of minimum median branch length in the galpy traceback results of each sample. The dynamical age and uncertainty are set by the average and standard deviation of those times, respectively, excluding times less than 1 Myr for PARSEC ages of 10-20 Myr, and less than 5 Myr for PARSEC ages larger than 20 Myr. Samples that fail these cuts need not contain bad velocities, but could consist of stars with very similar velocities that have dispersed little since formation, making their branch lengths uninformative.

\begin{deluxetable*}{cccccccccccccccccc}
\tablecolumns{17}
\tablewidth{0pt}
\tabletypesize{\scriptsize}
\tablecaption{Lithium depletion, dynamical, and isochronal ages for all associations and subgroups in our sample. All values are in Myr. PARSEC isochronal ages are adopted as the best age solutions in Section \ref{sec:agesynth}.}
\label{tab:ages}
\tablehead{
\colhead{Association} &
\colhead{SG} &
\multicolumn{6}{c}{Lithium Depletion\tablenotemark{a}} &
\multicolumn{3}{c}{Dynamical} &
\multicolumn{6}{c}{Isochrones\tablenotemark{b}} \\
\colhead{} &
\colhead{} &
\multicolumn{3}{c}{Analytical} &
\multicolumn{3}{c}{ML} & 
\multicolumn{3}{c}{} & 
\multicolumn{2}{c}{PARSEC\tablenotemark{c}} &
\multicolumn{2}{c}{BHAC15} &
\multicolumn{2}{c}{DSEP-Mag.}\\
\colhead{} &
\colhead{} &
\colhead{val} &
\colhead{lerr} &
\colhead{uerr} &
\colhead{val} &
\colhead{lerr} &
\colhead{uerr} &
\colhead{val} &
\colhead{err} &
\colhead{corr\tablenotemark{d}} &
\colhead{val} &
\colhead{err} &
\colhead{val} &
\colhead{err} &
\colhead{val} &
\colhead{err}
}
\startdata
AndS &  & 41.2 & 10.7 & 18.4 & 44.7 & 12.3 & 23.7 & 37.2 & 8.5 & 3.2 & 42.8 & 2.4 & 18.2 & 1.4 & 34.3 & 3.3 \\
AqE &  & 18.0 & 2.0 & 2.0 & 20.0 & 2.4 & 2.4 & 19.4 & 6.6 & 1.9 & 18.5 & 1.3 & 10.1 & 0.4 & 16.3 & 1.1 \\
AriS &  & 14.3 & 6.5 & 9.7 &  &  & 29.9 &  &  &  & 15.6 & 1.7 & 9.3 & 0.6 & 15.8 & 1.3 \\
CasE &  & 14.6 & 1.7 & 2.0 & 14.8 & 3.3 & 3.0 & 10.3 & 5.5 & 0.8 & 9.8 & 0.4 & 6.0 & 0.1 & 10.2 & 0.5 \\
CMaN &  & 26.3 & 2.0 & 2.2 & 24.5 & 1.6 & 2.4 & 23.7 & 4.9 & 3.0 & 26.0 & 1.2 & 13.0 & 0.3 & 23.0 & 1.1 \\
LeoE &  & 34.7 & 5.5 & 6.5 & 31.3 & 6.1 & 9.5 & 15.8 & 7.6 & 1.9 & 21.7 & 2.1 & 10.7 & 0.6 & 19.6 & 1.3 \\
LeoC &  &  &  & 17.2 & 12.2 & 4.9 & 6.0 & 15.6 & 10.8 & -0.1 & 12.4 & 1.5 & 7.8 & 0.6 & 13.9 & 1.2 \\
Theia 72 &  & 43.7 & 6.9 & 9.4 & 35.9 & 6.7 & 12.0 & 28.0 & 7.2 & 2.3 & 26.5 & 1.2 & 13.1 & 0.4 & 22.3 & 0.7 \\
Theia 72 & 0 &  &  &  &  &  &  &  &  &  & 25.7 & 1.0 & 10.8 & 1.1 & 19.4 & 2.7 \\
Theia 72 & 1 &  &  &  &  &  &  & 23.0 & 10.4 & 2.0 & 30.8 & 2.2 & 14.4 & 0.6 & 27.5 & 1.5 \\
Theia 72 & 2 &  &  &  &  &  &  & 19.5 & 10.4 & 2.9 & 24.7 & 0.6 & 12.4 & 0.6 & 21.3 & 0.8 \\
OphSE &  &  &  & 7.2 &  &  & 15.8 & 14.8 & 5.1 & 3.6 & 13.8 & 1.2 & 7.5 & 0.7 & 12.4 & 1.0 \\
ScuN & & 19.5 & 2.3 & 2.6 & 19.5 & 2.9 & 2.4 &  &  &  & 9.5 & 1.9 & 5.9 & 0.8 & 11.5 & 2.4 \\
TOR1A &  &  &  & 15.5 & 11.5 & 4.6 & 5.3 & 4.1 & 2.9 & 0.1 & 6.9 & 0.5 & 4.7 & 0.3 & 8.7 & 0.6 \\
Theia 78 &  & 24.8 & 3.5 & 3.7 & 23.4 & 4.2 & 4.1 & 21.7 & 7.5 & 2.2 & 17.1 & 1.5 & 9.5 & 1.1 & 14.9 & 1.9 \\
TOR1B &  & 50.7 & 22.2 & 38.4 & 37.6 & 23.8 & 45.6 &  &  &  & 7.1 & 2.1 & 4.7 & 0.8 & 9.1 & 0.8 \\
VulE &  & 57.5 & 10.8 & 16.6 & 63.1 & 18.9 & 29.2 &  &  &  & 30.7 & 3.9 & 15.5 & 2.7 & 27.4 & 3.0 \\
SCYA-54 &  & 51.3 & 13.3 & 15.5 & 41.7 & 13.5 & 23.6 & 23.3 & 12.5 & 6.2 & 29.8 & 1.9 & 15.5 & 0.8 & 27.7 & 1.8 \\
SCYA-54 & 0 &  &  &  &  &  &  & 31.2 & 18.7 & 5.4 & 36.3 & 5.2 & 16.0 & 1.1 & 29.0 & 2.8 \\
SCYA-54 & 1 &  &  &  &  &  &  &  &  &  & 28.4 & 1.5 & 14.4 & 1.6 & 26.4 & 2.1 \\
SCYA-79\tablenotemark{e} &  & 35.5 & 12.6 & 17.0 & 29.9 & 11.0 & 16.9 & 21.4 & 10.5 & 1.9 & 46.1 & 10.5 & 19.0 & 2.1 & 41.7 & 4.4 \\
SCYA-79 & 0 &  &  &  &  &  &  &  &  &  & 50.0 & 18.0 & 18.7 & 2.2 & 39.2 & 5.7 \\
SCYA-79 & 1 &  &  &  &  &  &  & 34.4 & 15.4 & 3.2 & 31.5 & 6.1 & 15.1 & 3.3 & 30.3 & 5.6 \\
\enddata
\tablenotetext{a}{lerr and uerr denote upper and lower 1-$\sigma$ uncertainties. The uerr value denotes an upper limit in cases where only that column is filled.}
\tablenotetext{b}{Uncertainties in isochronal ages are statistical, and therefore do not consider systematic uncertainties.}
\tablenotetext{c}{The PARSEC isochronal ages are our preferred solutions.}
\tablenotetext{d}{Debiasing factor to dynamical age described in Section \ref{sec:dynagecalc}.}
\tablenotetext{e}{The field sequence and pre-main sequence separate poorly in SCYA-79, resulting in unreliable isochronal ages. Our adopted age provides only an upper limit based on the Lithium depletion age and dynamical age for SCYA-79-1.}
\vspace*{0.1in}
\end{deluxetable*}

We then compute bias corrections to account for observational uncertainties in our uncorrected dynamical age results, roughly following the routine outlined in \citet{Couture23}. We first use galpy to trace each star in the dynamical age calculation back to its PARSEC isochronal age, and construct a model of the association at that time. We center the model at the association's mean X, Y, Z, and corresponding velocities at its PARSEC age, and set the velocity dispersion to the average dispersion across $\Delta v_{T, l}$ and $\Delta v_{T, b}$. The initial size of the association is more difficult to estimate, as molecular clouds vary in size, and traceback uncertainties are too large to resolve most parent cloud sizes. The mean standard deviation across X, Y, and Z at the time of its uncorrected dynamical age serves as a plausible upper limit, but the cloud size is otherwise largely unconstrained. Known clouds have scales ranging from around 1 pc, like Rho Oph and NGC 1333, up to tens of pc, like some filaments in the Taurus complex \citep[e.g.,][]{Ladjelate20, Krolikowski21, Soler23}. Without a strong reason to favor a small, compact cloud over a larger filamentary structure, we Monte Carlo sample the initial size with a value drawn from a flat distribution between 0.5 pc and the position standard deviation for each population at its uncorrected dynamical age, which ranges from 3-12 pc. 

For each population, we generate 1000 initial sizes, and for each size, we select positions and velocities from gaussian distributions centered on the average values above, and using the random population size and velocity standard deviations to set the corresponding widths. We then trace each model population forward to the present day, convert to sky coordinates, add random uncertainties based on the mean distance, proper motion, and RV uncertainties in the sample, and then apply our dynamical age routine to those generated stars. The dynamical age bias correction is set by the difference between the average result and the input PARSEC age, and the uncertainty is set by the corresponding standard deviation. The resulting offsets range from $\sim 0$ to $6$ Myr. The final dynamical age solution is the sum of these corrective values and the uncorrected dynamical age. We show the final dynamical age solutions for all populations where a fit is possible in Figure \ref{fig:dynages}, and summarize the results in Table \ref{tab:ages}. We do not report ages where the 1-$\sigma$ lower limit is within 1 Myr of the limit used for cutting stars or samples. 

\citet{Couture23} find that the initial size does not substantially affect the dynamical age bias, but this conclusion assumes a velocity dispersion of 1 km s$^{-1}$. Many of our associations have $\sigma_v < 0.5$ km s$^{-1}$, and we find that in this regime, the size of this bias is sensitive to the initial size of the population. In the case of CMaN, an initial size of 1 pc produces a mean offset of $\sim 2$ Myr, while a 10 pc size produces an offset of 6 Myr. Populations derived from filamentary cloud complexes may therefore have their dynamical ages systematically underestimated. 

\subsubsection{Lithium} \label{sec:liages}

\begin{figure*}
\centering
\setlength{\tabcolsep}{0pt}
\begin{tabular}{cccc}
    \includegraphics[height=2.4cm]{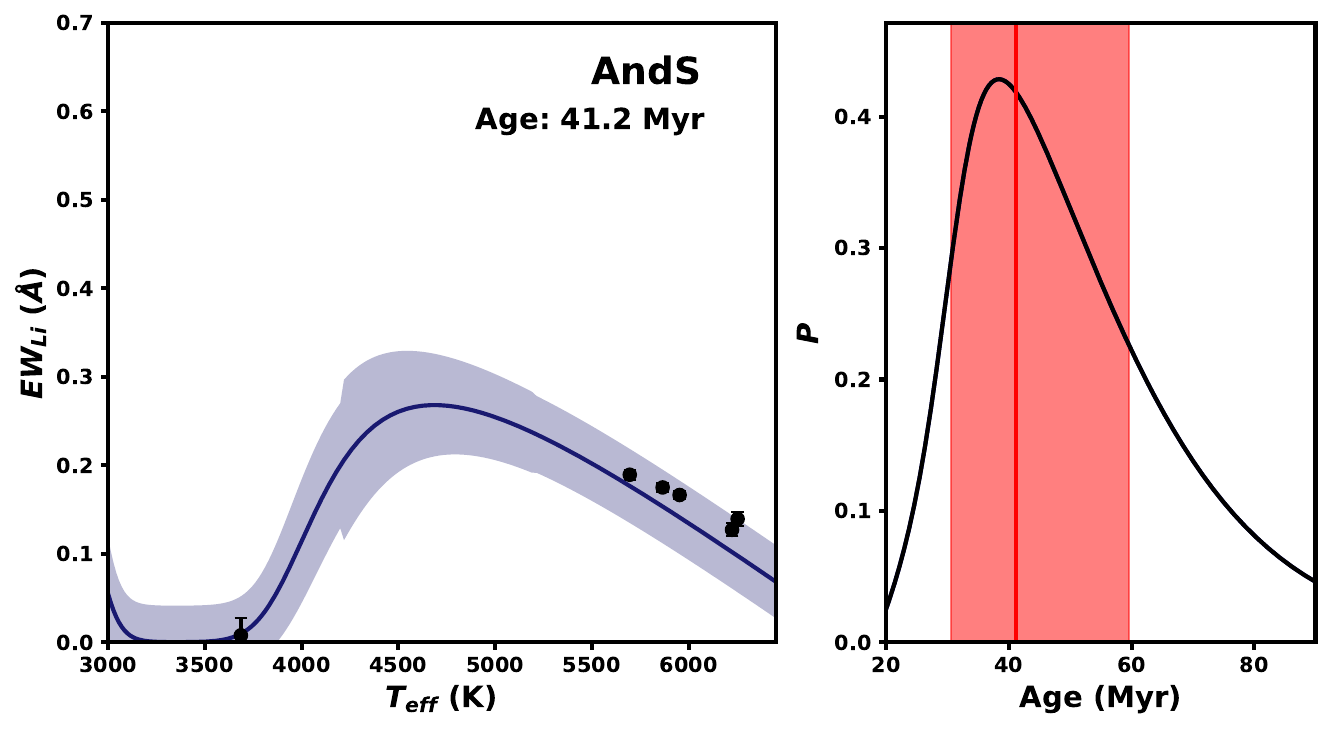}
    & \includegraphics[height=2.4cm]{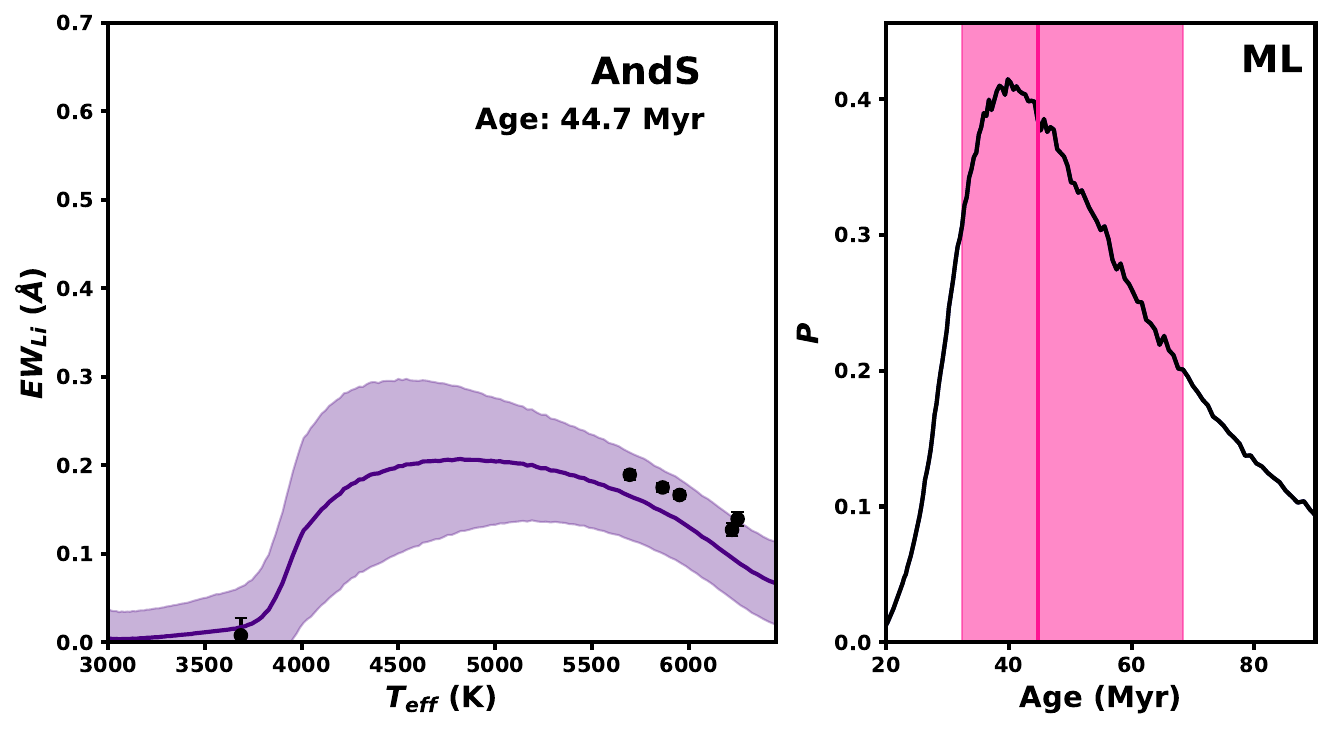}
    & \includegraphics[height=2.4cm]{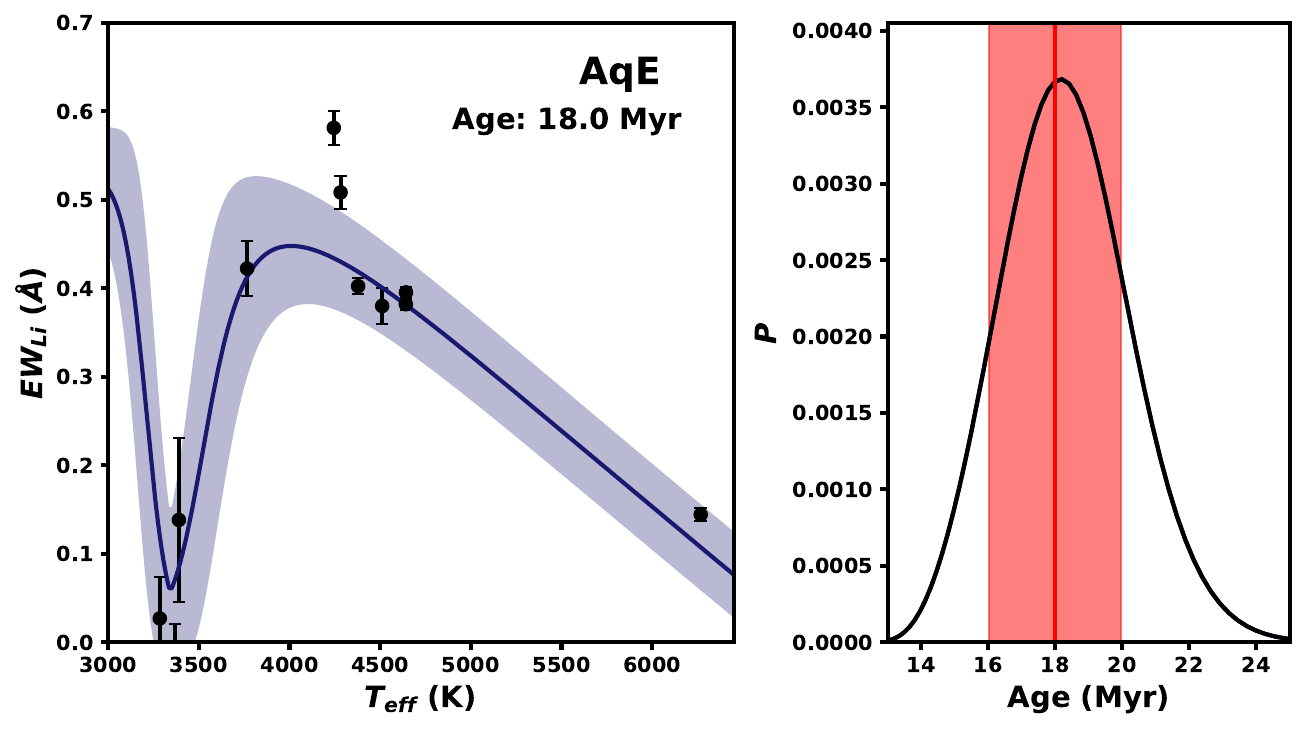}
    & \includegraphics[height=2.4cm]{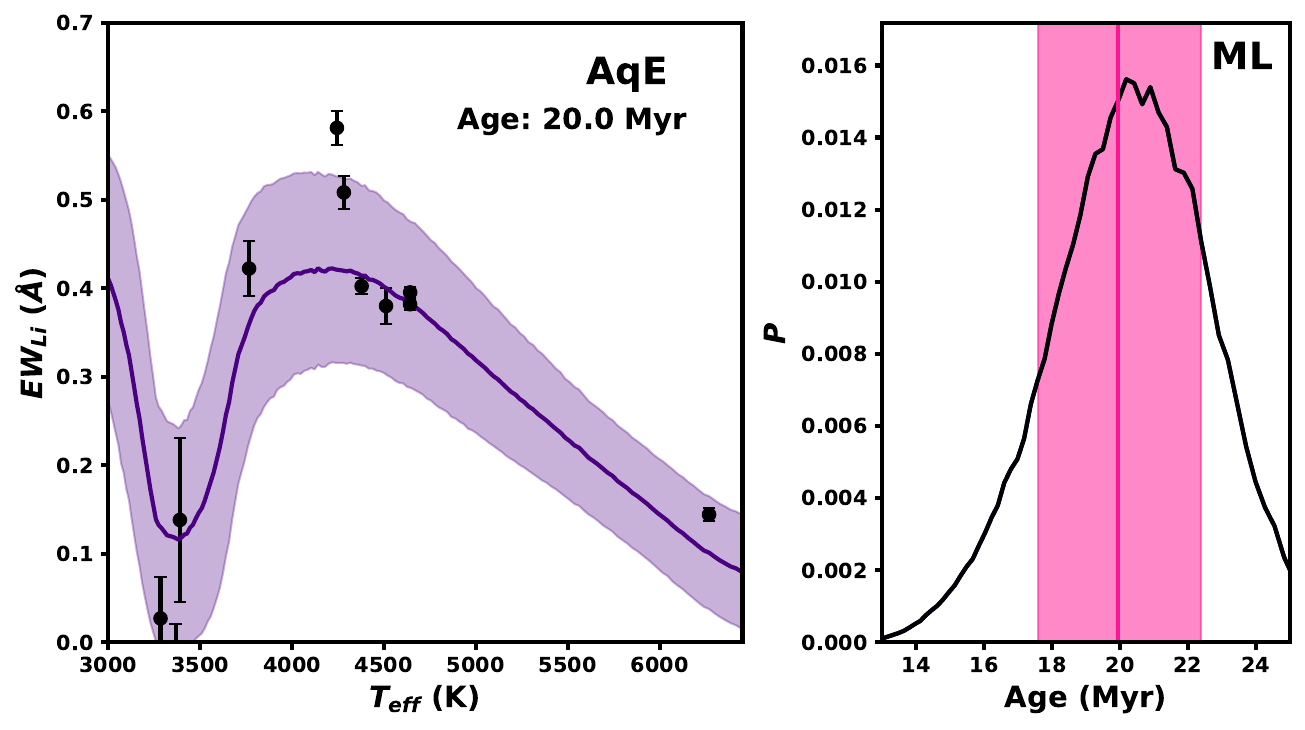} \\[-4pt]
    \includegraphics[height=2.4cm]{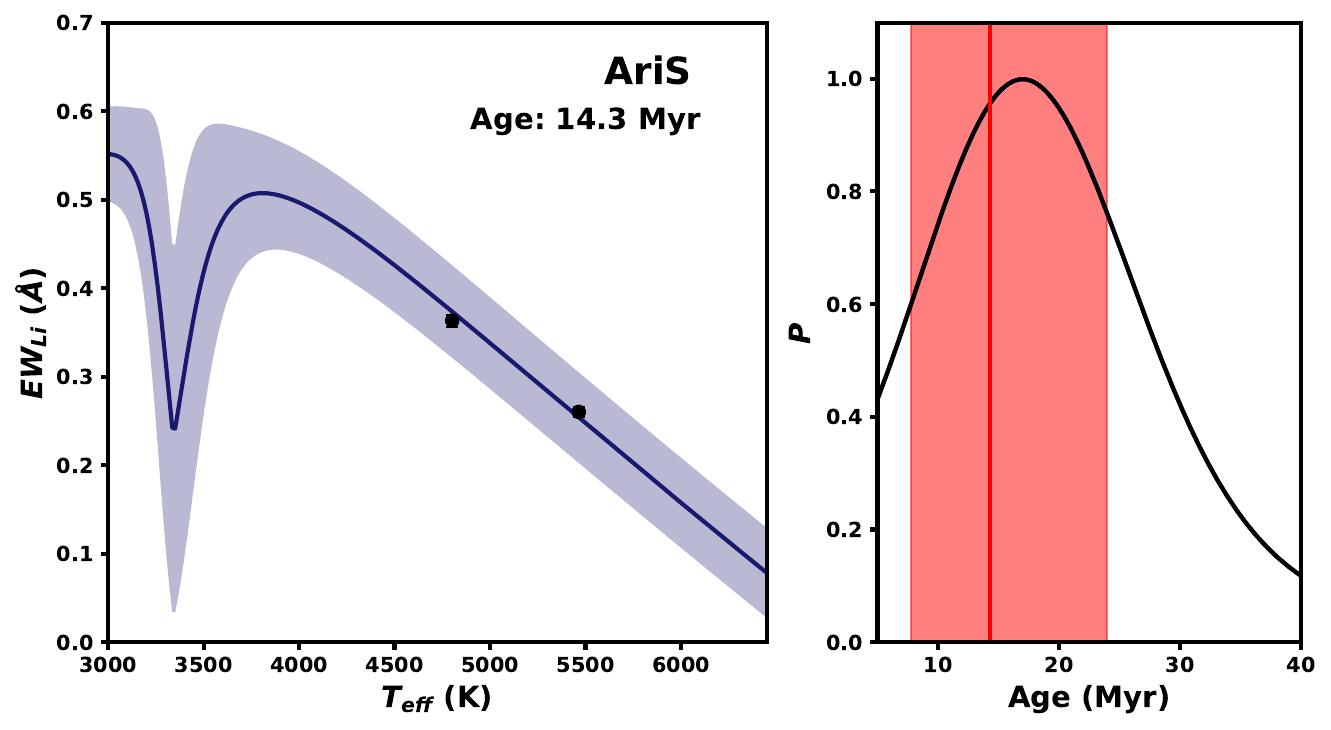}
    & \includegraphics[height=2.4cm]{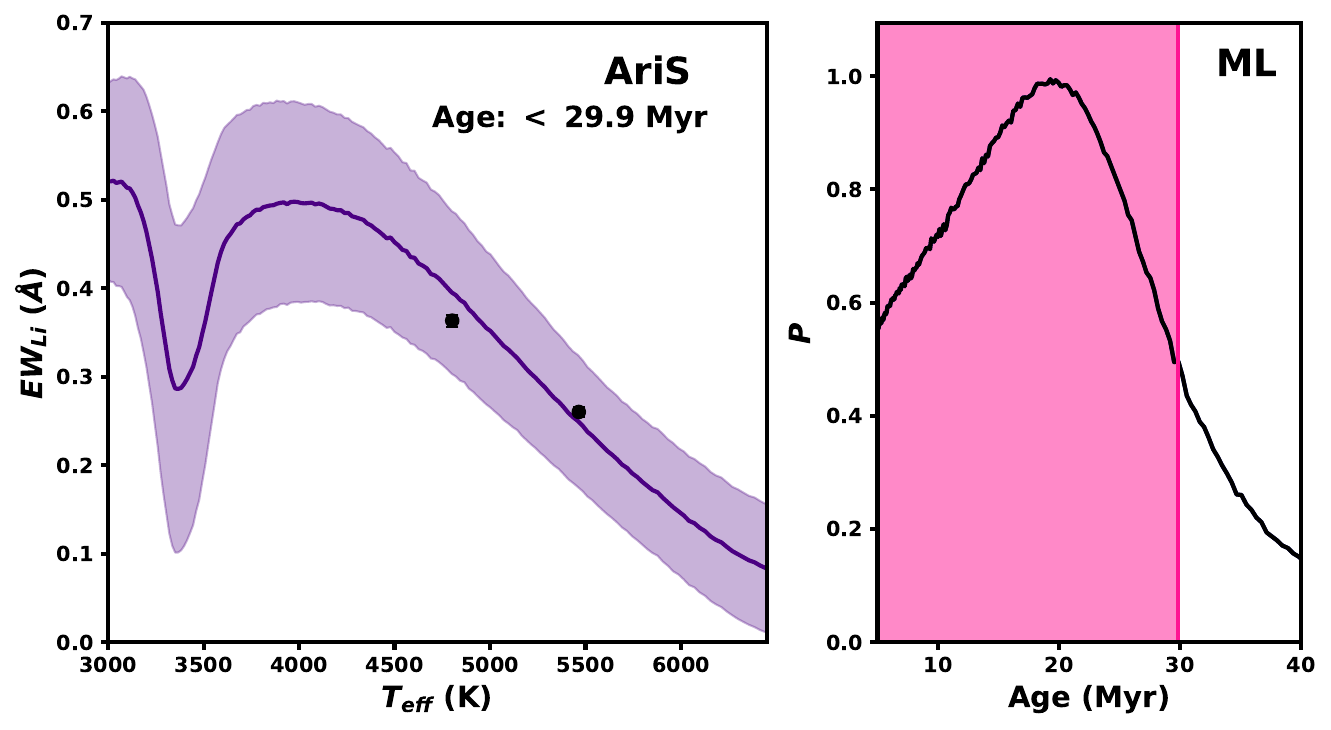}
    & \includegraphics[height=2.4cm]{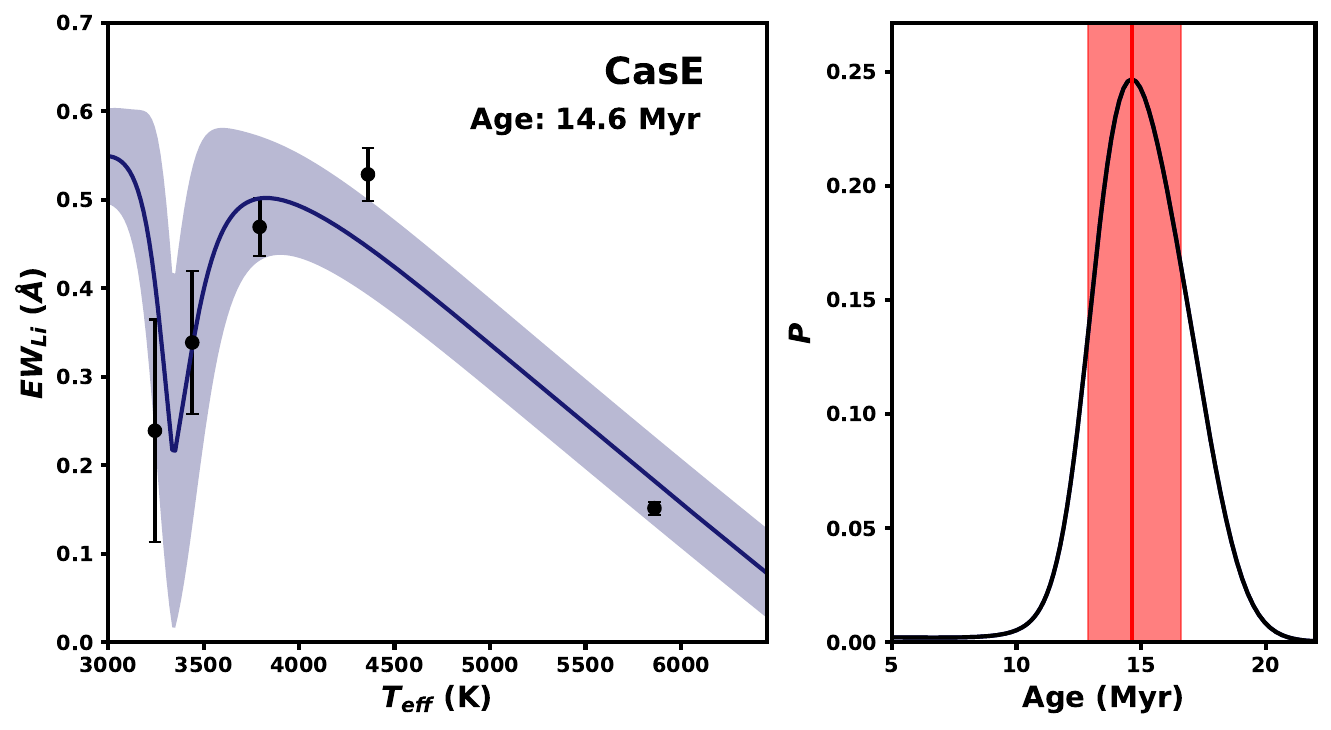}
    & \includegraphics[height=2.4cm]{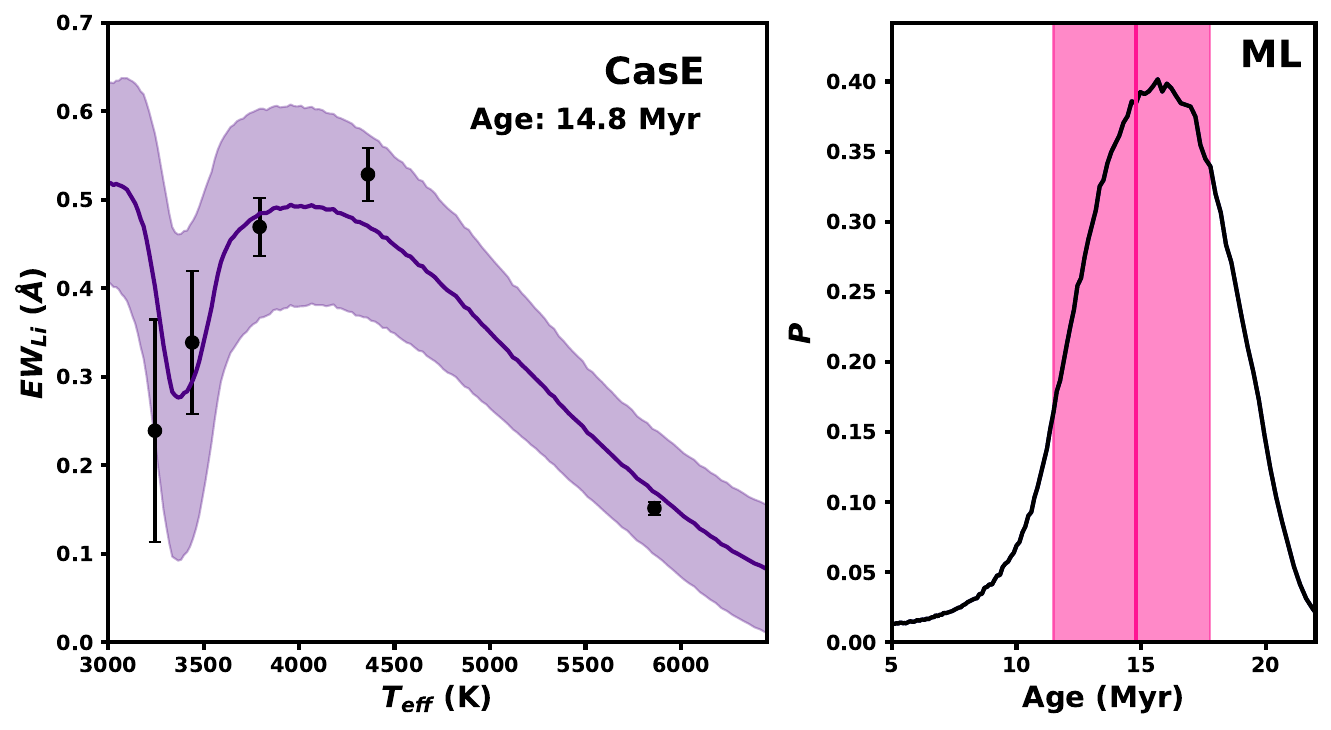}\\[-4pt]
    \includegraphics[height=2.4cm]{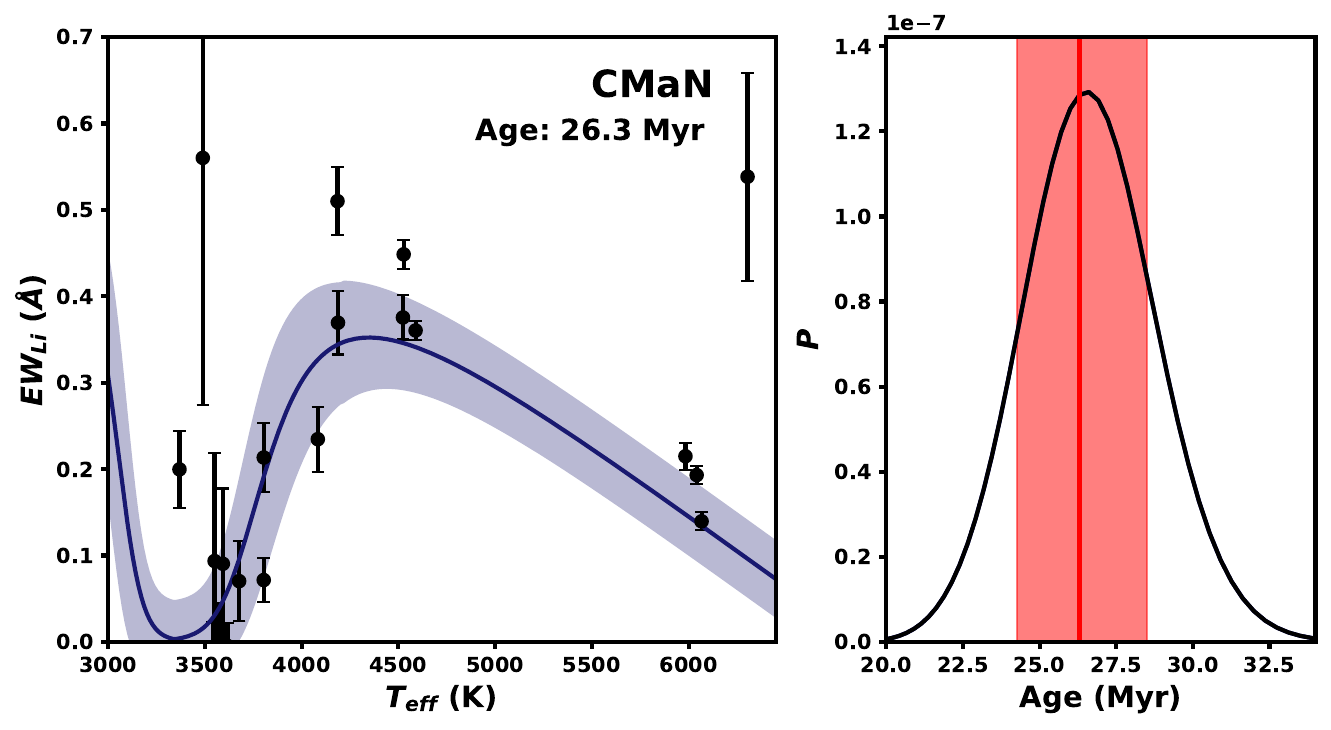}
    & \includegraphics[height=2.4cm]{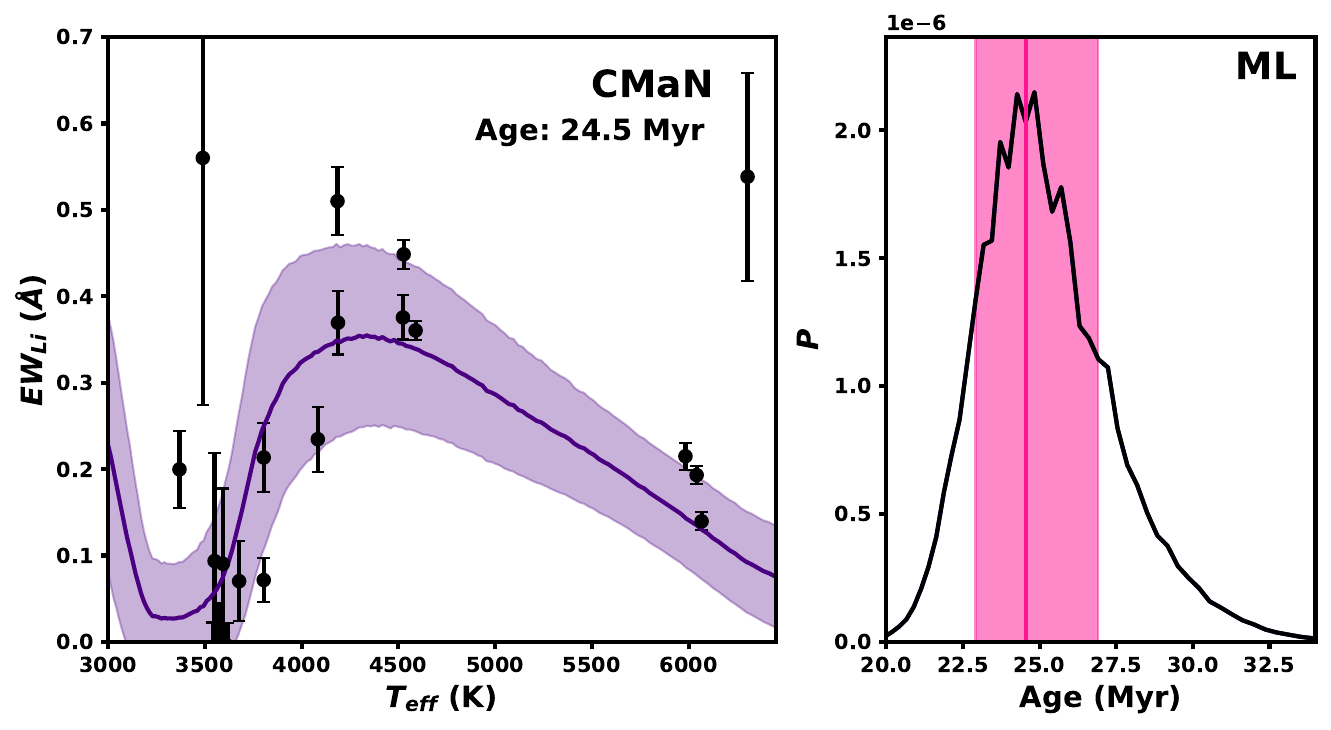}
    & \includegraphics[height=2.4cm]{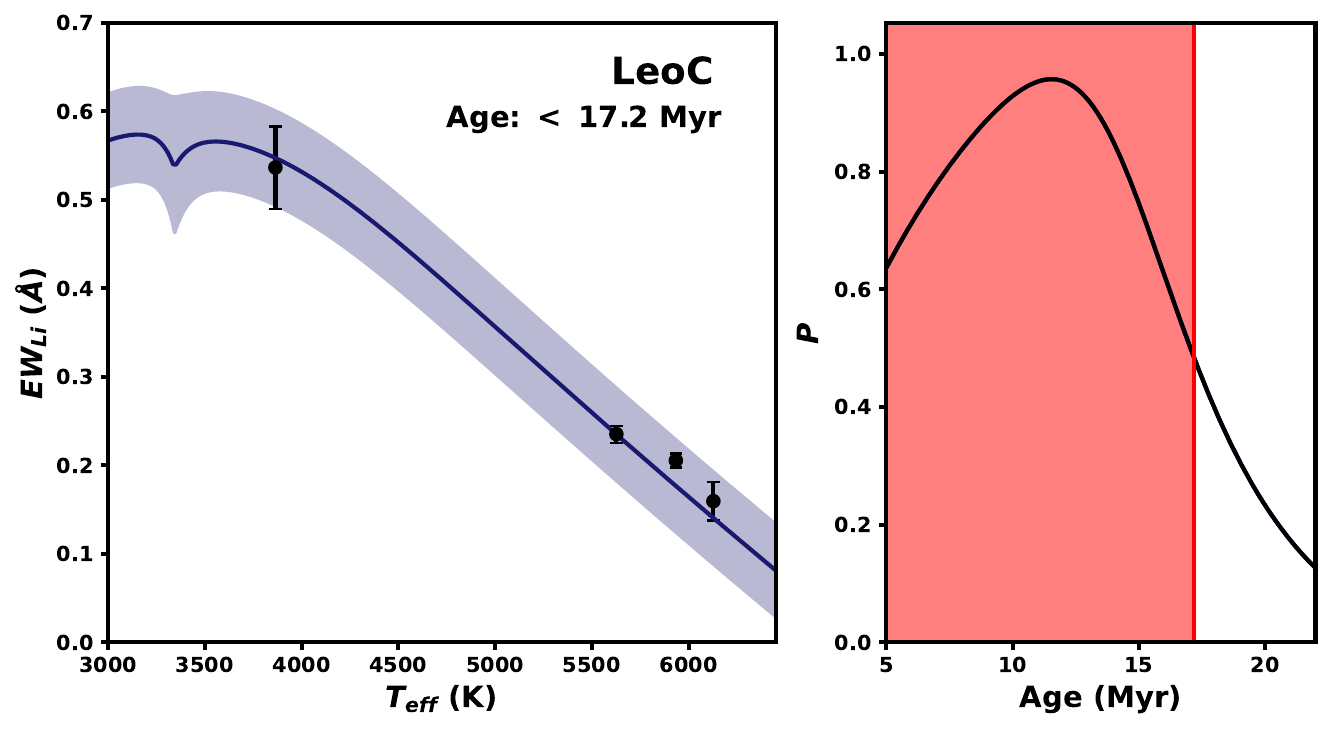}
    & \includegraphics[height=2.4cm]{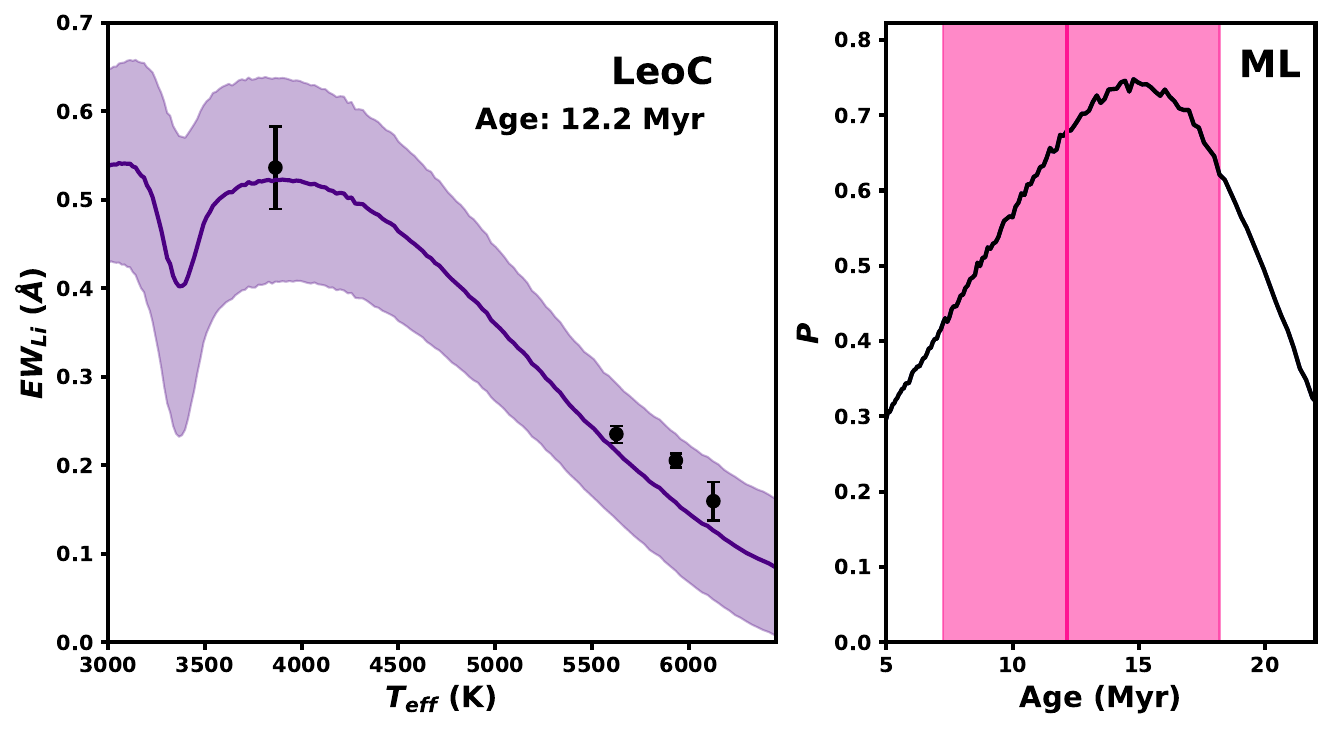}\\[-4pt]
    \includegraphics[height=2.4cm]{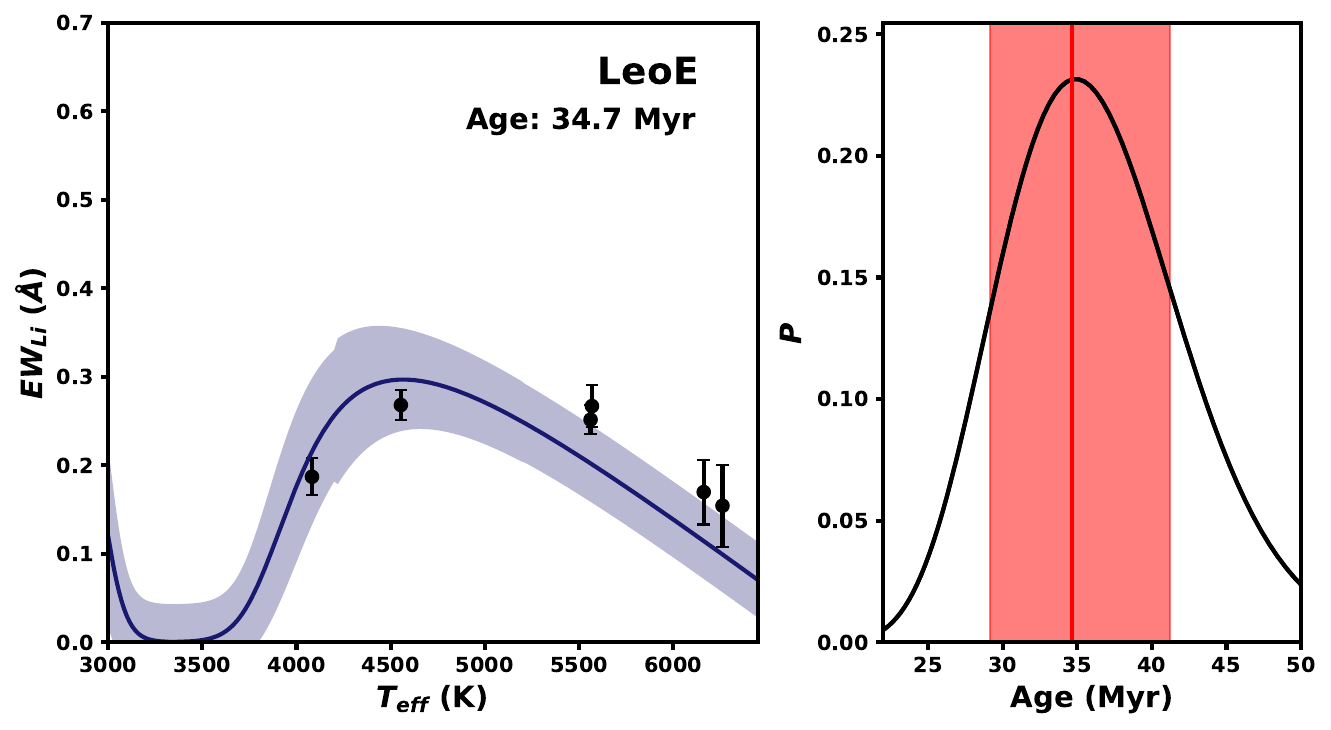}
    & \includegraphics[height=2.4cm]{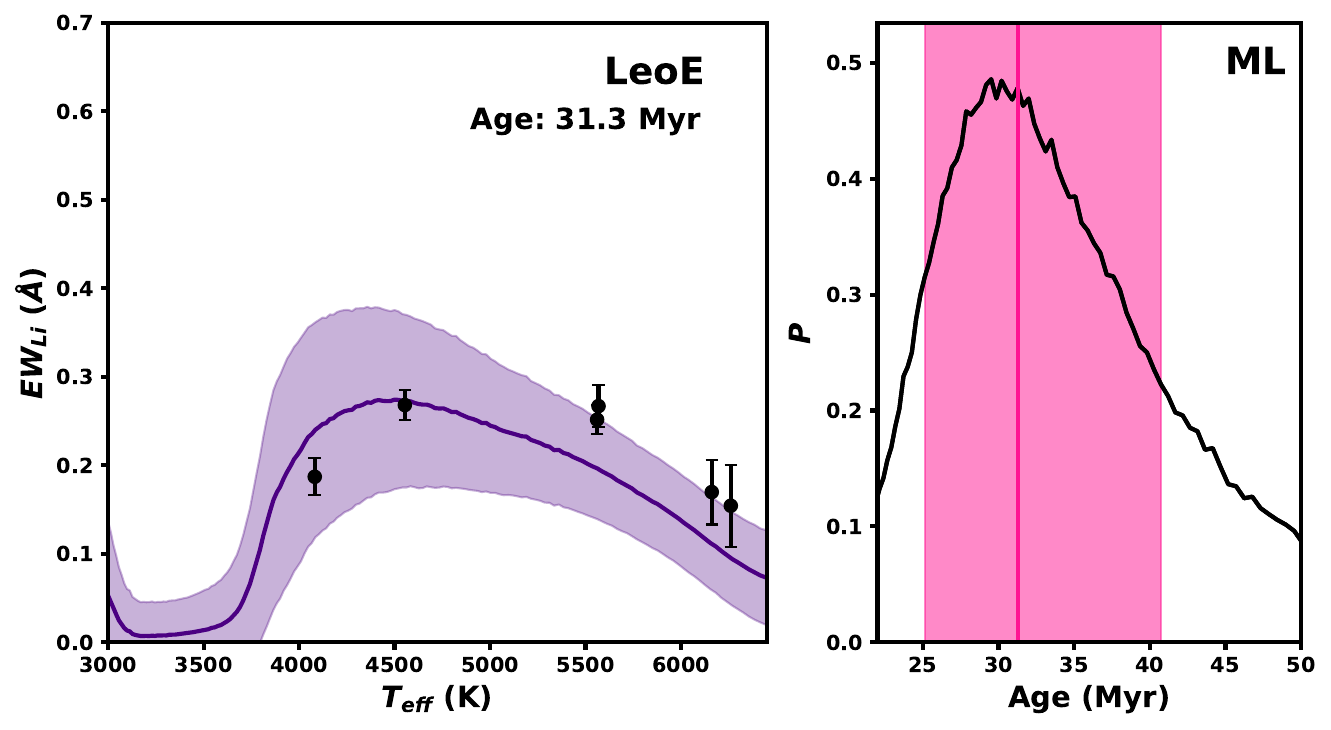}
    & \includegraphics[height=2.4cm]{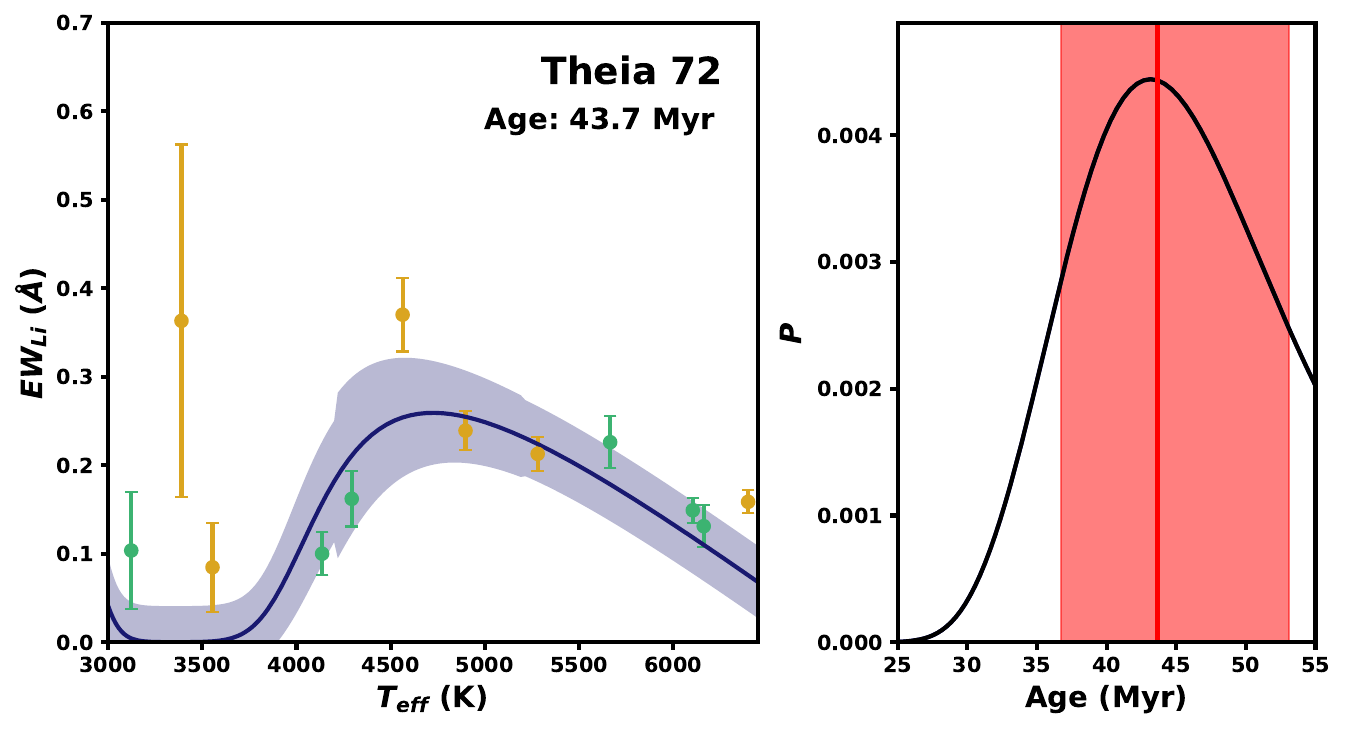}
    & \includegraphics[height=2.4cm]{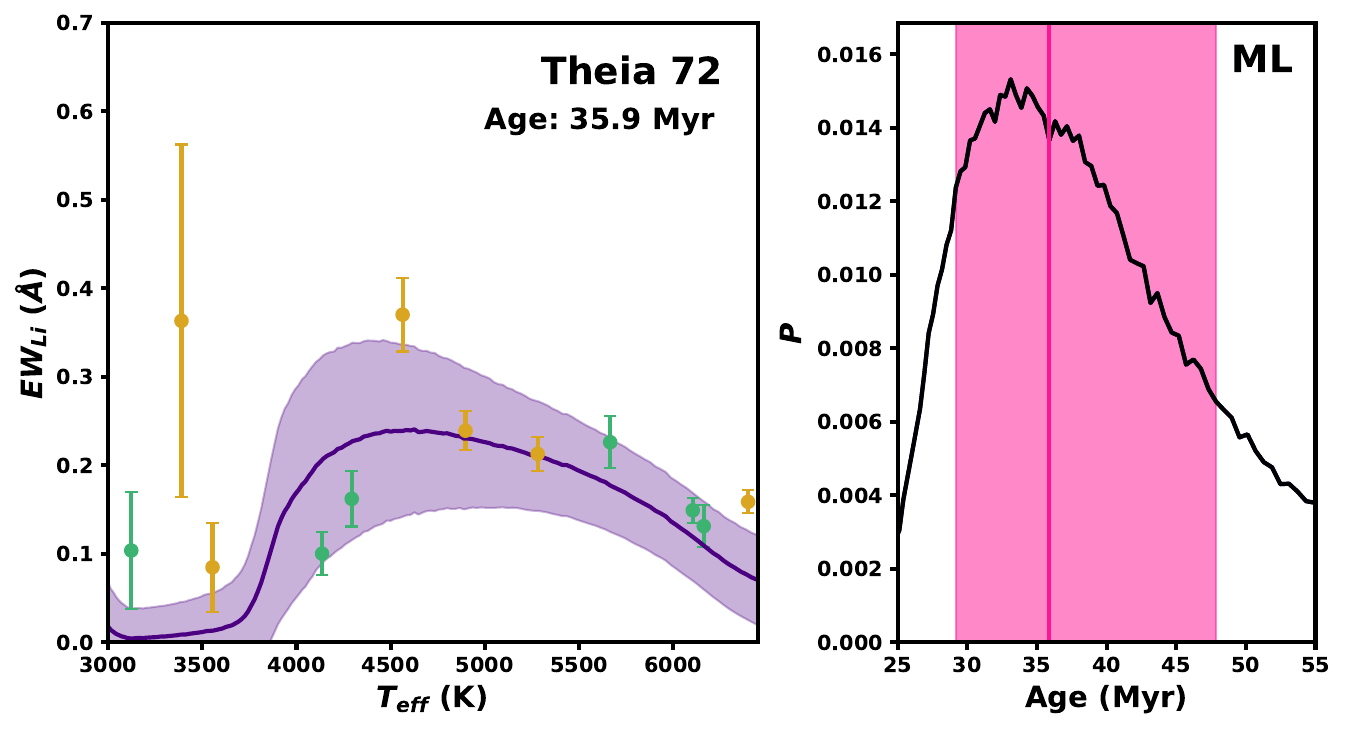}\\[-4pt]
    \includegraphics[height=2.4cm]{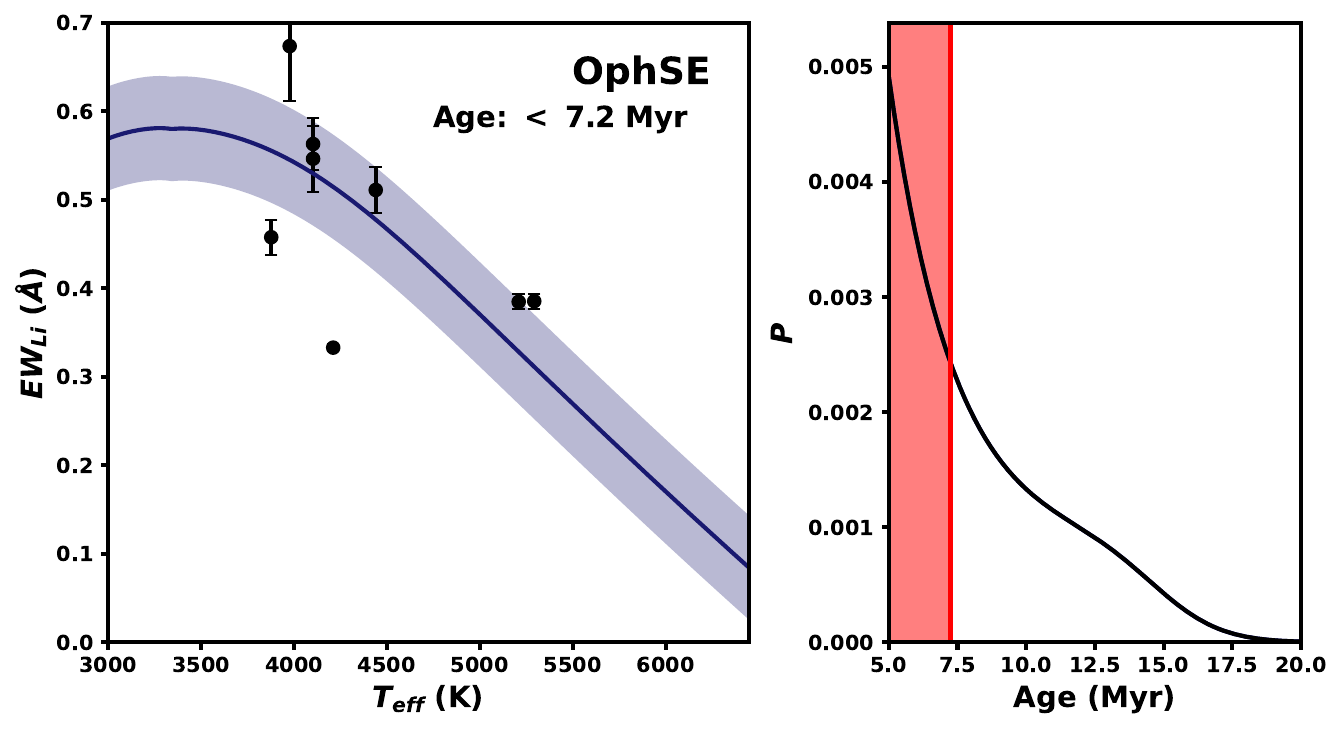}
    & \includegraphics[height=2.4cm]{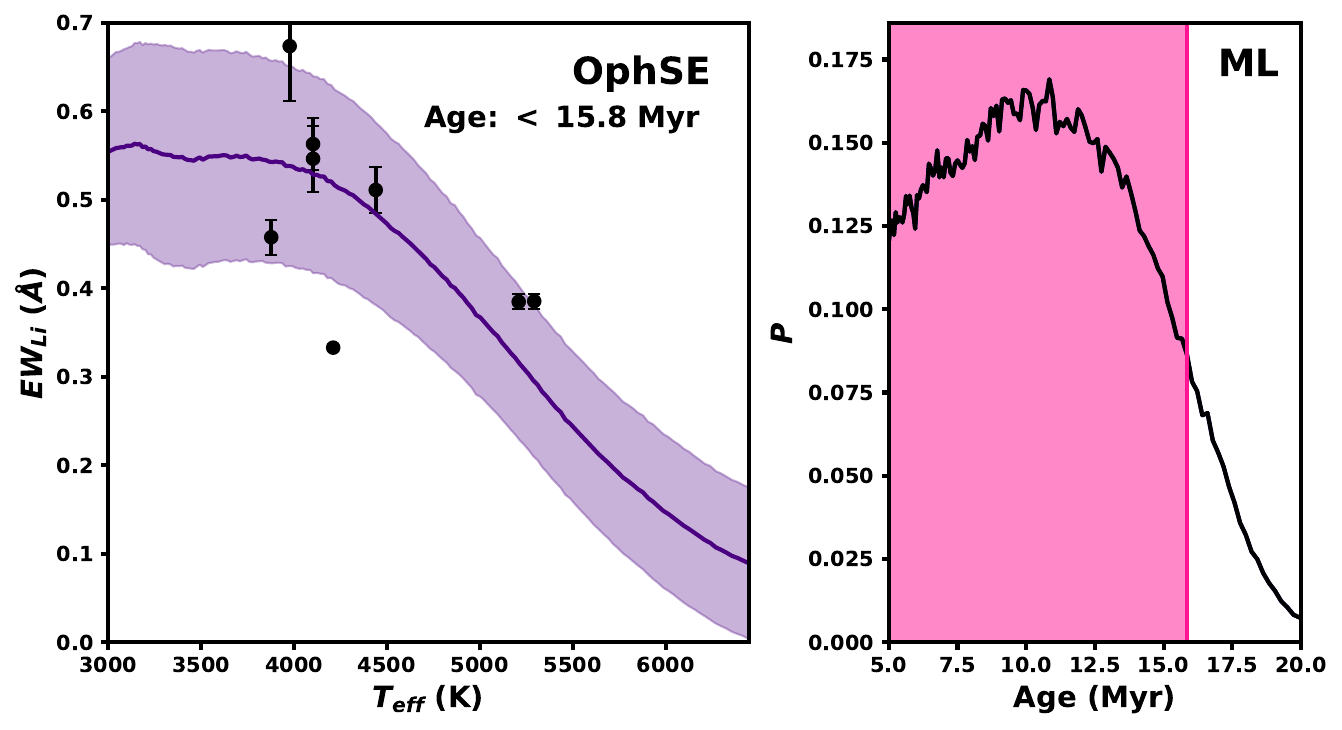}
    & \includegraphics[height=2.4cm]{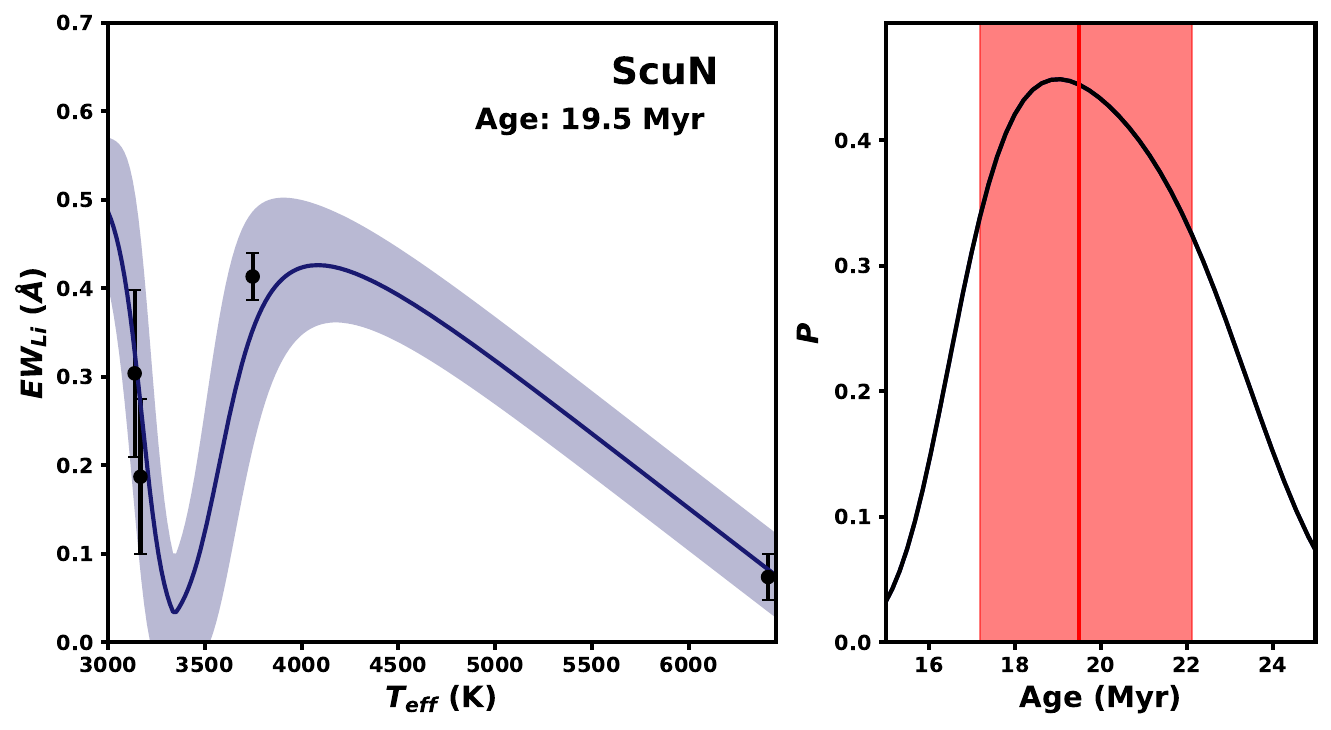}
    & \includegraphics[height=2.4cm]{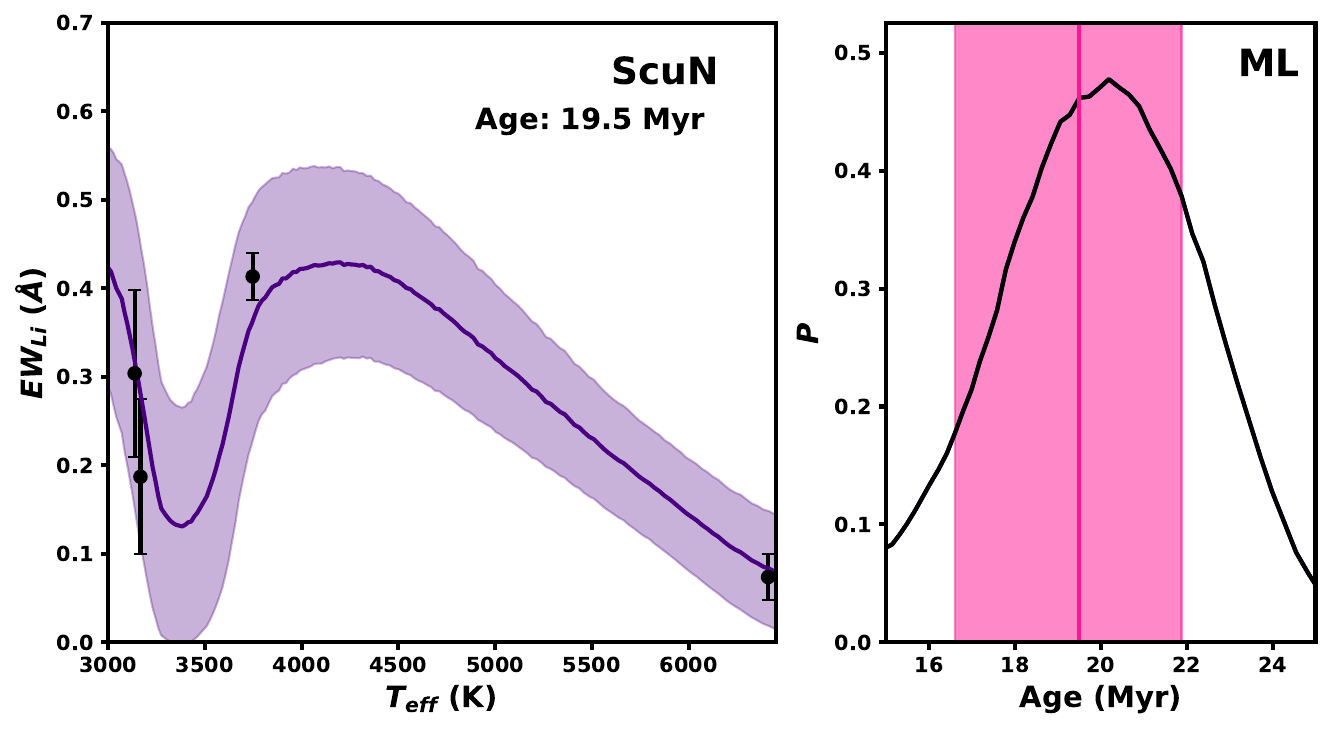}\\[-4pt]
    \includegraphics[height=2.4cm]{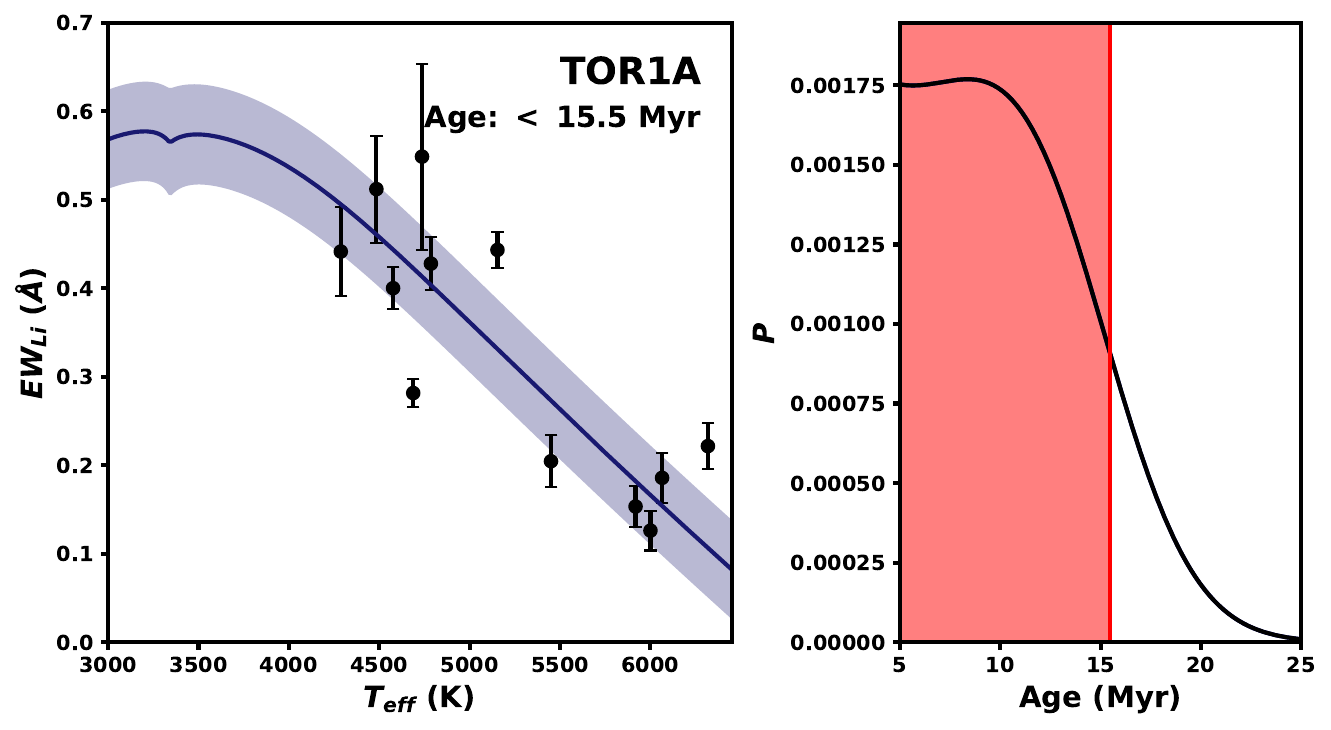}
    & \includegraphics[height=2.4cm]{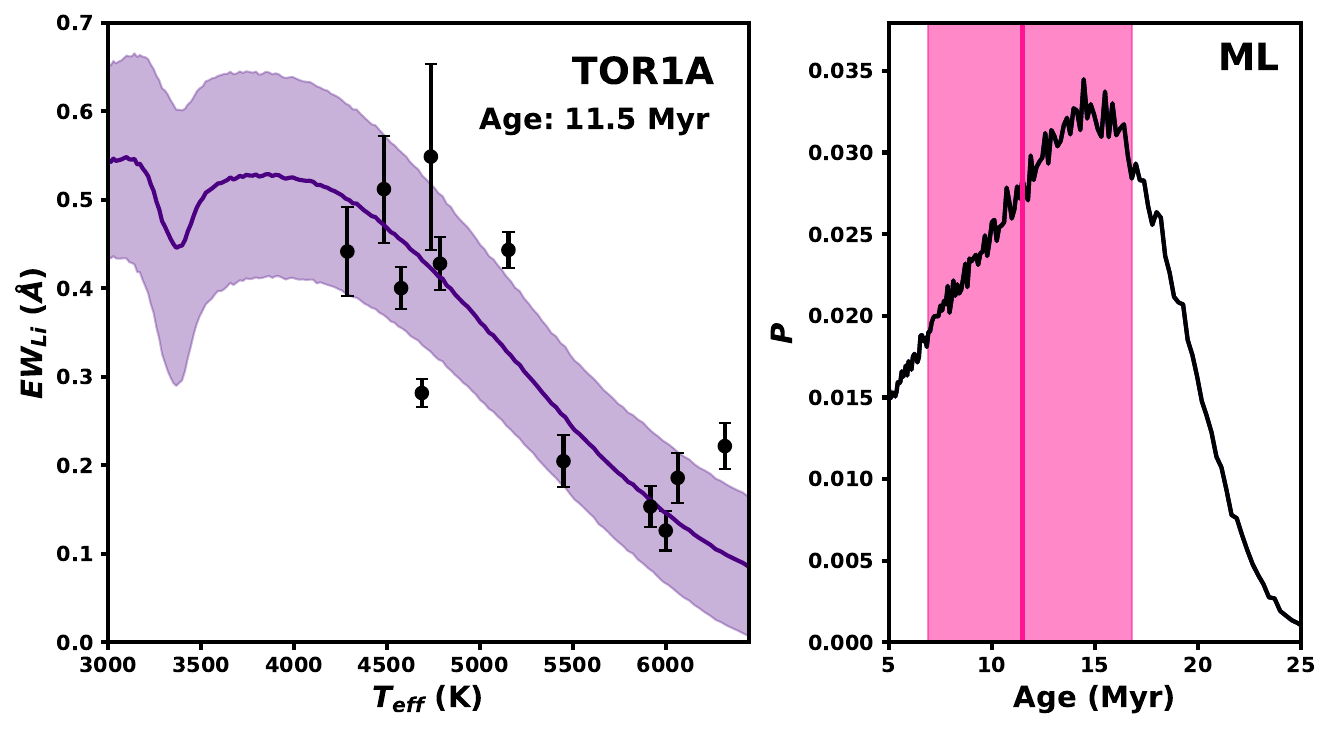}
    & \includegraphics[height=2.4cm]{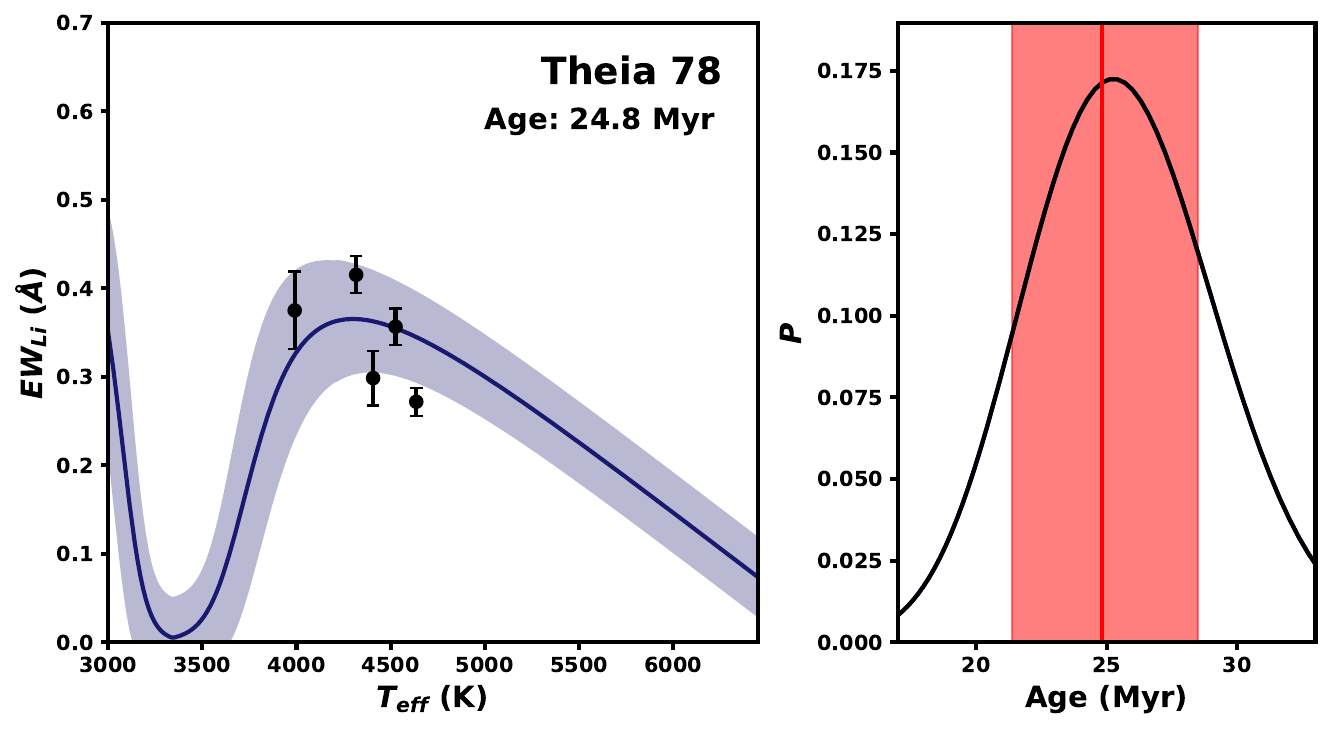}
    & \includegraphics[height=2.4cm]{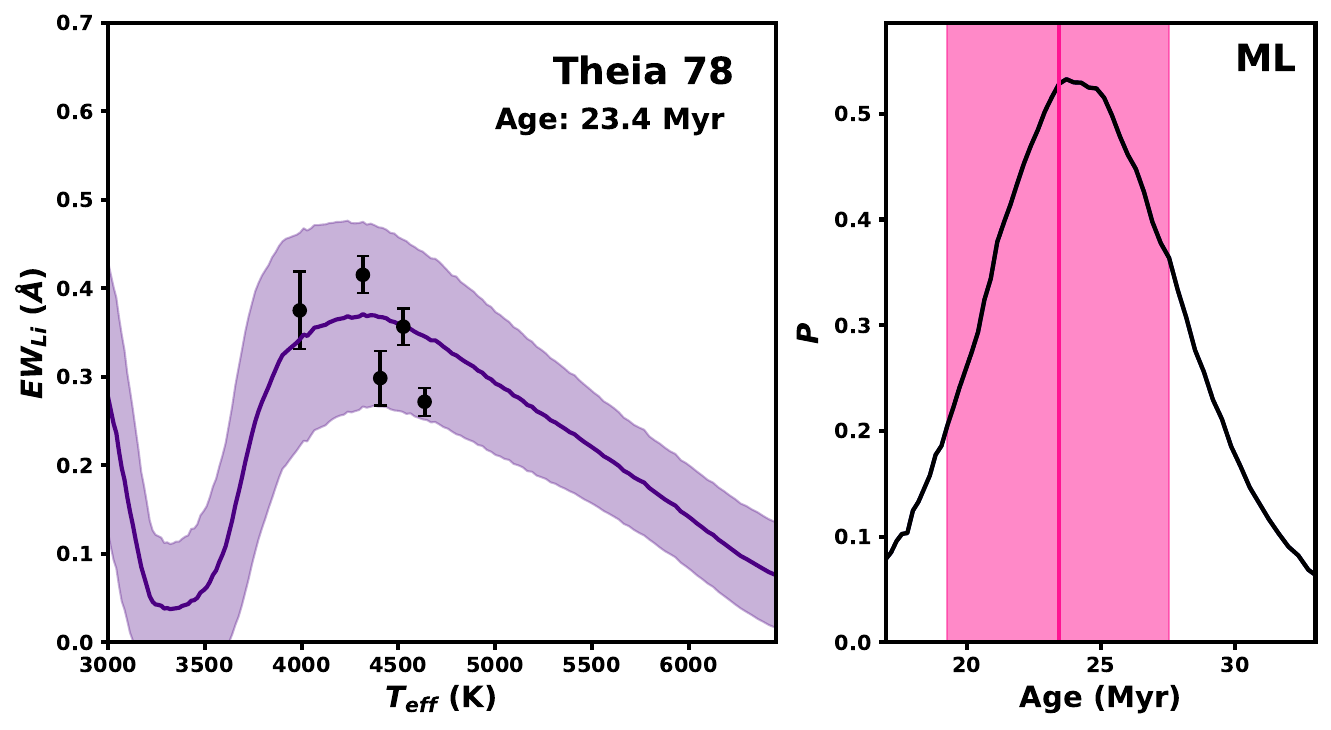}\\[-4pt]
    \includegraphics[height=2.4cm]{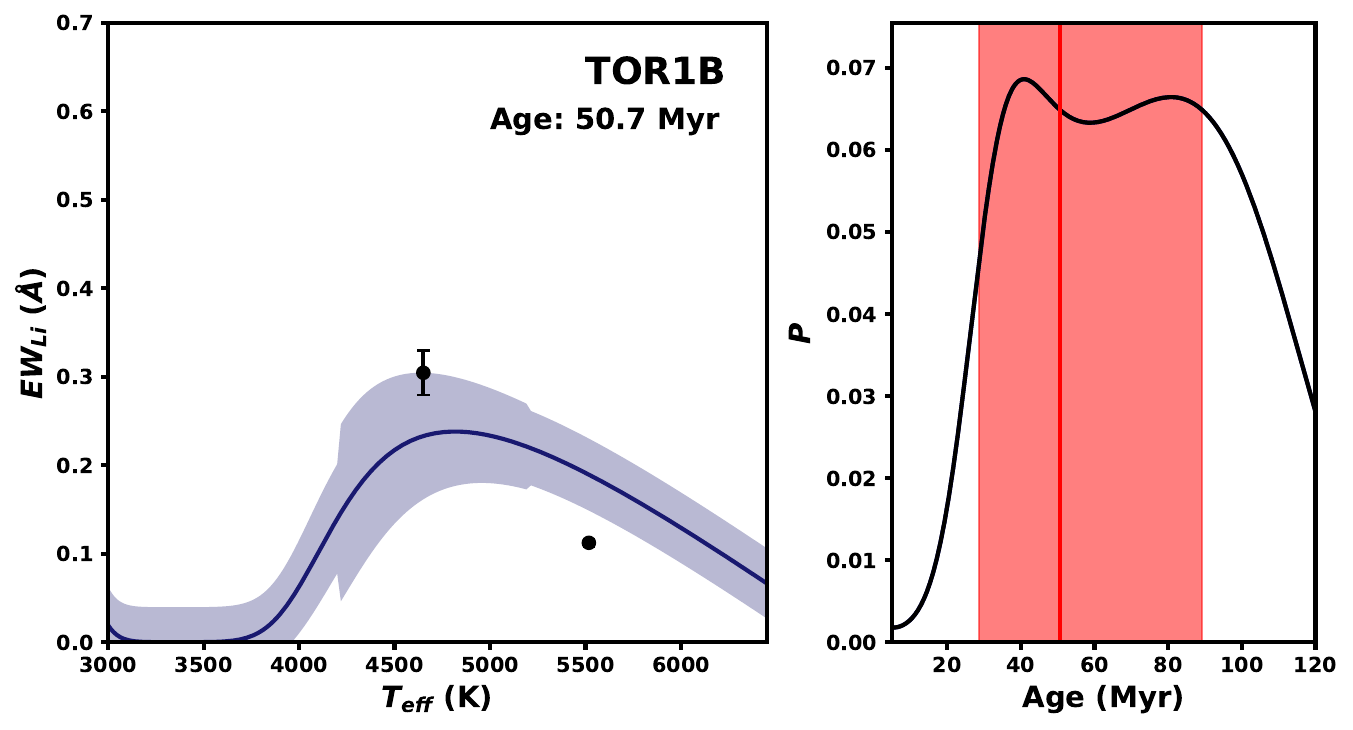}
    & \includegraphics[height=2.4cm]{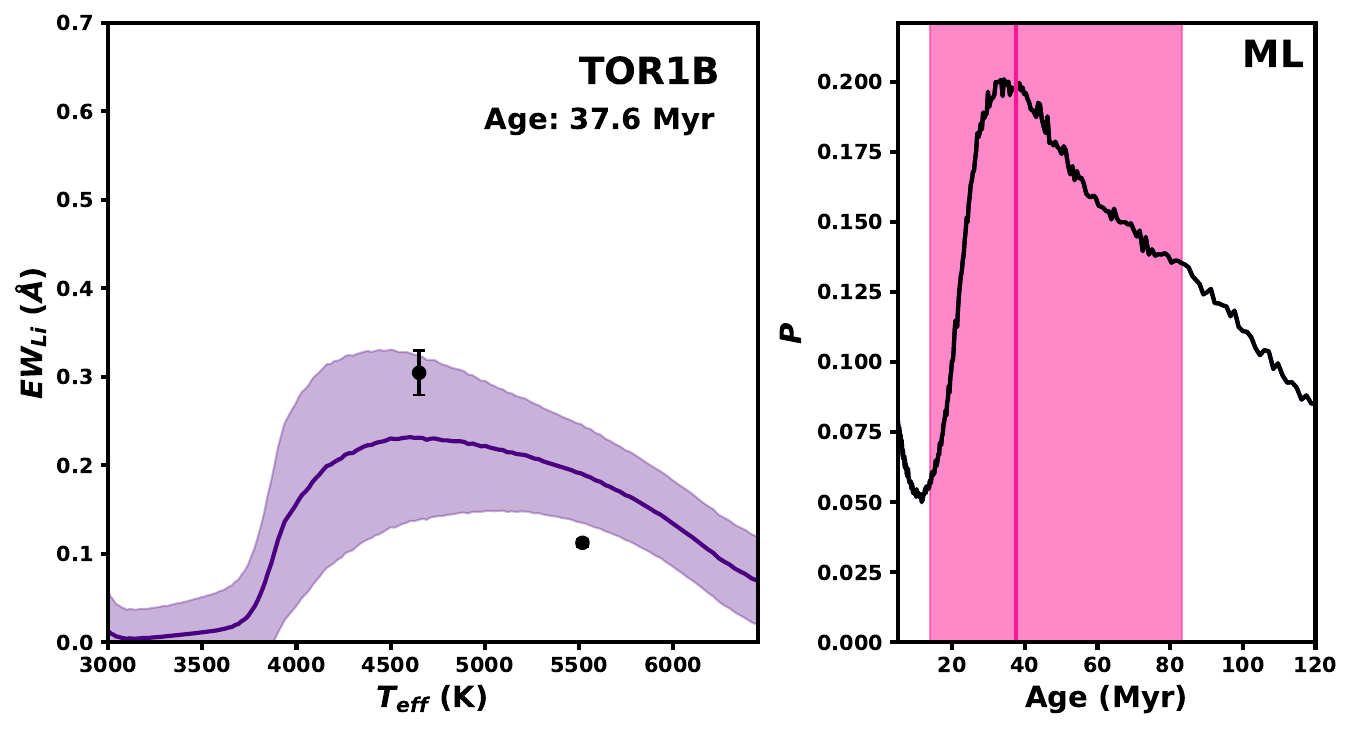}
    &\includegraphics[height=2.4cm]{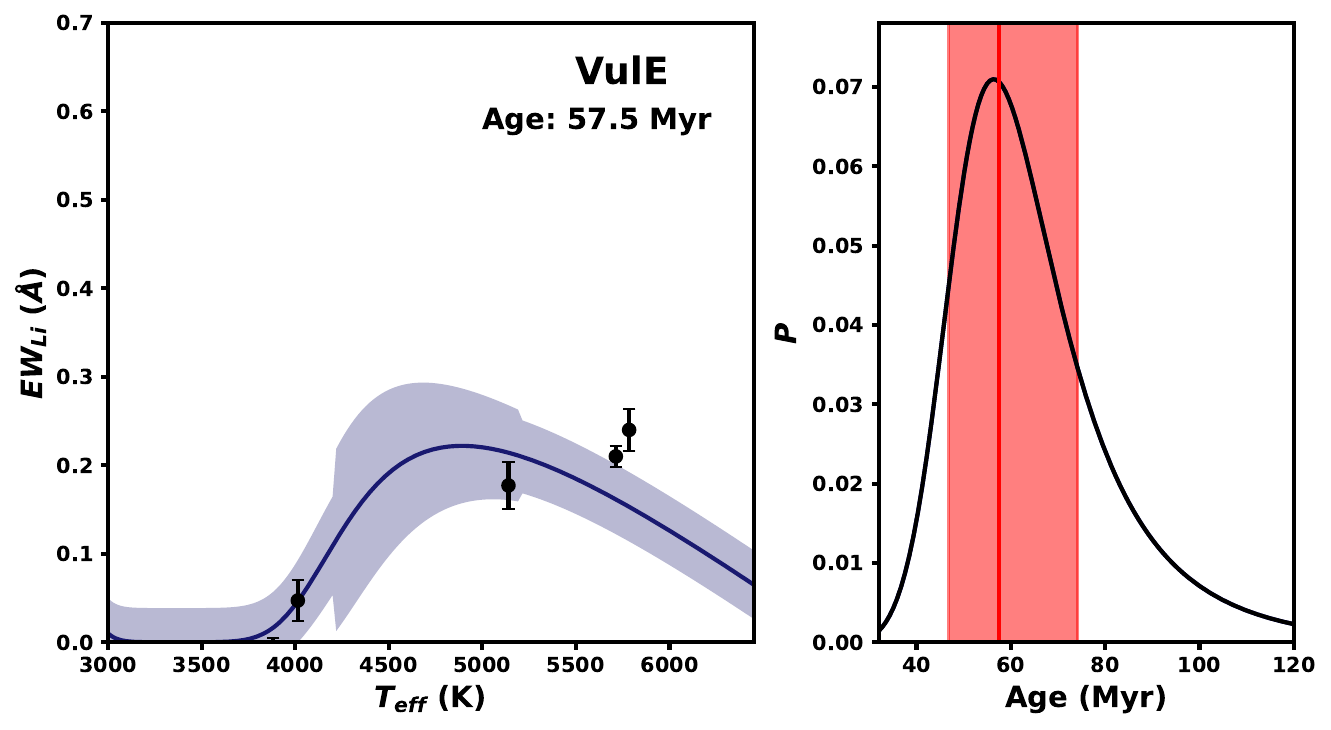}
    &\includegraphics[height=2.4cm]{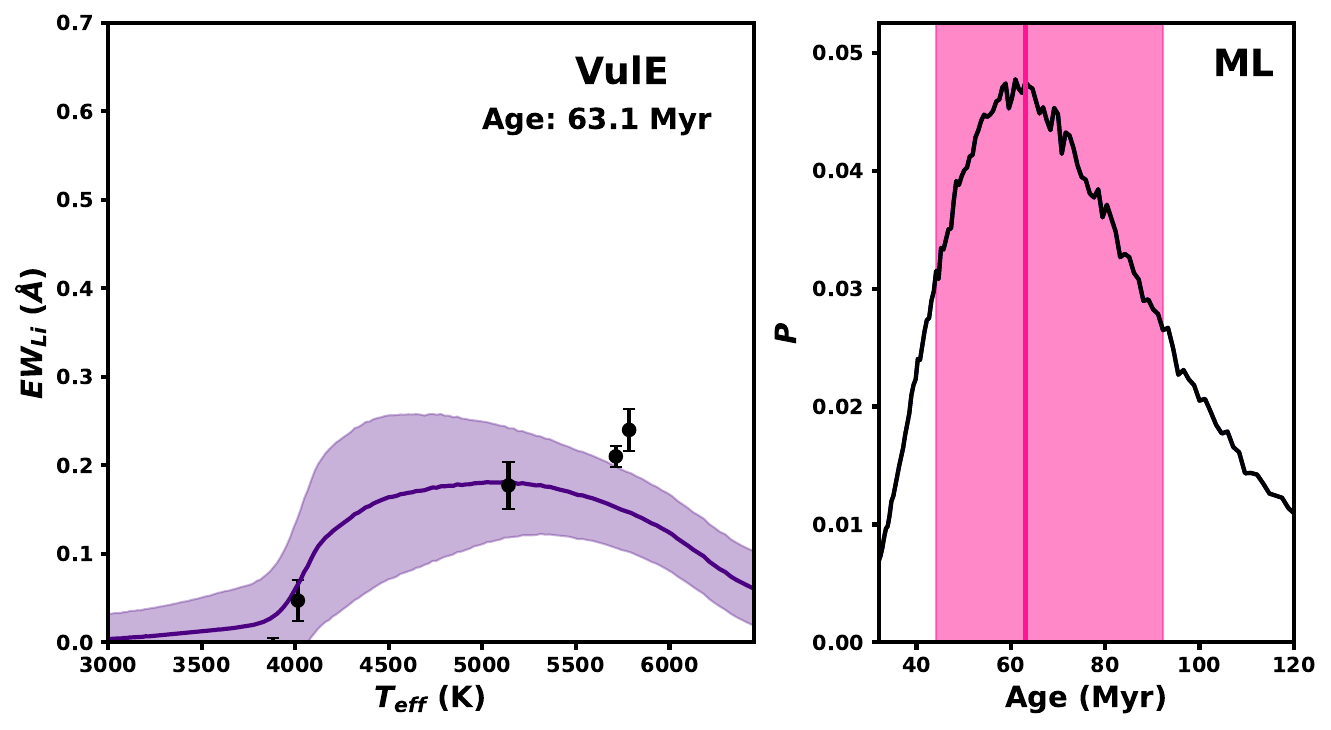}\\[-4pt]
    \includegraphics[height=2.4cm]{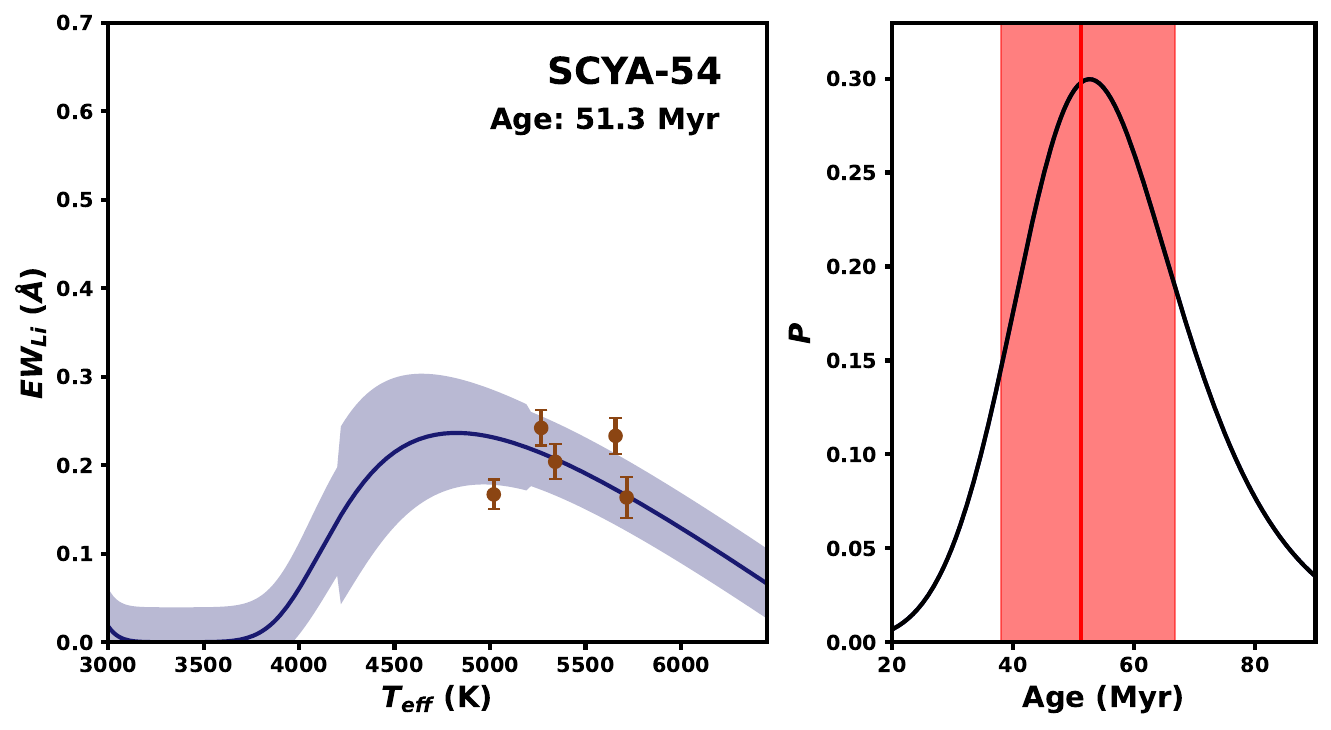}
    & \includegraphics[height=2.4cm]{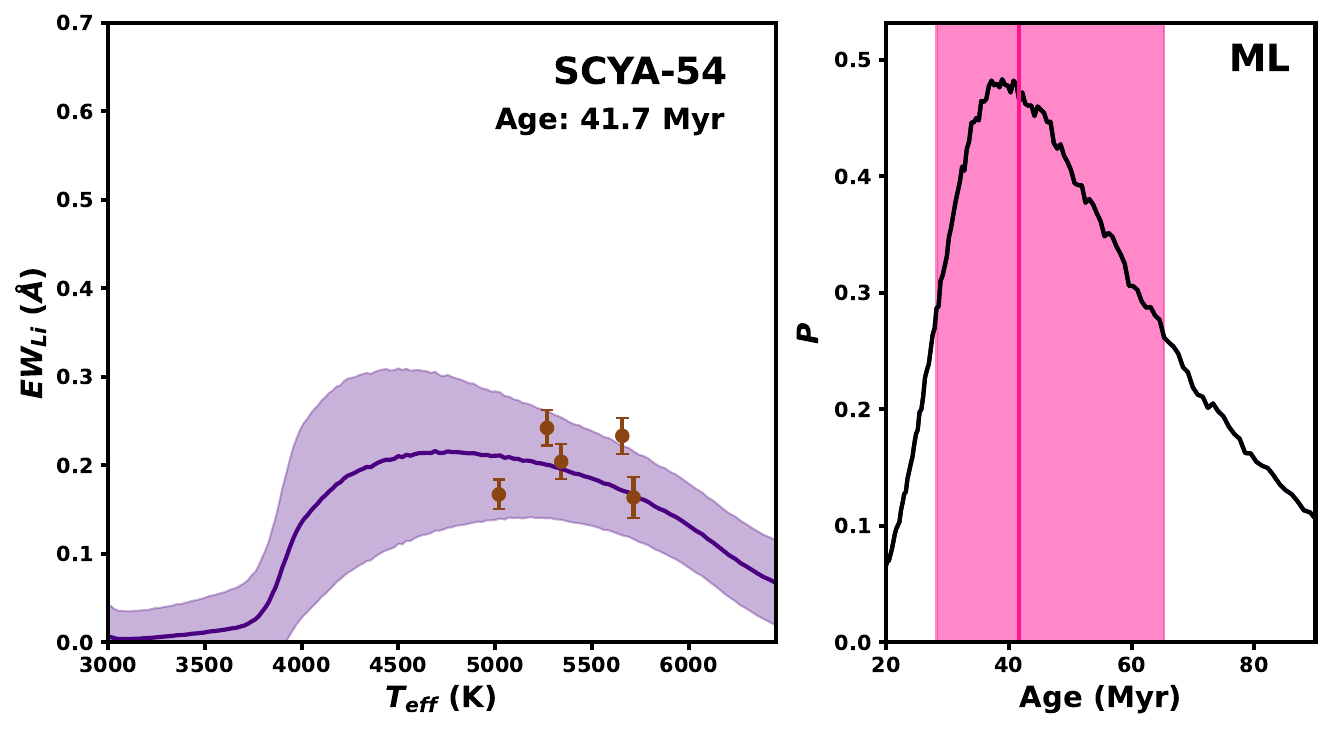}
    & \includegraphics[height=2.4cm]{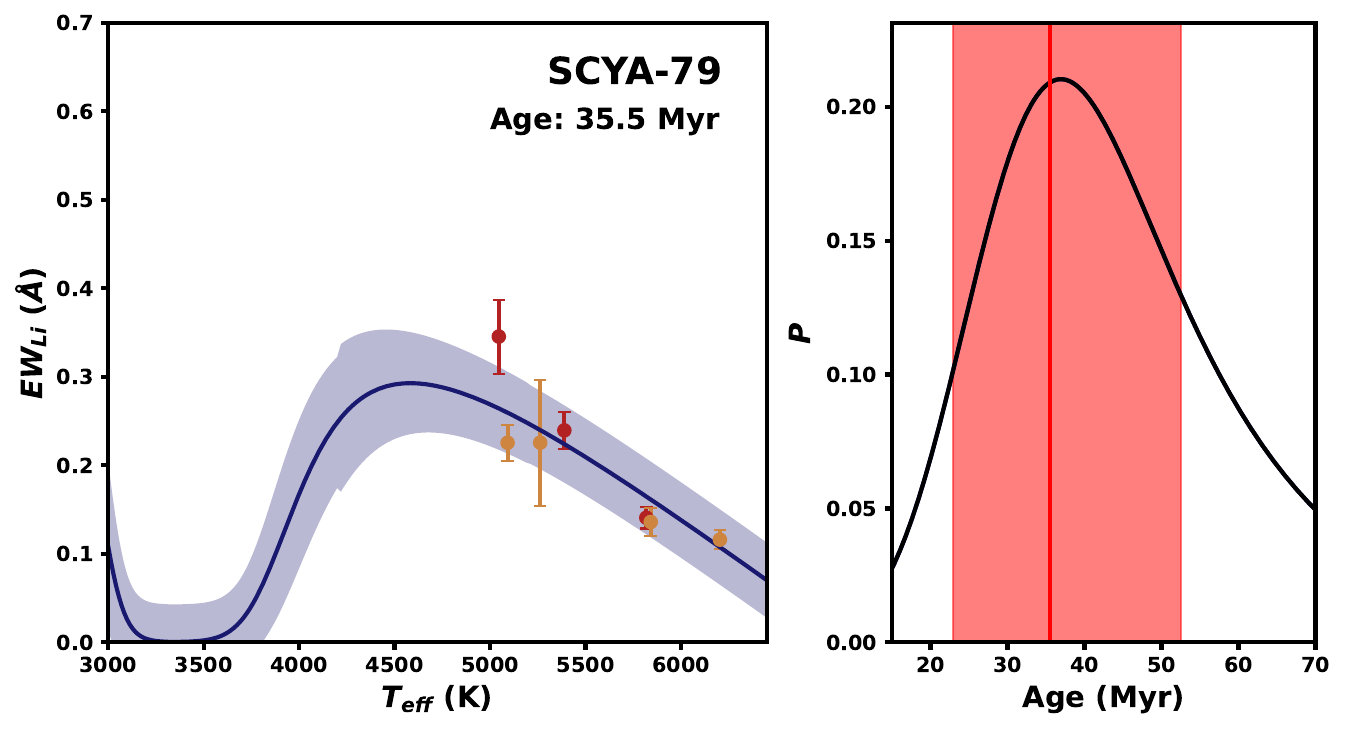}
    & \includegraphics[height=2.4cm]{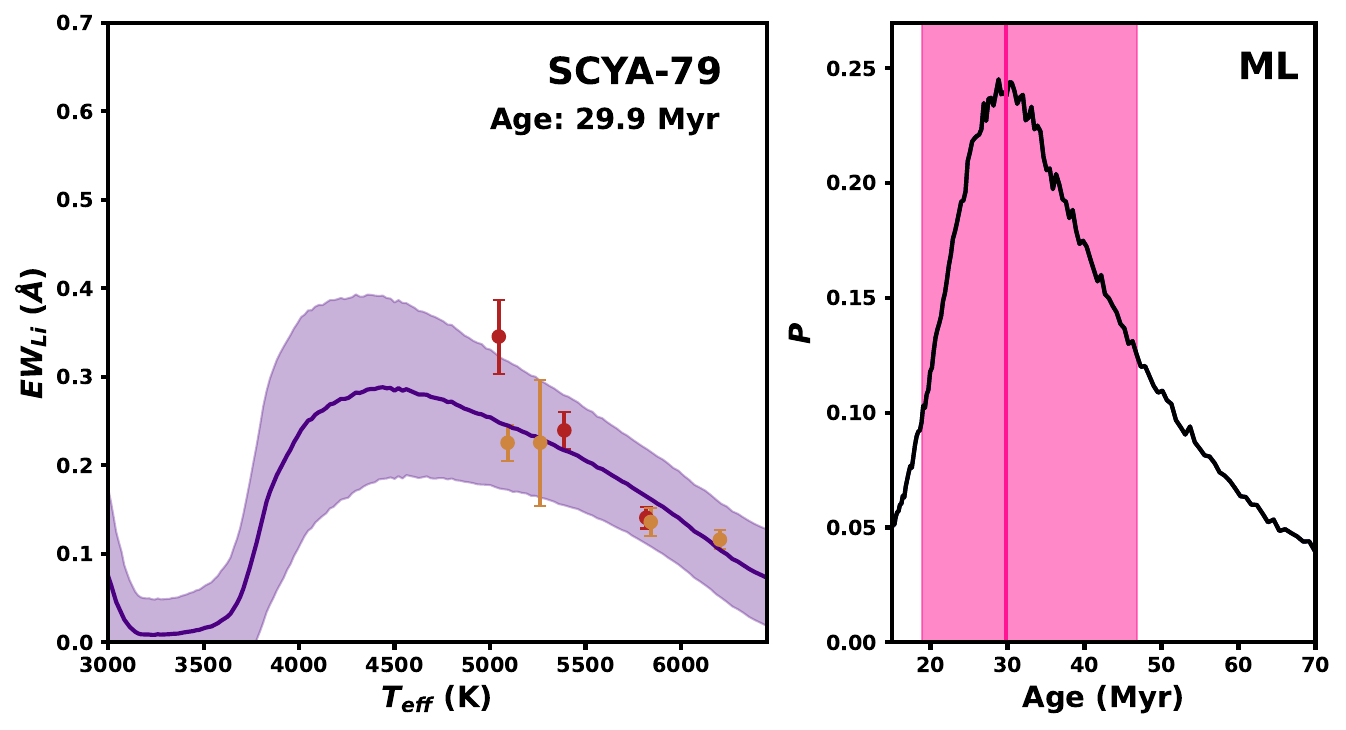}\\[-4pt]
\end{tabular}
\caption{EAGLES Lithium depletion fits for each of the 16 associations covered in this work, including both the original and ML EAGLES modules. For each association and EAGLES module, we show two panels; the left one shows the best fit to the Lithium sequence, and the right panel shows the combined posterior for all stars included in the fit. For populations with substructure, we color the Li measurements by their subgroup, following the color scheme in Figure \ref{fig:substructures}. Fits that use the EAGLES ML module are labeled in the top-right of the age posterior plot, and these are shown in purple and pink rather than the blue and red color scheme for the analytical results.}
\label{fig:liagefits}
\end{figure*}




The depletion of the Li-6707\AA~line can be used not just for youth indication as discussed in Section \ref{sec:specyouth}, but also as a powerful tool for age measurement. The EAGLES package provides empirical models of the lithium depletion curve as a function of age, as well functions for computing an age probability posterior for a given stellar Li EW and effective temperature $T_{eff}$. These functions use both the original analytical model (v1; \citealt{EAGLES23}) that we used in Section \ref{sec:specyouth}, and a new machine learning (ML) neural network version (v2; \citealt{Weaver24}). 

We start by limiting our sample to stars in EAGLES' $3000<T_{eff}<6500~K$ model range with Li EWs. We impose a membership cut of $P_{fin} > 0.8$, and remove binaries with $RUWE > 1.2$. We run EAGLES v1 for all stars that pass these conditions using their $T_{eff}$ and Fe I-deblended Li EWs computed in Section \ref{sec:specyouth}. We remove any stars with an EAGLES 3$\sigma$ lower age limit above the highest isochronal age solution in section \ref{sec:isoages}. This removes any remaining non-members or stars with contamination from an unseen companion. We only use the analytical EAGLES module for this step so that the sample is the same across each module, enhancing comparability. 

We multiply the age likelihoods of all remaining stars to produce aggregate posteriors for the entire population, and we do this separately for both the original and ML EAGLES modules. We set the age to the 50th percentile of the posterior, and set the lower and upper uncertainty intervals to the 16th and 84th percentiles, respectively. We exclude ages under 5 Myr, where EAGLES lacks age discrimination and therefore sets the likelihood to its value at 5 Myr \citep{EAGLES23}. We only report an upper age limit for populations where the posterior at 5 Myr exceeds half of its peak, which we set to the 95th percentile of the probability posterior. We plot the Lithium depletion age solutions in Figure \ref{fig:liagefits}, showing the Li EWs used in the calculation, the best-fit EAGLES Li curves, and the resulting combined age posterior. 

The two different EAGLES modules produce broadly similar results. The main difference is in the ``Li dip'', which reaches zero about 10 Myr earlier and transitions more sharply analytical version. In populations like AqE, where this dip is well-covered, the analytical model is in close agreement with both the nearly Li-free stars in the dip and the Li-rich stars to the right of it, while the ML version does not capture this sharp transition as closely. However, the ML version seems to produce more consistent results in populations with less coverage, especially populations with a significant Li scatter. The large scatter and Li dip shape in the ML module are produced by an observed scatter in real Li observations \citep{Weaver24}. Our harsh quality cuts on binarity and membership may improve this scatter beyond the expectations of the ML model in populations like AqE, but this does not appear to be universal. We therefore include both the analytical and ML results in Figure \ref{fig:liagefits} and Section \ref{sec:agesynth}, but use only the analytical results for membership assessment and quality checks for consistency with previous work (e.g., \citetalias{Kerr24}).

\subsubsection{Final Ages} \label{sec:agesynth}

We have produced isochronal ages using several different models, as well as dynamical and lithium depletion ages. The isochronal ages provide a consistent relative age scale derived from the height of the PMS that is available for all populations, unlike the dynamical and lithium depletion methods. However, isochronal ages are often inconsistent with each other and with other age calculation methods in absolute terms \citep[e.g.,][]{Herczeg15, Rottensteiner24}. The accuracy of these isochronal solutions can be assessed by comparison with the non-isochronal methods.

\begin{figure*}
\centering
\includegraphics[height=10cm]{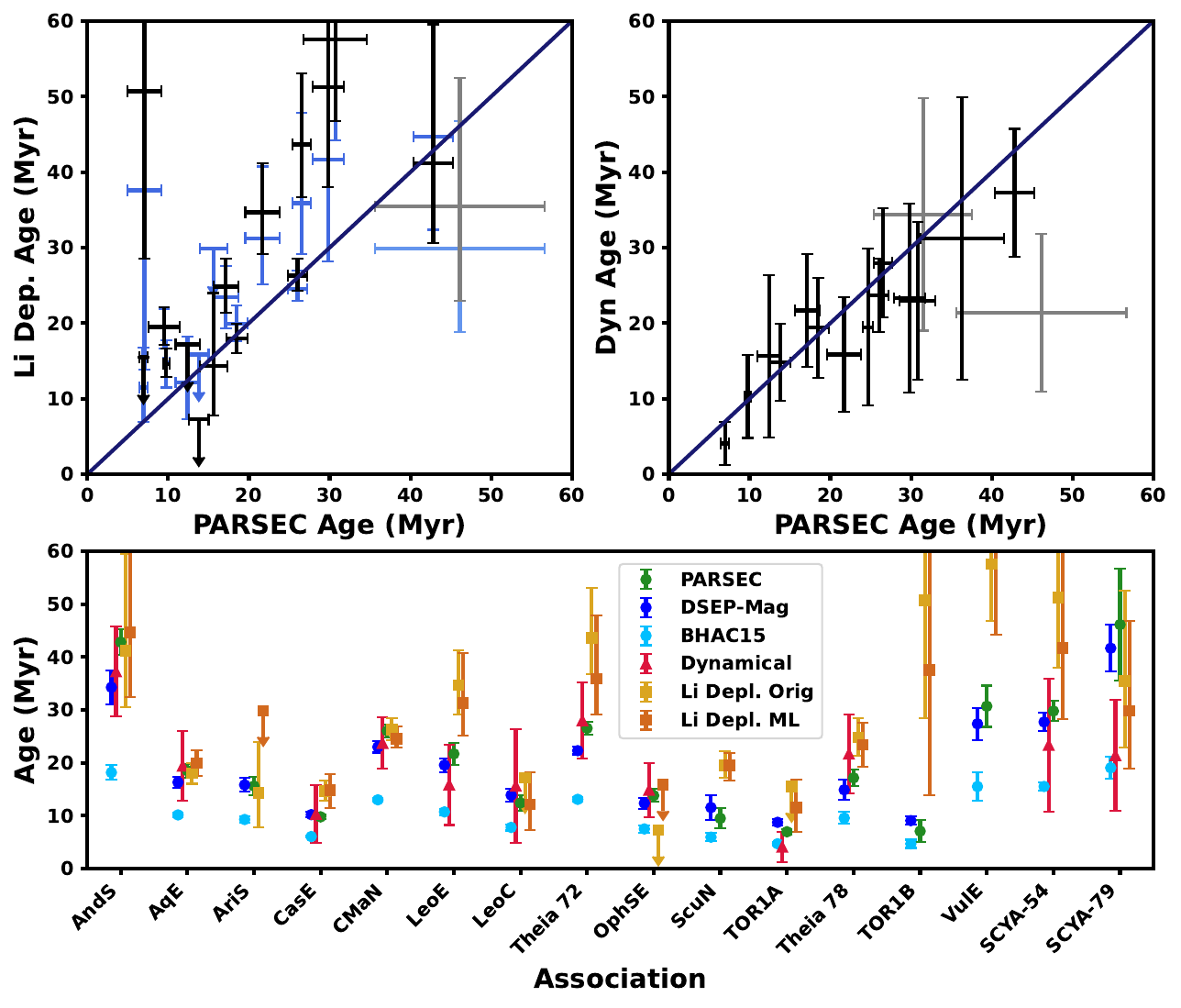}
\caption{In the top panels, PARSEC isochronal ages plotted against the lithium depletion (left) and dynamical (right) ages, with a 1:1 line of equivalence for reference. In the lithium depletion panel, we plot solutions that use the analytical eagles modules in black and the ML solutions in blue. Some lithium depletion ages provide only upper limits, which are plotted accordingly. We mark SCYA-79 with lighter shades, as its isochronal ages are made uncertain by a weak separation from the field. Both the Lithium and Dynamical results show broad agreement with the PARSEC isochronal results. The lithium depletion ages often exceed the PARSEC isochronal ages, however they agree closely for some populations, especially populations with the most complete coverage of their lithium sequences like AqE and CMaN. Nearly all dynamical ages are consistent with the PARSEC isochronal age solutions within uncertainties. In the bottom panel, we show all age solutions for each association, including the three isochrone models, dynamical ages, and lithium depletion ages from the analytical and ML EAGLES modules. There we show that most methods often agree, although the BHAC15 models frequently disagree with non-isochronal methods.}
\label{fig:ageagg}
\end{figure*}

We summarize our age results in Figure \ref{fig:ageagg}, plotting all age solutions for all populations in the bottom panel. In most cases, the BHAC15 isochronal age results are inconsistent with the non-isochronal ages, allowing us to readily dismiss those age solutions. The DSEP-Magnetic and PARSEC ages agree more closely with both each other and with the non-isochronal age methods, however they tend to diverge for older populations like AndS, Theia 72, and CMaN, where the PARSEC ages are consistently higher. It is rare that a well-constrained lithium depletion or dynamical age agrees with one isochronal age solution and not the other, however the older ages produced by PARSEC put that model in marginally better agreement with the lithium and dynamical age solutions for populations like Theia 72, OphSE, and Theia 78. CMaN has among the most complete lithium sequences of any population in our sample and its analytical lithium depletion age agrees closely with PARSEC, but is not within uncertainties of the DSEP-Magnetic age. 

While agreement with other methods may marginally favor the PARSEC over DSEP-Magnetic, the use of PARSEC is perhaps better motivated by the consistent agreement between that model and our pre-main sequences. In Figure \ref{fig:isochronefits}, more luminous stars tend to fall below the best-fit DSEP-Magnetic isochrone by up to 0.5 magnitudes, while PARSEC can fit both the upper and lower PMS simultaneously. Among those higher-mass stars, the best-fit ages produced by the DSEP-magnetic models can be up to double the best-fit age for the association as a whole, while PARSEC shows no such discrepancy. The DSEP-Magnetic ages may therefore be biased old in populations with more high-mass PMS stars, making the age results produced by those models a function of not just the height of the PMS, but also the mass function. We therefore conclude that PARSEC produces the most reliable age scale for our analysis.

In the top two panels of Figure \ref{fig:ageagg}, we compare the PARSEC isochronal age results to the dynamical and lithium depletion ages, with a 1:1 trend for reference. The lithium depletion ages often skew high, but many of the populations above the line of equivalence have very few Li measurements in the most informative part of the Li sequence, like in VulE and SCYA-54. Others have outliers that, if removed, would substantially change their age solutions, such as in TOR1B, where one star with a low Li EW only marginally passes our outlier cut in Section \ref{sec:liages}. The age solutions in AqE and CMaN, which have some of the most complete Li sequences of any populations in our sample, agree closely with their corresponding PARSEC age solutions. 

The ML module, which is designed to better capture natural scatter in these measurements, produces results that align more closely with PARSEC in the populations with high analytical Li ages relative to PARSEC. The apparent high bias in the Li age fits may therefore be driven by coverage that is too incomplete to reject unrealistically old age solutions. Isochrones that do not account for magnetically-driven radius inflation have been shown to produce age results far younger than suggested by Lithium depletion \citep{Jeffries17, Franciosini22}, and the BHAC15 isochrone fits appear to exemplify that effect. The PARSEC and DSEP-Magnetic fits, however, both include corrections that modify the mass-radius relationship \citep{Chen14, Feiden16}, and our results using those models show no conclusive evidence of systematic age underestimation relative to the Lithium Depletion age.

All dynamical ages are consistent with our PARSEC ages within uncertainties, except in SCYA-79, where the PARSEC ages are poorly constrained. There is no clear evidence of an offset between the dynamical and isochronal ages like what has previously been used to indicate a gas dispersal timescale \citep[e.g.,][]{Kerr22a, MiretRoig24}. We discuss this point further in Section \ref{sec:discussion}. 






Due to the agreement between our PARSEC ages and these other methods, we adopt the PARSEC isochronal age for most of our sample. We further discuss this choice in Section \ref{sec:discussion}. SCYA-79 has a strong Li sequence consistent with a 36 Myr age but lacks a clearly separable field sequence, which may be caused by uncorrected extinction in the crowded region of northern Cygnus where it resides. We therefore adopt only an upper limit of $60$ Myr for SCYA-79 and its subgroups, which is supported by the dynamical and lithium depletion ages. This entanglement with the field makes demographic studies unreliable, so we exclude it from Section \ref{sec:demographics}.

\subsection{Demographics} \label{sec:demographics}

All of the populations we explore in this paper have yet to have their sizes, masses, and total membership comprehensively investigated. \citetalias{Kerr23}'s candidate member lists provided the first look at sizes of many of these associations, while some larger populations have also been covered by the \citep{Kounkel19} membership lists. However, these publications did not address all potential sources of bias. Most of these populations represent an understudied demographic of low-mass, isolated populations that has yet to be explored, so detailed demographic work is essential to identify differences and similarities between these populations and the much larger populations that dominate the current literature on young associations. 

\subsubsection{Stellar Masses} \label{sec:mass}

We compute the masses of individual stars by comparing the stellar photometry to the PARSEC v1.2S isomass track grid from \citetalias{Kerr24}, which uses mass sampling every 0.005 $M_{\odot}$ for $0.09 M_{\odot}<M< 1 M_{\odot}$, every 0.01 $M_{\odot}$ between $1.0 M_{\odot}<M< 2.0 M_{\odot}$, every 0.02 $M_{\odot}$ for $2 M_{\odot}<M< 4 M_{\odot}$, and every 0.05 $M_{\odot}$ for $4 M_{\odot}<M< 20 M_{\odot}$. We set the mass to that of the nearest isomass track. Most populations have a maximum stellar mass between 2 and 4 $M_{\odot}$, with no candidates more massive than 4 $M_{\odot}$. These populations therefore lack stars that could constrain the main sequence turn-off ages, which is not surprising given their small total masses \citep[e.g.,][]{Chabrier05}.

\subsubsection{Correction for Low-Mass Stars} \label{sec:IMFcorr}

Our \textit{Gaia}-based stellar samples include most stars in the $M>0.09 M_{\odot}$ mass range covered by the PARSEC isochrones. However, stars near \textit{Gaia}'s $G=21$ magnitude limit often lack high-quality astrometry or photometry, resulting in a lower completeness rate at low masses, especially for more distant populations. We must therefore correct our stellar samples for \textit{Gaia} completeness to accurately estimate their demographics

\begin{figure*}
\centering
\begin{tabular}{ccccc}
    \includegraphics[width=3.3cm]{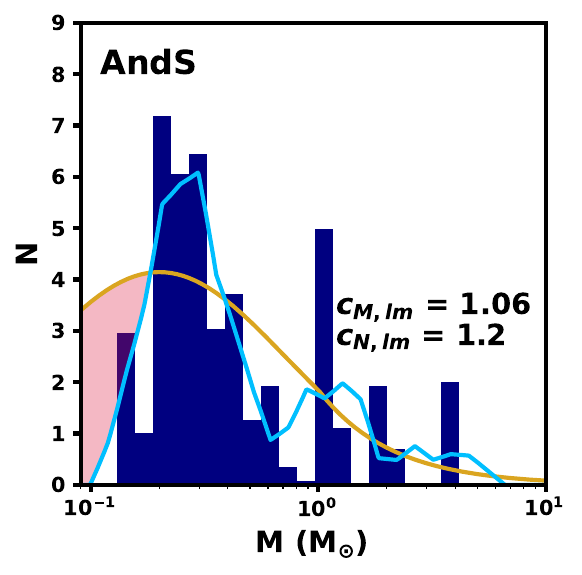}
    & \includegraphics[width=3.4cm]{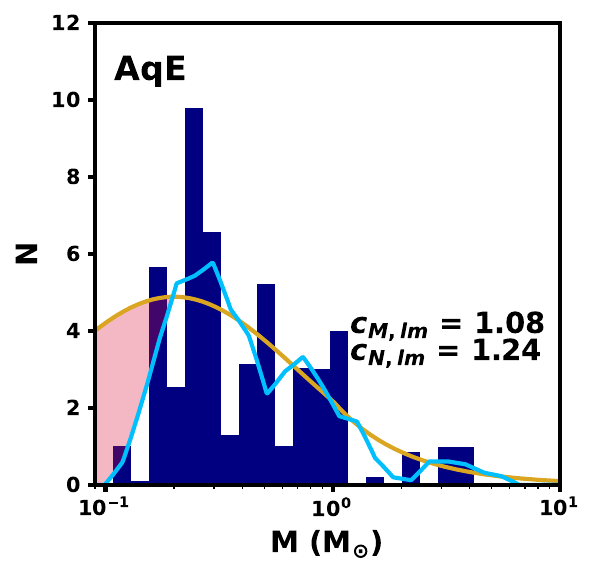}
    & \includegraphics[width=3.3cm]{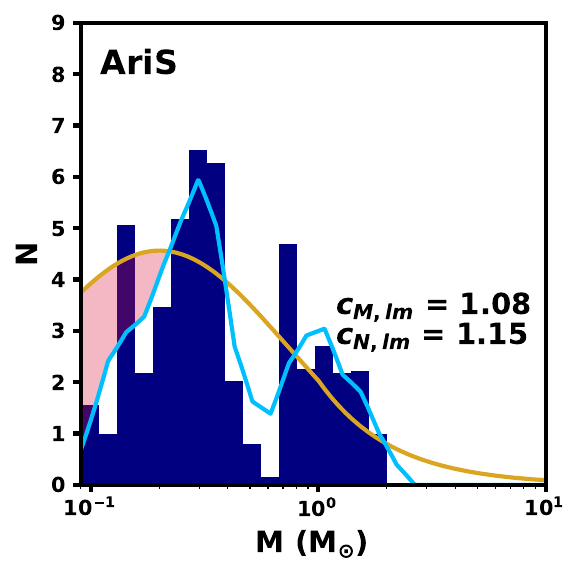}
    & \includegraphics[width=3.3cm]{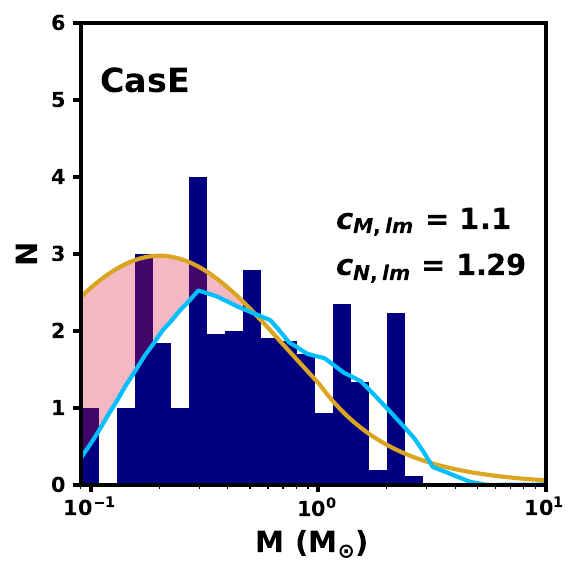}
    & \includegraphics[width=3.5cm]{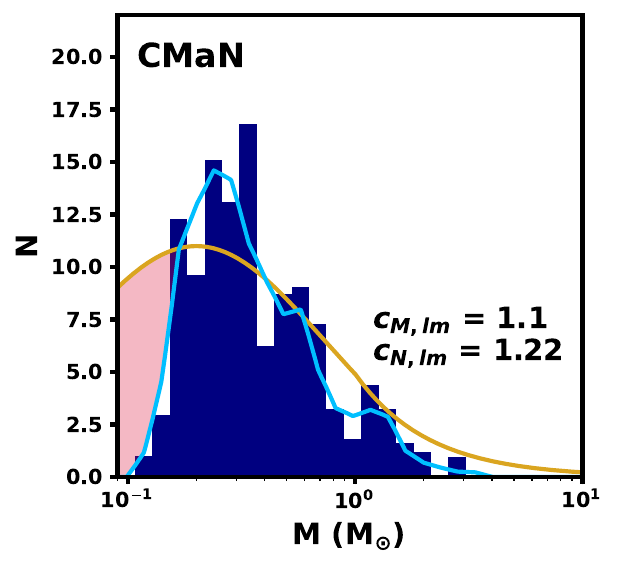}\\[-4pt]
    \includegraphics[width=3.3cm]{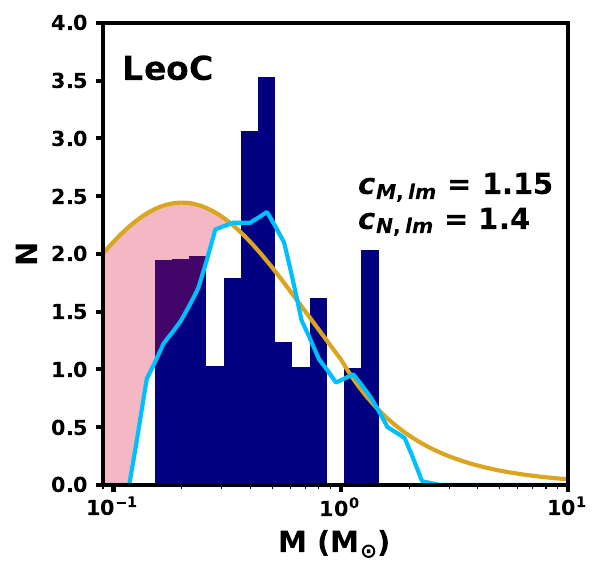}
    & \includegraphics[width=3.3cm]{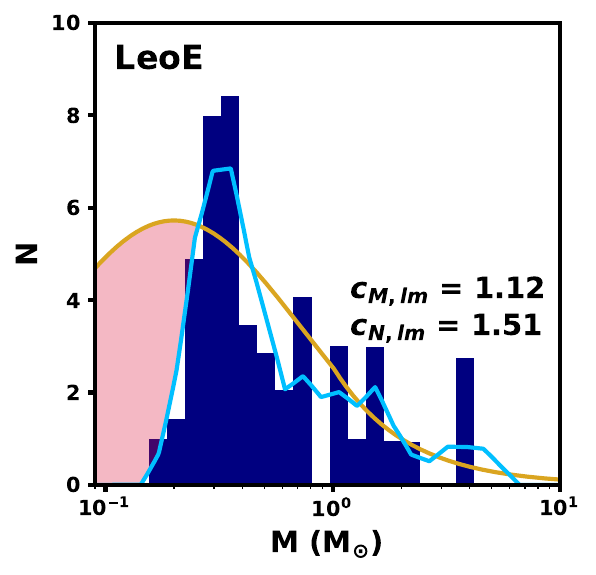}
    & \includegraphics[width=3.3cm]{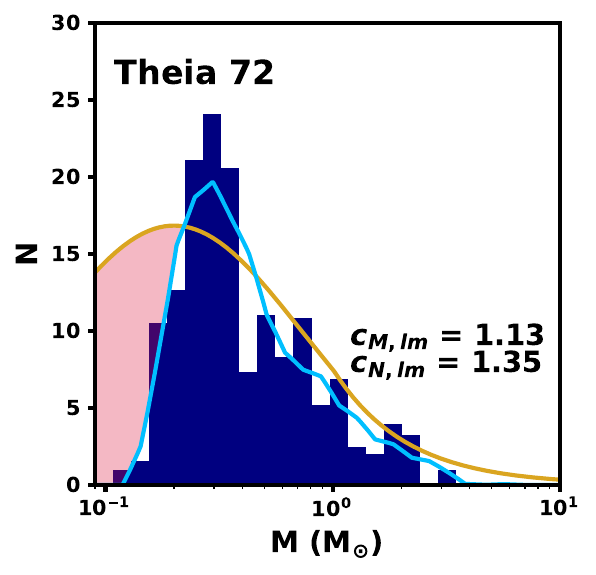}
    & \includegraphics[width=3.3cm]{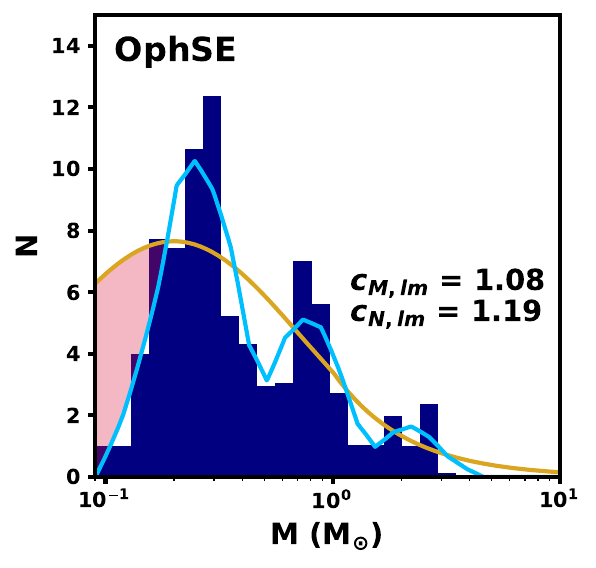}
    & \includegraphics[width=3.3cm]{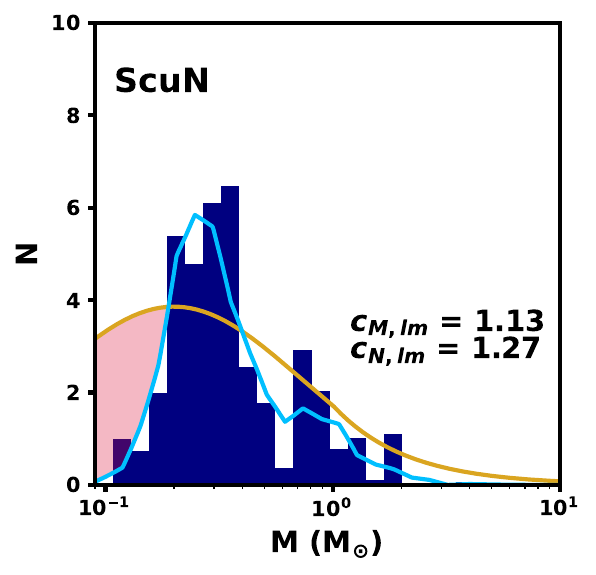}\\[-4pt]
    \includegraphics[width=3.3cm]{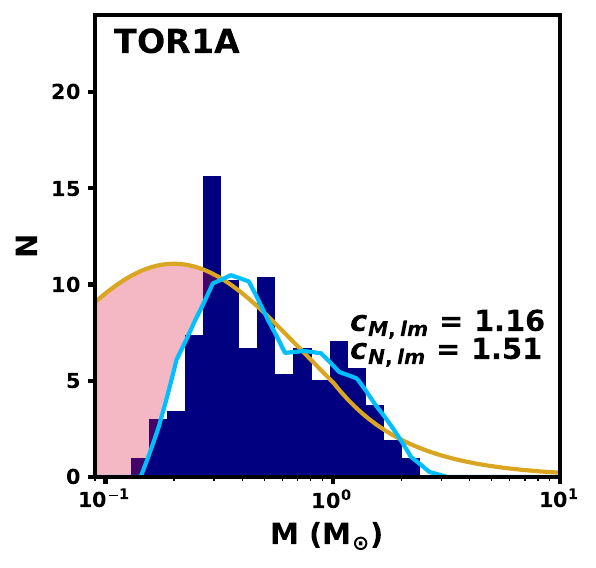}
    & \includegraphics[width=3.3cm]{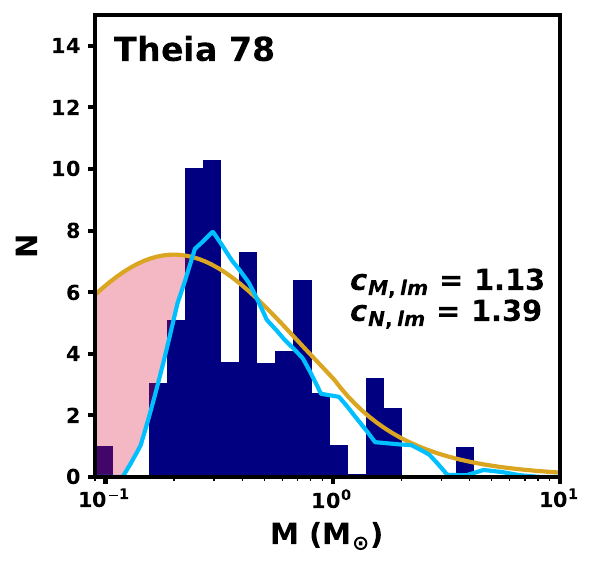}
    & \includegraphics[width=3.3cm]{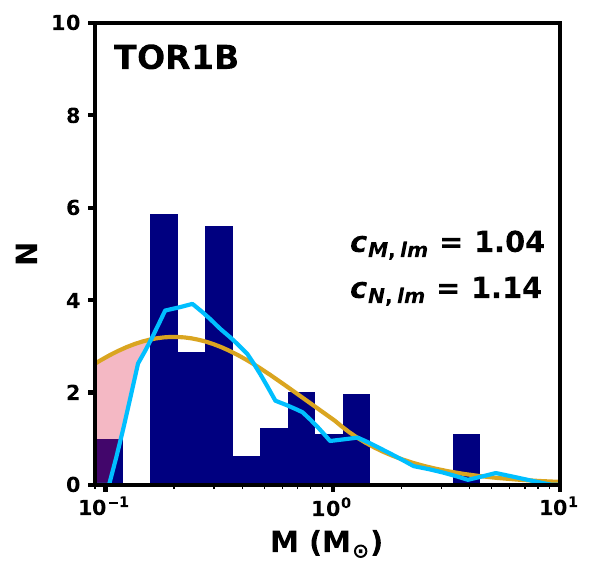}
    & \includegraphics[width=3.3cm]{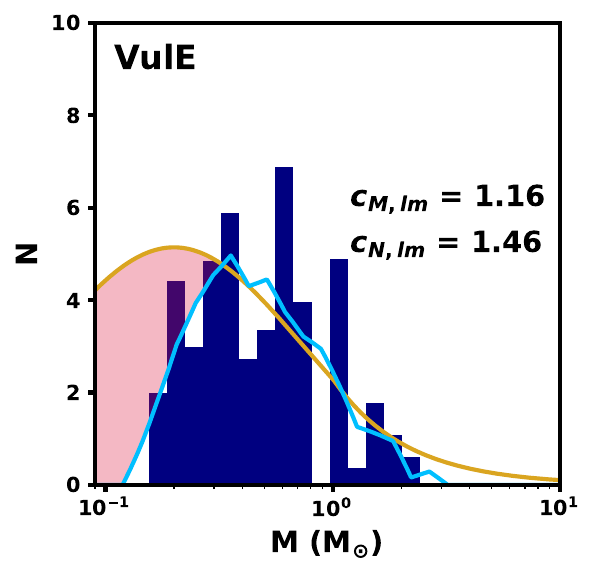}
    & \includegraphics[width=3.3cm]{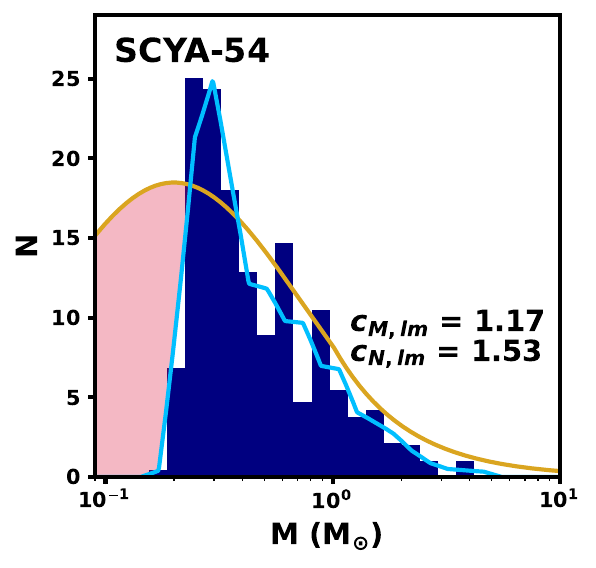}\\[-4pt]
\end{tabular}
\caption{IMF-based mass corrections to all 15 associations where demographics are possible. The dark blue histogram shows the distribution of association members, and a smoothed version of that is shown in light blue. The orange curve shows the \citet{Chabrier05} IMF. The missing mass, shown in red, is computed using the volume between the \citet{Chabrier05} IMF and the smoothed version of the stellar mass distribution curve. This is used to compute the corrective factors for mass and number ($C_{M,lm}$ and $C_{N,lm}$, respectively), which are annotated on each panel.}
\label{fig:IMFcorr}
\end{figure*}

We show a histogram of stars that pass the astrometric and photometric quality flags in Fig. \ref{fig:IMFcorr}, with bins evenly log-sampled between 0.09 and 15 M$_{\odot}$, and with each object weighed by $P_{fin}$. We smooth the result with a savitsky-golay filter to produce a mass frequency curve, which we also show in Fig. \ref{fig:IMFcorr}, alongside the \citet{Chabrier05} IMF scaled to best fit the mass frequency curve for $M > 0.25 M_{\odot}$ via least-squares optimization. Our mass distributions generally follow the \citet{Chabrier05} IMF, although there is some variation, most notably a stellar deficit at $M \sim 0.5$ and an excess at $M \sim 0.25$. This pattern has been shown in several prior publications, and is usually attributed to model innacuracies that displace some stars to lower masses \citep[e.g.,][]{Kraus14, Kerr24}.

All mass frequency curves show fewer stars in the low-mass regime compared to the \citet{Chabrier05} IMF, indicating a drop in \textit{Gaia} completeness there. We take the difference between the mass distribution curve and the scaled IMF to the left of the leftmost crossover point in Figure \ref{fig:IMFcorr} as the volume missed by \textit{Gaia}. The corrective factor to number of members is the sum across the entire stellar histogram, plus the sum over the area to the left of the crossover point, divided by the sum over the entire stellar histogram. We limit this calculation to stars with masses $M > 0.09 M_{\odot}$, or above the PARSEC lower mass limit, which is used as the mass cutoff for inclusion in our stellar sample. The mass corrective factor is computed the same way, but with each histogram and scaled IMF multiplied by the mass at each bin element. We did not bound this calculation in mass, as low-mass and sub-stellar objects can still contribute to the virial state of a stellar population. The resulting corrective factors range from 1.06 to 1.18 in mass and from 1.14 to 1.53 in number. The size of these corrections correlate with distance, with four of five populations with a mass correction less than 1.1 having $d<220$pc. SCYA-54, which contains the most distant subgroup we cover here, has the largest correction in both mass and number.

\subsubsection{High-mass stars} \label{sec:highmasscorr}

Some of our mass histograms in Fig. \ref{fig:IMFcorr} show a deficit of stars with $M\gtrsim2 M_{\odot}$, which may be caused by a lack of quality \textit{Gaia} astrometry and photometry for stars that saturate ($G \la 3$) \citep{GaiaDR1}. However, Hipparcos provides complementary coverage that is essentially complete for stars with $G \la 7$ \citep{Hipparcos97, Hipparcos07}. We therefore search each population for Hipparcos stars missed by \textit{Gaia}. We mark any Hipparcos star as a potential member if its distance to the tenth nearest association member in \citetalias{Kerr23}'s space-transverse velocity coordinate space ($d_{10}$) is less than that of the member with the greatest $d_{10}$, following previous SPYGLASS publications. We only consider stars without \textit{Gaia} astrometry to exclude stars that have already had their membership assessed.

Across all 16 populations, we identify only 5 credible additional members in the Hipparcos catalog across 4 of our associations. Only 3 stars: Mirzam ($\beta$ CMa) and HIP 32848 in CMaN and HIP 4285 in SCYA-54 have masses in the $M \ga 2 M_{\odot}$ range where we see a stellar deficit in Fig. \ref{fig:IMFcorr}. We therefore conclude that there is no evidence of our \textit{Gaia}-based sample systematically excluding high-mass stars, and choose not to apply a broad mass correction. It is possible that high-mass stars are intrinsically more difficult to detect, perhaps due to an increased velocity dispersion produced by the higher binarity rates of high-mass stars \citep{Sullivan21}. Follow-up work will therefore be necessary to determine whether some populations truly have mass distributions that differ from a standard IMF. 

\subsubsection{Binaries} \label{sec:binaries}

Binaries are common, and many of them are unresolved. Unresolved companions get hidden in the light of the primary, resulting in mass estimates that most closely reflect the primary rather than the entire system. We must therefore correct for the mass contribution of these unresolved companions. 

We estimate the mass contribution from unresolved companions by applying literature binary statistics to our detected primaries. We first identify resolved binaries to avoid double-counting companions when applying binary statistics. Following \citetalias{Kerr24}, we search a radius of $10^4$ AU on-sky around each star in \textit{Gaia}. We mark stars with $\Delta v_T < 3$ km s$^{-1}$ and $\frac{\Delta\pi}{\pi} < 0.2$ as in the same stellar system, with the brightest member marked as the primary, and dimmer members considered companions. These conditions roughly encapsulate the full range of expected velocities for stars near \textit{Gaia}'s roughly 1 arcsecond separation limit, given typical uncertainties \citep{rizzuto_zodiacal_2018}. Unresolved companions move the photometry of a star largely perpendicular to the pre-main sequence, changing its mass estimate relatively little. For measuring demographics, probable unresolved binaries with RUWE$>1.2$ can therefore be treated as if they are system primaries without substantially increasing the systematic uncertainties in our association mass estimates. 

We follow the methods applied in \citetalias{Kerr25} to estimate the mass in binaries, sampling $10^4$ stars from the \citet{Chabrier05} IMF at random for $0.09<M<4 ~M_{\odot}$, and computing the expected mass and number of companions for each sampled star according to the mean mass ratios from \citet{Sullivan21} and the companion fractions from \citet{Sullivan21} and \citet{Duchene13}. The expected binary contribution computed from this sampling increases population mass by a factor of 1.23, and increases number of stars by a factor of 1.39. These values produce an average correction to the contribution from binaries for any given stellar primary, including low-mass stars that we correct for in Section \ref{sec:IMFcorr}.

\begin{deluxetable*}{ccccccccccccccccccc}
\tablecolumns{19}
\tablewidth{0pt}
\tabletypesize{\scriptsize}
\tablecaption{Demographics and mean properties of the 16 associations covered in this paper. Where regions host substructure, we show both properties for the entire population and the properties of any subgroups.}
\label{tab:demographics}
\tablehead{
\colhead{ASSOC\tablenotemark{a}} &
\colhead{SG} &
\colhead{N\tablenotemark{b}} &
\colhead{M\tablenotemark{b}} &
\colhead{RA} &
\colhead{Dec} &
\colhead{l} &
\colhead{b} &     
\colhead{d} &
\colhead{$\overline{\mu_{RA}}$} &
\colhead{$\overline{\mu_{Dec}}$} &
\colhead{$\overline{v_{T,l}}$} &  
\colhead{$\overline{v_{T,b}}$} &  
\colhead{$\overline{v_r}$} &  
\colhead{$R_{hm}$} &  
\colhead{$\sigma_{1D}$} &  
\colhead{$\sigma_{vir}$} &
\multicolumn{2}{c}{Vir. Rat.}  \\
\colhead{} &
\colhead{} &
\colhead{} &
\colhead{(M$_{\odot}$)} &
\multicolumn{2}{c}{(deg)} &
\multicolumn{2}{c}{(deg)} &
\colhead{(pc)} &
\multicolumn{2}{c}{(mas yr$^{-1}$)} &
\multicolumn{3}{c}{(km s$^{-1}$)} &
\colhead{(pc)} &
\multicolumn{2}{c}{(km s$^{-1}$)} &
\colhead{val}&
\colhead{err}}
\startdata
AndS &  & 73 & 37.0 & 6.4 & 30.4 & 116.3 & -32.0 & 169 & 14.5 & -3.1 & 11.2 & -3.9 & 8.7 & 11.1 & 0.222 & 0.053 & 2.9 & 1.3 \\
AqE &  & 80 & 33.1 & 297.5 & -8.7 & 31.7 & -16.8 & 135 & 8.3 & -26.0 & -12.6 & -12.0 & -7.8 & 12.2 & 0.245 & 0.048 & 3.6 & 1.8 \\
AriS &  & 70 & 30.7 & 43.5 & 17.6 & 159.8 & -35.9 & 142 & 12.1 & -7.2 & 9.4 & 0.5 & 10.7 & 18.0 & 0.782 & 0.038 & 14.4 & 10.5 \\
CasE &  & 53 & 27.5 & 32.6 & 65.6 & 130.9 & 4.1 & 289 & 10.3 & -8.2 & 16.9 & -6.4 & -4.2 & 21.8 & 0.704 & 0.033 & 15.1 & 9.9 \\
CMaN &  & 191 & 73.0 & 102.1 & -15.1 & 226.1 & -7.5 & 183 & -0.1 & -5.3 & 4.0 & -2.1 & 24.6 & 8.6 & 0.209 & 0.086 & 1.7 & 0.5 \\
LeoE &  & 100 & 51.8 & 173.5 & 17.2 & 238.0 & 69.5 & 308 & -9.2 & -1.5 & -5.3 & -12.5 & 13.6 & 6.9 & 0.323 & 0.08 & 2.8 & 1.2 \\
LeoC &  & 43 & 16.2 & 157.8 & 14.5 & 227.0 & 55.1 & 271 & -9.4 & -1.9 & -3.3 & -11.8 & 16.8 & 14.2 & 0.246 & 0.031 & 5.5 & 1.9 \\
Theia72 &  & 273 & 107.8 & 102.2 & -10.2 & 221.7 & -5.2 & 278 & -1.1 & -4.5 & 4.6 & -3.9 & 28.6 & 12.0 & 0.33 & 0.088 & 2.7 & 0.8 \\
Theia72 & 0 & 68 & 26.4 & 104.6 & -12.9 & 225.3 & -4.4 & 265 & -1.7 & -4.1 & 3.7 & -4.2 & 28.3 & 5.7 & 0.255 & 0.063 & 2.9 & 1.1 \\
Theia72 & 1 & 100 & 34.4 & 101.4 & -9.3 & 220.5 & -5.5 & 298 & -0.9 & -4.3 & 4.9 & -3.8 & 29.2 & 7.7 & 0.289 & 0.062 & 3.3 & 1.5 \\
Theia72 & 2 & 105 & 47.0 & 101.4 & -9.2 & 220.4 & -5.5 & 266 & -0.8 & -4.9 & 5.0 & -3.7 & 28.2 & 5.3 & 0.312 & 0.088 & 2.5 & 1.0 \\
OphSE &  & 127 & 55.0 & 257.7 & -18.4 & 4.5 & 12.4 & 213 & -5.1 & -11.4 & -12.4 & -2.3 & -14.7 & 4.9 & 0.326 & 0.098 & 2.3 & 1.0 \\
ScuN &  & 60 & 22.8 & 278.3 & -8.2 & 23.6 & 0.3 & 205 & 0.3 & -18.7 & -15.9 & -8.6 & -7.5 & 14.0 & 0.81 & 0.037 & 15.3 & 6.1 \\
TOR1A &  & 185 & 79.1 & 66.4 & 13.3 & 182.0 & -24.2 & 311 & 4.8 & -5.7 & 11.0 & 0.3 & 10.8 & 17.5 & 0.475 & 0.062 & 5.4 & 1.7 \\
Theia78 &  & 120 & 49.9 & 80.3 & -4.6 & 206.6 & -22.2 & 320 & 0.2 & 5.2 & -6.8 & 3.9 & 26.3 & 11.2 & 0.309 & 0.062 & 3.5 & 1.6 \\
TOR1B &  & 34 & 16.2 & 66.2 & 15.5 & 180.0 & -22.9 & 290 & 3.3 & -3.9 & 7.1 & 0.2 & 21.0 & 8.2 & 0.43 & 0.041 & 7.3 & 3.0 \\
VulE &  & 89 & 36.1 & 309.2 & 21.6 & 64.7 & -11.5 & 297 & 10.0 & -5.5 & 1.8 & -15.9 & -1.7 & 8.1 & 0.301 & 0.062 & 3.5 & 1.3 \\
SCYA-54 &  & 306 & 119.4 & 16.5 & 51.4 & 125.3 & -11.4 & 319 & 15.5 & -6.9 & 24.0 & -9.0 & -8.8 & 14.4 & 0.414 & 0.084 & 3.5 & 1.5 \\
SCYA-54 & 0 & 177 & 70.1 & 14.3 & 51.5 & 123.9 & -11.3 & 324 & 15.2 & -6.3 & 23.6 & -9.2 & -9.0 & 10.3 & 0.485 & 0.076 & 4.5 & 1.6 \\
SCYA-54 & 1 & 129 & 49.3 & 19.6 & 51.2 & 127.3 & -11.4 & 311 & 16.0 & -7.8 & 24.6 & -8.8 & -8.5 & 6.0 & 0.216 & 0.084 & 1.8 & 1.0 \\
SCYA-79 &  &  &  & 310.7 & 45.1 & 84.3 & 1.7 & 438 & 3.7 & 0.2 & 5.0 & -5.8 & -5.9 & 22.5 & 0.581 &  &  &  \\
SCYA-79 & 0 &  &  & 308.8 & 43.5 & 82.2 & 1.8 & 447 & 3.2 & -0.2 & 3.7 & -5.8 & -4.9 & 11.6 & 0.504 &  &  &  \\
SCYA-79 & 1 &  &  & 312.0 & 46.2 & 85.7 & 1.7 & 431 & 4.0 & 0.5 & 5.8 & -5.8 & -6.1 & 12.7 & 0.309 &  &  &  \\
\enddata
\tablenotetext{a}{Unique choices and properties for specific associations: in CasE, we set stellar membership probabilities below a cut in Fig. \ref{fig:isochronefits} to zero to exclude an elevated field sequence. TOR1A and TOR1B are components of the TOR1 association that have velocities inconsistent with common formation. We provide basic position and velocity data for SCYA-79, but do not report population size, mass, or any values deriving from them due to weak separation with the field. }
\tablenotetext{b}{For our mass and stellar population measurements, we adopt a 10\% systematic uncertainty.}
\vspace*{0.1in}
\end{deluxetable*}

\subsubsection{Final Population Demographics} \label{sec:finaldemographics}

Considering our final stellar membership probabilities in addition to the corrections for binaries and low-mass stars, the membership contribution that is implied by the presence of a given star is provided by the following:
\begin{equation}
    N_x = P_{fin} \times c_{N,BIN} \times c_{N,lm}
\end{equation}

where $c_{N,BIN}$ and $c_{N,lm}$ are the number corrections for binaries and low-mass stars that lack complete \textit{Gaia} coverage, respectively. We exclude objects flagged as companions in a binary or multiple system from this calculation and the mass calculation, as their demographics are included in the binarity correction. The corresponding expected mass is provided by:

\begin{equation}
    M_x = M_{*} \times P_{fin} \times c_{M,BIN} * c_{M,lm}
\end{equation}

where $M_{*}$ is the stellar mass, $c_{M,BIN}$ is the binarity mass correction, and $c_{M,lm}$ is the corrective factor to account for the mass of objects below our detection limit. The total population size $N_{tot}$ and total mass $M_{tot}$ are provided by the sums of $N_x$ and $M_x$, respectively, across all system primaries and stars without a visible companion. We summarize the results in Table \ref{tab:demographics}. The uncertainties in mass and number are largely systematic, and we adopt a 10\% uncertainty following \citet{Kerr24}, which accounts for the typical variation between literature IMF and binarity rates \citep[e.g.,][]{Duchene13, Sullivan21}. 

The populations in our sample have masses ranging from \added{16.2} $M_{\odot}$ and \added{34} expected members in TOR1B to \added{119.4} $M_{\odot}$ and \added{306} expected members in SCYA-54. Of the 15 top-level populations we provide masses for, \added{nine} have \added{$M < 50 M_{\odot}$}, and \added{four} (TOR1B, LeoC, ScuN, and CasE) have \added{$M< 30~M_{\odot}$} ($\lesssim 60$ stars), making them smaller than even populations like the 67-member TW Hydrae association \citep{Luhman23}. These associations are therefore among the smallest ever characterized \citep{Gagne18BXIII}. 

\subsubsection{Virial States} \label{sec:virialstates}

The virial state of stellar populations is often used to differentiate between young associations and open clusters, with the latter containing gravitationally bound populations that may not disperse, potentially skewing the dynamical ages. 
Bound stellar populations satisfy $\sigma_{1D}<\sqrt2\sigma_{virial}$, where $\sigma_{1D}$ is the 1D velocity dispersion, and $\sigma_{virial}$ is the virial velocity, which is defined as
\begin{equation}
    \sigma_{virial} = \left(\frac{G M_{cl}}{r_{hm}\eta}\right)^{1/2} 
\end{equation}
where $M_{cl}$ is the cluster mass, $r_{hm}$ is the half-mass radius, and $\eta$ is a factor related to the mass profile \citep{PortegiesZwart10}. We use the masses computed in Section \ref{sec:finaldemographics} for $M_{cl}$, and set $\eta = 5$, which is a value consistent with the broadest association density profiles \citep{Kuhn19, Wright24}. A small $\eta$ favors high $\sigma_{virial}$ and therefore more bound states, allowing us to test whether any of these sparse populations could be bound given the most favorable assumptions. 

We compute $r_{hm}$ following \citetalias{Kerr24}, which uses $l$ and $b$ galactic sky coordinates to avoid larger uncertainties in the radial direction, and permit the modeling of small-scale dissolving clusters by fitting bi-variate gaussian to the stellar distribution according to a Kernel Density Estimator. We convert the result from sky coordinates to on-sky distance in pc, and convert from 2D to 3D using the \citet{Wolf10} conversion factor for a Gaussian. We use the total mass of the population as $M$ in all cases. In populations with a dense core and broad halo, this assumes that all halo stars were at one point in the core. In the sparse populations like those we investigate here, these compact dissolving clusters are often the only plausibly bound structures \citep[e.g.][]{Kerr22b,Kerr22a}. We compute $\sigma_{1D}$ using the clipped transverse velocity anomalies of stars within $r_{hm}$ with $P_{fin}>0.5$ that have no evidence of binarity. We subsample that dataset, selecting half of the stars at random, compute a 2-$\sigma$ clipped median, and then set $\sigma_{1D}$ and its uncertainty to the mean and standard deviation of the results, de-convolved with the average uncertainties.

The virial ratios are defined as $\sigma_{1D}/(\sqrt2\sigma_{virial})$ and recorded in Table \ref{tab:demographics}. A virial ratio less than 1 is considered bound. We find that none of these populations are likely bound, although the virial ratio of the SCYA-54-1 subgroup is with uncertainties of 1, making it plausibly bound. The dynamical age of that subgroup is, however, consistent with its isochronal age, so there is no evidence for this potential boundedness affecting dispersal. 

\section{Discussion} \label{sec:discussion}

Our adopted age measurements use the PARSEC isochronal models, which produce ages broadly consistent with our dynamical and lithium depletion age solutions. However, the uncertainties in the non-isochronal ages, systematic and otherwise, leave the age scale weakly constrained in absolute terms. In our stellar model-independent dynamical ages, which have the largest uncertainties of any method we employ, that uncertainty is partially intrinsic, however the initial size of the population also contributes to the uncertainty, as populations that start out larger produce larger dynamical age biases (see Sec \ref{sec:dynagecalc}). While small star-forming clouds may be assumed for some small associations we cover, the $6.9\pm0.5$ Myr old TOR1A has \added{$r_{hm} = 17.5$} pc, the \added{fourth} largest in our entire sample. With a velocity dispersion of \added{$\sigma_{1D} = 0.48$ pc Myr$^{-1}$} and a dynamical age of only $4.3\pm2.9$ Myr, stars in this population would need to be nearly an order of magnitude older than any estimate of its age for stellar dispersal to produce its large size, indicating that its scale is primarily primordial. The existence of such a large population containing only \added{79} $M_{\odot}$ of stellar mass indicates that a compact initial configuration cannot be relied upon, even in low-mass associations. This motivates future dynamical studies of near-newborn ($<10$ Myr) young associations to determine whether the initial scale correlates with any dynamical or morphological patterns that persist after formation. 

Unlike many recent studies, our dynamical ages lack a clear offset relative to the isochronal ages, which has previously been viewed as a gas dispersal timescale \citep{MiretRoig24}. The PARSEC ages are only 1.6 $\pm$ 2.3 Myr older than the dynamical ages on average, and the fact that this value is positive is in large part driven by the older populations in our sample, where PARSEC ages are more consistently older. However, older populations require more precise measurements to achieve the same traceback accuracy at formation, and as such, they are more vulnerable to poor velocity measurements, which produce near-zero times of closest approach and bias the dynamical age young. While we use cuts to mitigate this issue in Section \ref{sec:dynagecalc}, it is difficult to remove all bad stars, especially in older populations with less complete datasets. This may explain the lower dynamical ages relative to PARSEC among older populations without the need for any systematic offset. The dynamical age bias corrections inspired by \citet{Couture23} are of a similar order to the offset between the dynamical and adopted ages in \citetalias{Kerr22a}, suggesting that the offset computed there arose from a lack of this correction. However, more robust offsets have been presented in more massive populations like Upper Sco \citep{MiretRoig24}, suggesting that our low-mass populations were either never globally bound, or became unbound during the star formation process. 

Some lithium depletion ages have small uncertainties, but they are highly sensitive to outliers, especially in populations with less complete coverage. This, combined with the uncertain dynamical ages, means that age inaccuracies of order 20\% or larger cannot be ruled out. This motivates future efforts to better anchor the absolute age scale across a wide range of young stellar associations, such as using the Lithium Depletion Boundary (LDB). The LDB is produced by the rapid onset of Li burning in fully convective stars, where nearly all Li is exhausted almost immediately after the onset of lithium burning in the core. More massive stars reach the necessary temperature for Li burning earlier in their evolution, resulting in high Li abundances below the boundary, and low Li abundances above it. The LDB mass at a given age has little dependence on the initial Li abundance and is consistent across different models, making the LDB a gold standard for association ages \citep{Wood23}. However, there remain relatively few LDB ages in the solar neighborhood \citep{Binks14, Binks21, Couture23}, due to the need for medium to high resolution spectra of dim stars. It is therefore impractical to compute LDB ages for all associations, however new LDB ages can anchor the isochronal ages, either by re-scaling the isochronal results to align with the LDB age scale, or by comparing the CMD of a new population to that of populations with robust LDB ages. \citet{Rottensteiner24} produced a suite of new empirical isochrones that can be used for this latter purpose, however the ages used in that work are largely based on the same PARSEC isochrones that we consider here. The combination of empirical isochrones like these with reliable LDB ages could be a powerful tool for age measurement. 

Most of our populations show an under-abundance of stars more massive than 2-4 M$_{\odot}$. While a bottom-heavy IMF may exist in these low-mass associations, \textit{Gaia} surveys of several much more massive populations show a similar pattern. This may indicate that high-mass stars are more difficult to assign to parent associations, likely due to the high binarity rate that often introduces large orbital velocities \citep{Sullivan21}. We therefore cannot rule out the presence of high-mass stars excluded by this survey. These stars may have a significant effect on boundedness, especially in small populations like TOR1B, which is less massive than many individual stars like Mirzam, a CMaN candidate member \citep{Fossati15}. A survey for missing high-mass stars will be necessary to confirm our total population masses and determine whether our populations follow a standard \citet{Salpeter55} IMF in the high-mass regime. 

Many questions remain about the origins of these populations, such as whether they truly form in isolation, or out of leftover material from larger star-forming events. We address these issues in our companion paper, \citetalias{Kerr25c}, where we trace all associations back to their site at formation, identify co-natal structures that affect the formation environment, and search for small-scale dynamical patterns that connect to features seen in simulations. 

\section{Conclusion} \label{sec:conclusion}

We have produced the first demographic overview of 15 young associations identified in \citetalias{Kerr23}. Using \textit{Gaia} photometry and astrometry alongside ground-based spectra, we have detected new substructure, and calculated their stellar masses and virial states. We have also produced accurate, self-consistent ages that are supported by isochrones, lithium depletion, and dynamics. Our key findings can be summarized as follows: 

\begin{enumerate}
    \item The stellar masses in these associations range from \added{16.2} $M_{\odot}$ to \added{119.4} $M_\odot$, and the stellar populations range from \added{34} to \added{306} members. This makes these associations among the smallest ever discovered, especially among populations with no clear connection to a larger association.
    \item Theia 72, SCYA-54, SCYA-79, and TOR1 all show detectable substructure. We find that the velocities of TOR1's substructures are inconsistent with a common formation site, leading to the discovery of a new dynamically distinct low-mass association, TOR1B, which is distinct from the TOR1A association that contains most stars originally assigned to TOR1. 
    \item We find that the PARSEC isochronal ages produce an age sequence consistent with the lithium depletion and dynamical ages. These ages range from $6.9\pm0.5$ Myr in TOR1A to $42.8\pm2.4$ Myr in AndS.  
    \item We do not observe a systematic offset between the isochronal and dynamical ages, suggesting that these populations were either never globally bound or became unbound while they were actively forming. 
\end{enumerate}

These results provide the first in-depth look at this emerging demographic of low-mass populations, and follow-up work is underway to better understand their dynamics and relationship to larger young associations. Future work should further test our age results with LDB measurements, which will better anchor the absolute scaling of the PARSEC ages we have used in this work. 

\begin{acknowledgments}
The Dunlap Institute is funded through an endowment established by the David Dunlap family and the University of Toronto. Funding was received by the Heising-Simons Foundation. RMPK acknowledges the use of computational  resources  at  the  Texas  Advanced Computing Center (TACC) at the University of Texas at Austin, which was used for the more computationally intensive operations in this project. RMPK acknowledges the staff at the McDonald Observatory, who made the many observations presented in this paper possible. This research has made use of the SIMBAD database, operated at CDS, Strasbourg, France. This research has made use of the VizieR catalogue access tool, CDS, Strasbourg, France. The original description 
of the VizieR service was published in \citet{Vizier00}. RMPK thanks Rob Jeffries, who provided helpful advice on the use of the EAGLES module. JSS is grateful for support from NSERC Discovery Grant (RGPIN-2023-04849) and a University of Toronto Connaught New Researcher Award. J.C. acknowledges support from the Agencia Nacional de Investigacio\'n y Desarrollo (ANID) via Proyecto Fondecyt Regular 1231345, and by ANID
BASAL project FB210003.

The authors thank the Sloan Digital Sky Survey V for contributing data to this project. Funding for SDSS-V has been provided by the Alfred P. Sloan Foundation, the Heising-Simons Foundation, the National Science Foundation, and the Participating Institutions. SDSS acknowledges support and resources from the Center for High-Performance Computing at the University of Utah. SDSS telescopes are located at Apache Point Observatory, funded by the Astrophysical Research Consortium and operated by New Mexico State University, and at Las Campanas Observatory, operated by the Carnegie Institution for Science. The SDSS web site is \url{www.sdss.org}.

SDSS is managed by the Astrophysical Research Consortium for the Participating Institutions of the SDSS Collaboration, including Caltech, The Carnegie Institution for Science, Chilean National Time Allocation Committee (CNTAC) ratified researchers, The Flatiron Institute, the Gotham Participation Group, Harvard University, Heidelberg University, The Johns Hopkins University, L'Ecole polytechnique f\'{e}d\'{e}rale de Lausanne (EPFL), Leibniz-Institut f\"{u}r Astrophysik Potsdam (AIP), Max-Planck-Institut f\"{u}r Astronomie (MPIA Heidelberg), Max-Planck-Institut f\"{u}r Extraterrestrische Physik (MPE), Nanjing University, National Astronomical Observatories of China (NAOC), New Mexico State University, The Ohio State University, Pennsylvania State University, Smithsonian Astrophysical Observatory, Space Telescope Science Institute (STScI), the Stellar Astrophysics Participation Group, Universidad Nacional Aut\'{o}noma de M\'{e}xico, University of Arizona, University of Colorado Boulder, University of Illinois at Urbana-Champaign, University of Toronto, University of Utah, University of Virginia, Yale University, and Yunnan University.
\end{acknowledgments}

%

\vspace{5mm}
\facilities{Gaia, McDonald Observatory 2.7m Harlan J. Smith Telescope, Sloan Digital Sky Survey BOSS and APOGEE spectrographs at APO and LCO}


\software{astropy \citep{Astropy13, Astropy18, astropy22},  numpy \citep{numpy}, pandas \citep{pandas}, LMFIT \citep{lmfit}
          saphires \citep{Tofflemire19}, matplotlib \citep{Hunter07}}




\bibliography{rkerr_citations}{}
\bibliographystyle{aasjournal}



\end{CJK*}

\end{document}